\documentclass[a4paper,fleqn,usenatbib]{mnras}

\usepackage{newtxtext,newtxmath}
\usepackage{gensymb}

\usepackage[T1]{fontenc}
\usepackage{ae,aecompl}

\usepackage{graphicx}	
\usepackage{amsmath}	
\usepackage{longtable}
\usepackage{arydshln}
\usepackage{booktabs}
\usepackage{tabularx}
\usepackage{supertabular}

\title[Spin--filament alignment and bulge growth]{The SAMI Galaxy Survey: flipping of the spin--filament alignment correlates most strongly with growth of the bulge}

\author[S.\ Barsanti et al.]
{Stefania Barsanti$^{1,2}$, \thanks{E-mail: stefania.barsanti@anu.edu.au}
Matthew Colless,$^{1,2}$
Charlotte Welker,$^{3}$
Sree Oh,$^{1,2}$
Sarah Casura,$^{4}$
\newauthor
Julia J. Bryant,$^{2,5,6}$ 
Scott M. Croom,$^{2,5}$ 
Francesco D'Eugenio,$^{7}$
Jon S. Lawrence,$^{8}$ 
\newauthor
Samuel N. Richards,$^{5}$  
Jesse van de Sande$^{2,5}$
\\
\\
$^{1}${Research School of Astronomy and Astrophysics, Australian National University, Canberra, ACT 2611, Australia}\\
$^{2}${ARC Centre of Excellence for All Sky Astrophysics in 3 Dimensions (ASTRO 3D), Australia}\\
$^{3}${Department of Physics \& Astronomy, Johns Hopkins University, Baltimore, MD 21218, USA}\\
$^{4}${Hamburger Sternwarte, Universit\"{a}t Hamburg, Gojenbergsweg 112, 21029 Hamburg, Germany}\\
$^{5}${Sydney Institute for Astronomy (SIfA), School of Physics, The University of Sydney, NSW 2006, Australia}\\
$^{6}${AAO-USydney, School of Physics, University of Sydney, NSW 2006, Australia}\\
$^{7}${Cavendish Laboratory and Kavli Institute for Cosmology, University of Cambridge, Madingley Rise, Cambridge, CB3 0HA, United Kingdom}\\
$^{8}${AAO-Macquarie, Macquarie University, NSW 2109, Australia}\\ 
}

\date{Accepted XXX. Received YYY; in original form ZZZ}

\pubyear{2022}

\begin{document}
\label{firstpage}
\pagerange{\pageref{firstpage}--\pageref{lastpage}}
\maketitle

\begin{abstract}
We study the alignments of galaxy spin axes with respect to cosmic web filaments as a function of various properties of the galaxies and their constituent bulges and discs. We exploit the SAMI Galaxy Survey to identify 3D spin axes from spatially-resolved stellar kinematics and to decompose the galaxy into the kinematic bulge and disc components. The GAMA survey is used to reconstruct the cosmic filaments. The mass of the bulge, defined as the product of stellar mass and bulge-to-total flux ratio $M_{\rm bulge}=M_\star\times({\rm B/T})$, is the primary parameter of correlation with spin--filament alignments: galaxies with lower bulge masses tend to have their spins parallel to the closest filament, while galaxies with higher bulge masses are more perpendicularly aligned. $M_\star$ and B/T separately show correlations, but they do not fully unravel spin--filament alignments. Other galaxy properties, such as visual morphology, stellar age, star formation activity, kinematic parameters and local environment, are secondary tracers. Focussing on S0 galaxies, we find preferentially perpendicular alignments, with the signal dominated by high-mass S0 galaxies. Studying bulge and disc spin--filament alignments separately reveals additional information about the formation pathways of the corresponding galaxies: bulges tend to have more perpendicular alignments, while discs show different tendencies according to their kinematic features and the mass of the associated bulge. The observed correlation between the flipping of spin--filament alignments and the growth of the bulge can be explained by mergers, which drive both alignment flips and bulge formation. 
\end{abstract}




\section{Introduction}
\label{Introduction}

How galaxies acquire their angular momentum in the cosmic web is a crucial element in understanding galaxy formation and evolution. Since galaxies are not randomly distributed in the Universe but found along ordered filaments and walls, their properties are expected to be influenced by their host halos, and by the current location and past history of these halos in the evolving cosmic web. According to the tidal torque theory, galaxy spin is generated by the torques acting on the collapsing proto-halo, which gains angular momentum from the gravitational perturbations in the tidal field \citep{Hoyle1951,Peebles1969,Doroshkevich1970,White1984,Porciani2002,Schafer2009}. Thus, a correlation between the galaxy spin and the tidal field should likely exist. The alignment of the galaxy spin axis with respect to the orientation of the filament within which it resides represents a memory of the galaxy's formation. Therefore the signal is expected to be weak at low redshift ($z<0.1$; \citealp{Codis2018}).

Cosmological N-body simulations have predicted the orientation of the dark matter halo spin vector to be mass-dependent (\citealp{AragonCalvo2007,Codis2012,Trowland2013,GaneshaiahVeena2018}). The sense of this trend is that low-mass halos tend to have their spin aligned parallel with the closest filament, while the spin axis of high-mass halos tends to be orthogonal to the filament. This trend has also been seen for galaxies in large-scale cosmological hydrodynamical simulations \citep{Dubois2014,Laigle2015,Codis2018,Wang2018,Kraljic2020}. In the context of galaxy formation mechanisms, this suggests that low-mass galaxies are formed via gas accretion mechanisms, while high-mass galaxies are formed via mergers occurring along the filament within which they are embedded \citep{Dubois2014,Welker2014}. On the other hand, \citet{GaneshaiahVeena2019} found a preferential perpendicular alignment for galaxies at all masses, though they studied a relatively small volume. 

Apart from stellar mass, simulations have found that spin--filament alignments also depend on other galaxy properties, such as morphology \citep{Codis2018,GaneshaiahVeena2019}, colour and magnitude \citep{Tempel2015,Wang2018}, triaxiality \citep{Wang2018}, degree of rotation, star formation activity and HI mass \citep{Kraljic2020}. 

A significant effort has been devoted to detecting the galaxy spin--filament alignments in observations of low redshift galaxies. Most of these studies found preferentially perpendicular orientations for early-type galaxies \citep{Tempel2013a,Pahwa2016,Hirv2017,Chen2019,Kraljic2021}, while late-type galaxies have their spins preferentially parallel to the closest filament \citep{Tempel2013a,Tempel2013b,Hirv2017,BlueBird2020,Kraljic2021,Tudorache2022}. However, some disagreement exists regarding the late-types, with some studies finding a perpendicular orientation or no clear trend for these galaxies \citep{Jones2010,Zhang2015,Pahwa2016,Krolewski2019}. These discrepant findings could plausibly be explained by differences in the selection criteria for the galaxy samples and in the algorithms used to reconstruct the cosmic filaments, combined with the intrinsic weakness of the signal at low redshift.

The advent of integral field spectroscopy (IFS) has made possible a more precise measurement of the galaxy spin axes from spatially-resolved stellar kinematic maps and thus a more statistically significant detection of the spin--filament correlation signal with respect to photometric data. The mass-dependent trend for the galaxy spin--filament alignments was observed for the first time with >2$\sigma$ confidence by \citet{Welker2020}. They selected $\sim$1400 galaxies from the Sydney--AAO Multi-object Integral-field (SAMI) Galaxy Survey \citep{Croom2012,Bryant2015} with available spatially-resolved stellar kinematics and used the Galaxy And Mass Assembly (GAMA; \citealp{Driver2011}) survey to reconstruct the underlying cosmic filaments. The most recent results of \citet{Kraljic2021} for the IFS MaNGA survey \citep{Bundy2015} show that the 3D spin--filament alignment is preferentially parallel for spiral galaxies, with the correlation dominated by low-mass galaxies, and preferentially perpendicular for S0 galaxies. They also find a strong perpendicular alignment for low-mass S0 galaxies, at odds with the expected stellar mass-dependency of the signal. 

Most of these previous results, whether based on simulations or observations, suggest that the growth of a bulge is expected to affect the spin--filament alignment. According to the hierarchical formation scenario of structures in our Universe, galaxies build up their discs via accretion and their bulges via mergers (\citealp{Aguerri2001,Hopkins2010,Wilamn2013}). However, the formation of S0 galaxies, characterised by both bulge and disc components, is still debated, with multiple competing mechanisms depending on the environment and/or galaxy properties thought to be responsible for their origin (\citealp{Dressler1980,ElicheMoral2013,Johnston2014,FraserMcKelvie2018,Coccato2019,Barsanti2021a,Croom2021b}). 

Thus, an interesting question that arises is whether we can detect a correlation between the bulge properties and the spin--filament alignment trends in observations. We aim to address this hypothesis by investigating the signal according to the bulge-to-total flux ratio, as well as by exploring the separate bulge and disc spin--filament alignments. We take advantage of the SAMI Galaxy Survey to identify the spin axes of galaxies, bulges, and discs based on spatially-resolved stellar kinematics, and of the deep and highly-complete GAMA spectroscopic survey to reconstruct the cosmic web. These analyses will help to shed light on the formation mechanisms of galaxies, bulges, and discs.

This paper is organised as follows. We present our galaxy sample, spin proxies, and galaxy properties in Section~\ref{Data and Galaxy Sample}. In Section~\ref{Methods} we describe the methods used to identify the orientation of the spin axes and to reconstruct the cosmic filaments. In Section~\ref{Results} we present our results about the alignments and their correlations with galaxy properties. In Section~\ref{Discussion} we compare our findings to previous studies, and we discuss their physical interpretations. Finally, we summarise our findings and state our conclusions in Section~\ref{Summary and conclusions}. Throughout this work, we assume $\Omega_{m}=0.3$, $\Omega_{\Lambda}=0.7$ and $H_{0}=70$\,km\,s$^{-1}$\,Mpc$^{-1}$ as the cosmological parameters.

\section{Data and Galaxy Sample}
\label{Data and Galaxy Sample}

\subsection{The SAMI Galaxy Survey}
\label{SAMI galaxy survey}

The Sydney--AAO Multi-object Integral-field spectrograph (SAMI) was mounted on the 3.9\,m Anglo-Australian Telescope \citep{Croom2012}. The instrument has 13 fused optical fibre bundles (hexabundles), each containing 61 fibres of 1.6$^{\prime\prime}$ diameter so that each integral field unit (IFU) has a 15$^{\prime\prime}$ diameter \citep{Bland2011,Bryant2014}. The SAMI fibres feed the two arms of the AAOmega spectrograph \citep{Sharp2006}. The SAMI Galaxy Survey uses the 580V grating in the blue arm, giving a resolving power of $R=1812$ and wavelength coverage of 3700--5700\,\AA, and the 1000R grating in the red arm, giving a resolving power of $R=4263$ over the range 6300--7400\,\AA. The median full-width-at-half-maximum values for each arm are FWHM$_{\rm blue}$=2.65\,\AA\ and FWHM$_{\rm red}$=1.61\,\AA\  \citep{vandeSande2017}.

The SAMI Galaxy Survey is a spatially-resolved spectroscopic survey of more than 3000 galaxies with stellar mass range $\log(M_{\star}/M_{\odot})=8$--12 and redshift range $0.004<z\leq 0.115$ \citep{Bryant2015,Croom2021}. Most of the SAMI targets belong to the three equatorial fields GAMA G09, G12 and G15 of the Galaxy And Mass Assembly survey \citep{Driver2011}. Eight massive clusters were also observed to explore high galaxy density environments \citep{Owers2017}, however they are not included in this analysis. The data are reduced using the SAMI {\sc python} package \citep{Allen2014}, which uses the {\tt 2dfdr} package \citep{2015ascl.soft05015A}. A complete description of the data reduction from raw frames to datacubes can be found in  \citet{Sharp2015}, \citet{Allen2015}, \citet{Green2018} and \citet{Scott2018}. The final datacubes are characterised by a grid of 0.5$^{\prime\prime}\times0.5^{\prime\prime}$ spaxels, where the blue and red spectra have pixel scales of 1.05\,\AA\ and 0.60\,\AA\ respectively. 

\subsection{Spin proxies}
\label{Spin proxies}

\subsubsection{Stellar kinematics}
\label{Stellar kinematics}

To identify the spin axis of a galaxy, we take advantage of spatially-resolved stellar kinematics. A complete description of the stellar kinematics for the SAMI Galaxy Survey is presented in \citet{vandeSande2017}. In a nutshell, the line-of-sight velocity distributions are obtained from the Penalised Pixel-Fitting software (pPXF; \citealp{Cappellari2004,Cappellari2017}), where the red spectrum is smoothed with a Gaussian kernel to match the spectral resolution of the blue spectrum, then the combined blue and red spectrum is re-binned on a grid of uniform velocity spacing. For each bin, pPXF is run in a multistep process on each galaxy spaxel to estimate the noise from the fit residual, remove emission lines and extract velocity and velocity dispersion. The best-fitting templates are derived from the MILES library of stellar spectra \citep{SanchezBlazquez2006,FalconBarroso2011}. The stellar kinematic position angles (PA) are measured from the spatially-resolved stellar velocity maps using the {\sc fit\_kinematic\_pa} routine (see Appendix C of \citealp{Krajnovic2006}). Only spaxels with stellar continuum signal-to-noise ratio S/N$>$3 and velocity uncertainty $<$30\,km\,s$^{-1}$ are used in the fitting. 

The galaxies' semi-major axis effective radii ($R_e$) and ellipticities within $R_e$ ($\epsilon_e$) are measured using Multi-Gaussian Expansion (MGE; \citealp{Emsellem1994,Cappellari2002}). For the SAMI Galaxy Survey, the MGE technique is applied to $r$-band SDSS images \citep{York2000}; a detailed presentation of the MGE fits can be found in \cite{DEugenio2021}.

\subsubsection{Stellar kinematics of bulges and discs}
\label{Stellar kinematics of bulges and discs}

In this work we aim to explore the separate spin--filament alignments of bulges and discs. To disentangle the spin axes of the two components, we take advantage of the 2D kinematic bulge/disc decomposition performed by \citet{Oh2020}. They used pPXF to estimate simultaneously the spatially-resolved velocity and velocity dispersion of the bulge and the disc, using photometric weights and a new subroutine for dealing with degeneracy in the solutions. The photometric inputs are based on the 2D photometric bulge/disc decomposition presented in the next Section. \citet{Oh2020} found that the combination of these two components adequately reproduces the major kinematic features of galaxies over a wide range of morphologies.

\begin{figure*}
\includegraphics[width=12cm]{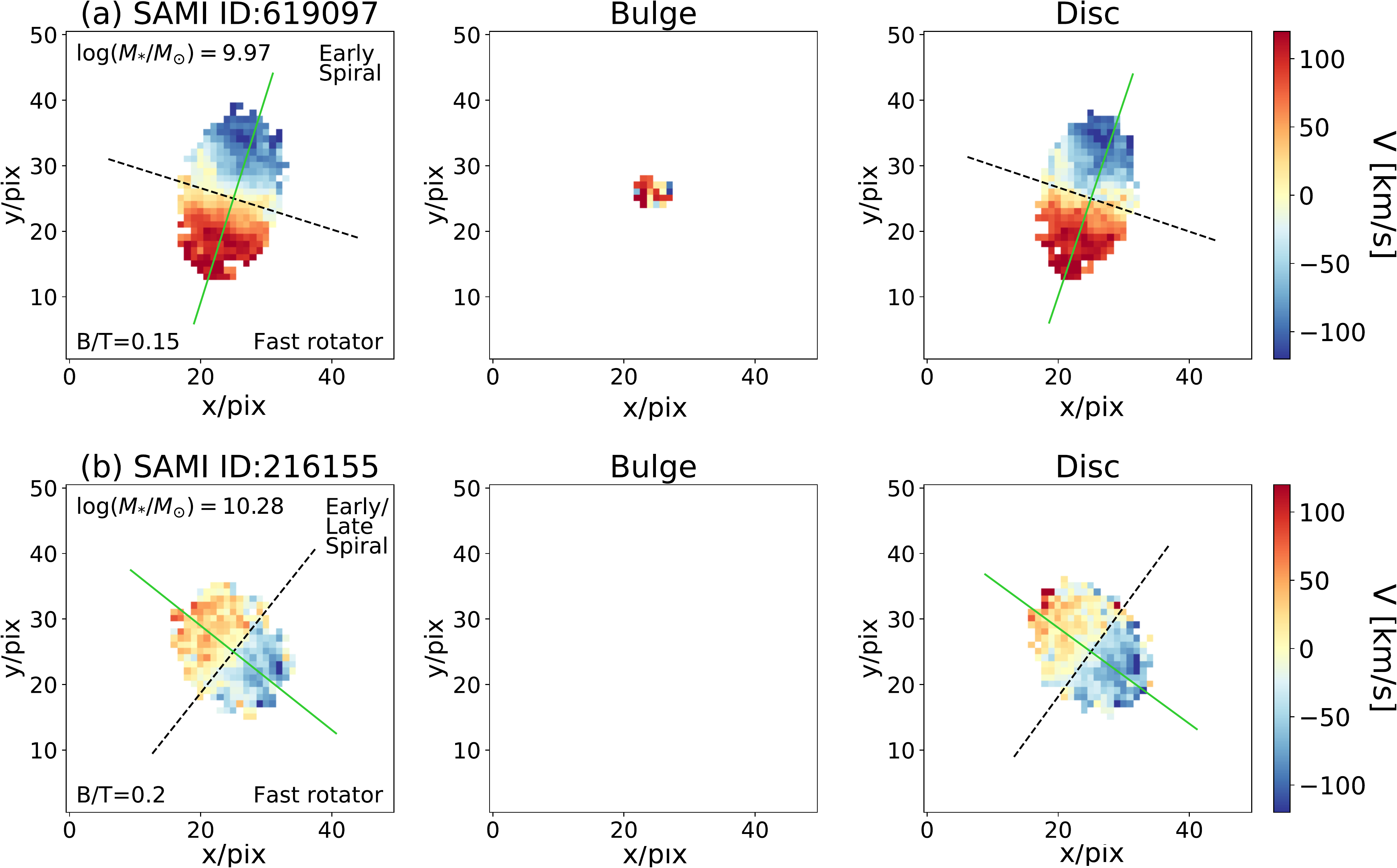}

\vspace{2mm}

\includegraphics[width=12cm]{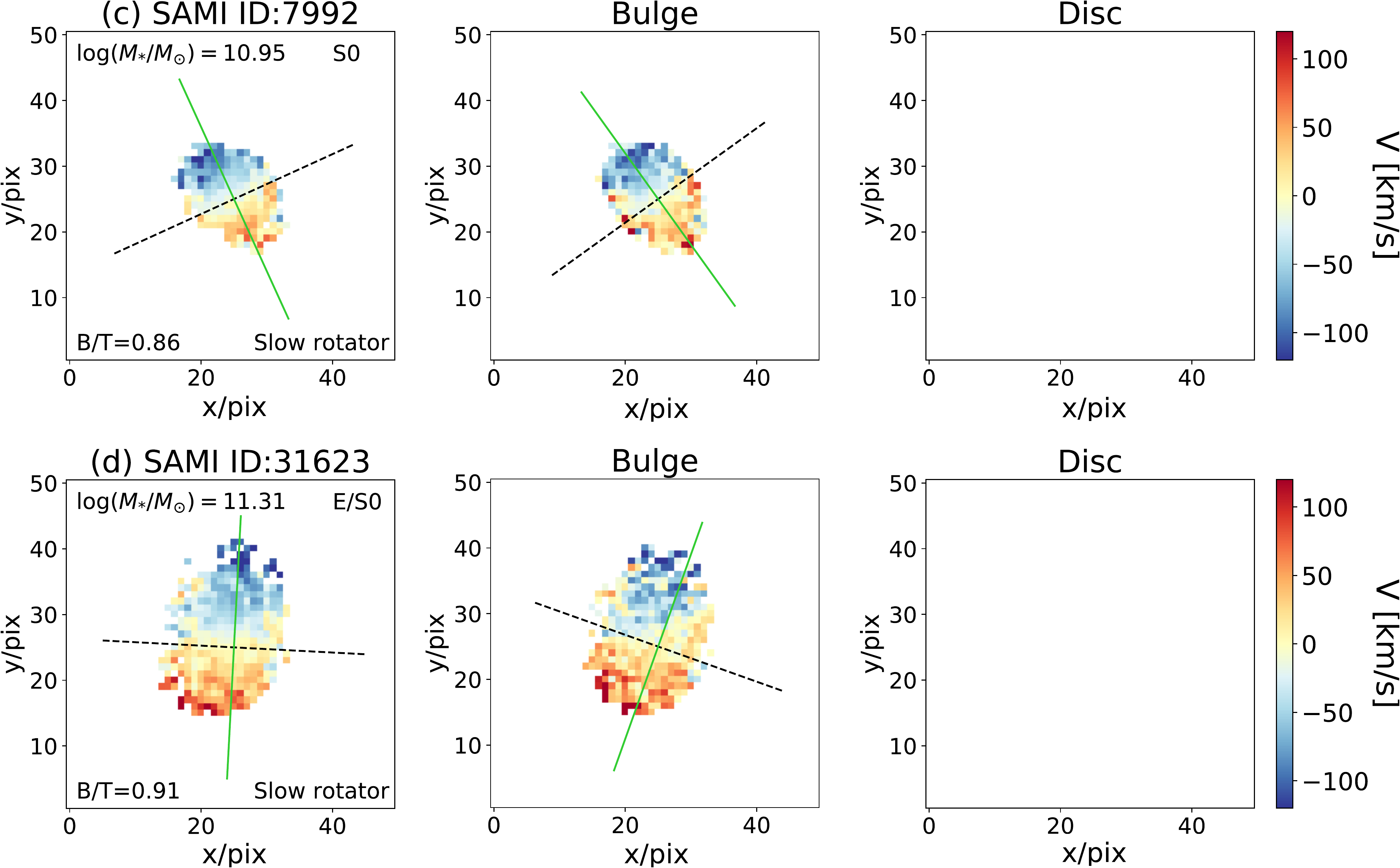}

\vspace{2mm}

\includegraphics[width=12cm]{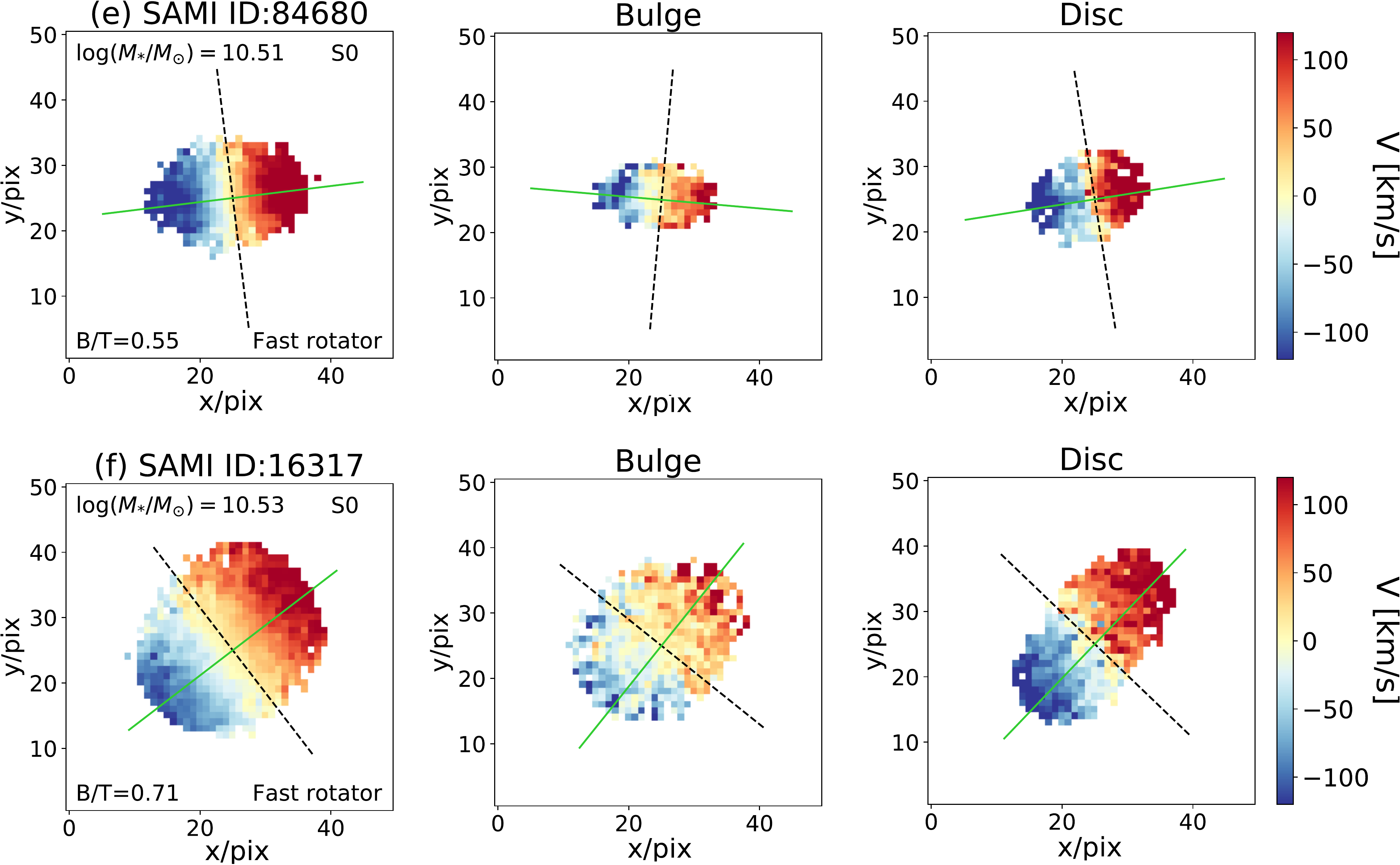}
\caption{Examples of stellar kinematic PAs for galaxies, bulges, and discs. For each galaxy (row), we show its velocity map (left) and the separate velocity maps of the bulge (centre) and disc (right). Only spaxels with S/N$>$3 and velocity uncertainty $<$30\,km\,s$^{-1}$ are shown. The green line shows the kinematic PA and the black dashed line marks the projected axis of rotation. For examples (a) and (b), only the discs have measured kinematic PAs; for examples (c) and (d), only the bulges have measured kinematic PAs; for examples (e) and (f), both components have measured kinematic PAs.}
\label{GalaxyBulgeDisc_Vel_Maps}
\end{figure*}

We estimate the separate kinematic PAs of the bulges and the discs from the corresponding spatially-resolved velocity maps using the {\sc fit\_kinematic\_pa} routine. Only spaxels with continuum S/N$>$3 for the respective component are used in the fitting. Moreover, we select only galaxies with velocity maps where at least 70\% of the spaxels within $1\,R_e$ have S/N$>$3 for the respective component. Figure~\ref{GalaxyBulgeDisc_Vel_Maps} shows some examples of velocity maps for galaxies, bulges and discs, where the green lines mark the kinematic PAs. We exhibit galaxies where only the disc component has measured kinematic PA (examples a \& b), where only the bulge component has measured kinematic PA (examples c \& d), and where measurements are available for both components (examples e \& f). 

\subsubsection{Photometric position angles}
\label{Photometric position angles}

The spatially-resolved kinematic bulge/disc decomposition for the SAMI survey galaxies is based on photometric weights taken from the 2D photometric bulge/disc decomposition of \citet{Casura2022}. This latter allows us to estimate separate photometric properties of bulges and discs: the disc is defined as the
exponential component, while the bulge corresponds to the
S\'ersic component representing the light excess over the exponential component. The photometric decomposition uses the image analysis package {\sc ProFound} and the photometric galaxy profile fitting package {\sc ProFit} \citep{Robotham2017,Robotham2018}. \citet{Casura2022} follow a similar method to that used for the SAMI cluster galaxies in \citet{Barsanti2021}, with some differences that are outlined below. 

The decomposition is performed on the $g$-, $r$-, and $i$-band images from the Kilo-Degree Survey (KiDS; \citealp{deJong2017}). The {\sc ProFound} steps include image segmentation, source identification, sky subtraction, initial parameter estimation and local PSF estimation. Then, each galaxy is fitted using {\sc ProFit} with three models based on S\'ersic profiles \citep{Sersic1963}, and a combination of downhill gradient and full MCMC algorithms: (i)~a single-component S\'ersic model; (ii)~a double-component S\'ersic bulge + exponential disc model; and (iii)~a double-component point source + exponential disc model. The single-component S\'ersic model has seven free parameters: $x$ and $y$ positions of the profile centre, magnitude, effective radius containing half of the flux, S\'ersic index, position angle of the major axis, and axial ratio. These parameters are also left free to vary for the S\'ersic bulge model, while the exponential disc has the S\'ersic index fixed to 1. The point source model is described by $x$ and $y$ coordinates and magnitude. Both double-component models have the positions of the two components tied together.

\citet{Casura2022} identify which of the three models best characterises each galaxy. This galaxy characterisation is based on Deviance Information Criterion cuts, which are calibrated against a visually inspected random sample of 1000 $r$-band objects. We make use of photometric PAs to identify the shape--filament alignments of galaxies, bulges, and discs for a comparison with the findings based on stellar kinematic PAs. The galaxy photometric PAs are derived from the $r$-band single-component S\'ersic profiles, while the bulge and disc photometric PAs are estimated from the $r$-band double-component S\'ersic bulge + exponential disc models.

\subsubsection{Gas kinematics}
\label{Gas kinematics}

We explore whether the kinematic misalignment between the stellar and gas components has an impact on the galaxy spin--filament alignment. The gas kinematic PAs are estimated from the SAMI spatially-resolved H$\alpha$ velocity maps using the {\sc fit\_kinematic\_pa} routine. The H$\alpha$ velocity maps are obtained from the emission-line fitting software {\sc lzifu} \citep{Ho2016}, where the stellar continuum is subtracted using pPXF. Only spaxels with H$\alpha$ S/N$>$5 are used to estimate the gas kinematic PA \citep{Bryant2019}. 

\subsection{Galaxy properties}
\label{Galaxy properties}
Our aim is to explore the galaxy spin--filament alignments in relation to various galaxy properties to understand which show the strongest correlations and therefore may be linked by physical processes. We focus on stellar mass and bulge-to-total flux ratio. These two parameters correlate with each other, however they provide different information about the mechanisms linked to the spin--filament alignments. Stellar mass traces the position of the galaxy in the cosmic web and it is related to the overall accretion of material (e.g., \citealp{Codis2015} and references within). The bulge-to-total flux ratio traces particularly gas-rich major mergers \citep{Welker2017}, which are identified to
most efficiently build-up the bulge component and to
produce the most striking changes in galaxy shape and angular momentum \citep{Welker2014,Lagos2018}.

Stellar masses ($M_\star$) are measured from K-corrected $g$--$i$ colours and $i$-band magnitudes \citep{Bryant2015,Taylor2011}. The bulge-to-total flux ratio (B/T) is estimated from the 2D photometric bulge/disc decomposition applied to the $r$-band KiDS images (see Section~\ref{Photometric position angles}). We make use of the B/T values extrapolated to infinity, but we find the same results using integrated quantities limited to a segment radius for the photometric fitting of the galaxies. The combination of the $M_\star$ and B/T parameters allows us to investigate the spin--filament alignments as a function of the mass of the bulge, defined as $M_{\rm bulge}=M_\star\times({\rm B/T})$.

We also analyse alignment trends according to morphological, star formation, kinematic and environmental properties. Previous studies based on simulations and/or observations have shown that the galaxy spin--filament alignments also depend on these properties \citep{Dubois2014}. The visual morphological classification is based on \citet{Cortese2016}, where the galaxy morphology is represented by: elliptical=0, elliptical/S0=0.5, S0=1, S0/early-spiral=1.5, early-spiral=2, early/late-spiral=2.5 and late-spiral=3. Galaxies are classified according to their star formation characteristics into passive, star-forming and in-transition (H$\delta$-strong) by \citet{Owers2019}. Single stellar population properties are parametrised by the average light-weighted age and metallicity \citep{Scott2017,Scott2018}. The degree of ordered stellar rotation versus random motions in the galaxy is represented by the spin parameter evaluated within an effective radius, $\lambda_e$ \citep{Emsellem2007,Emsellem2011,Cappellari2016,vandeSande2017}. We also measure $(V/\sigma)_e$ as the 
flux-weighted mean
within an effective radius \citep{Cappellari2007,vandeSande2017}. The kinematic morphology classification, which separates galaxies into slow rotators and fast rotators according to a Bayesian mixture model, is taken from \citet{vandeSande2021a}. The kinematic misalignment for a galaxy between the stellar and gas components is measured according to $\Delta{\rm PA}=|{\rm PA_{stars}-PA_{gas}}|$, where $\Delta{\rm PA}$ is the absolute difference between the stellar and gas kinematic position angles (e.g., \citealp{Bryant2019,Duckworth2019}). Finally, the local environment is characterised by the local galaxy density measured as the fifth-nearest neighbour surface density $\Sigma_5$ from \citet{Brough2013}. In addition, we make use of the GAMA galaxy group catalogue \citep{Robotham2011} to identify group central galaxies, group satellite galaxies, and isolated galaxies.

\subsection{Galaxy sample and selection criteria}
\label{Galaxy sample and selection criteria}

Stellar kinematics are available for 3070 SAMI galaxies. We exclude 917 galaxies that belong to the eight clusters observed by SAMI, since environmental processes might affect and decrease the spin--filament alignment signal in these regions of high galaxy density \citep{Dubois2014}. We select 1711 galaxies with $9<\log(M_\star/M_{\odot})<12$ for reliable stellar kinematic velocity maps \citep{vandeSande2017}. Following \citet{Welker2020}, we choose 1516 galaxies with measured ellipticity and kinematic PAs, and we select 1311 galaxies having 1$\sigma$ uncertainties $\delta{\rm PA}<25$\degree. This cut allows us to select converged fits and to recover the various galaxy morphologies at different stellar mass ranges. From a visual inspection of the spatially-resolved velocity maps after fitting the PAs, we flag $\sim$2\% of them (mostly slow rotators) as possibly not well-resolved; however, the exclusion of these galaxies does not change our results. Our final SAMI galaxy sample comprises 1121 galaxies with measured $M_\star$ and B/T. 

\begin{figure*}
\centering
\includegraphics[width=14cm]{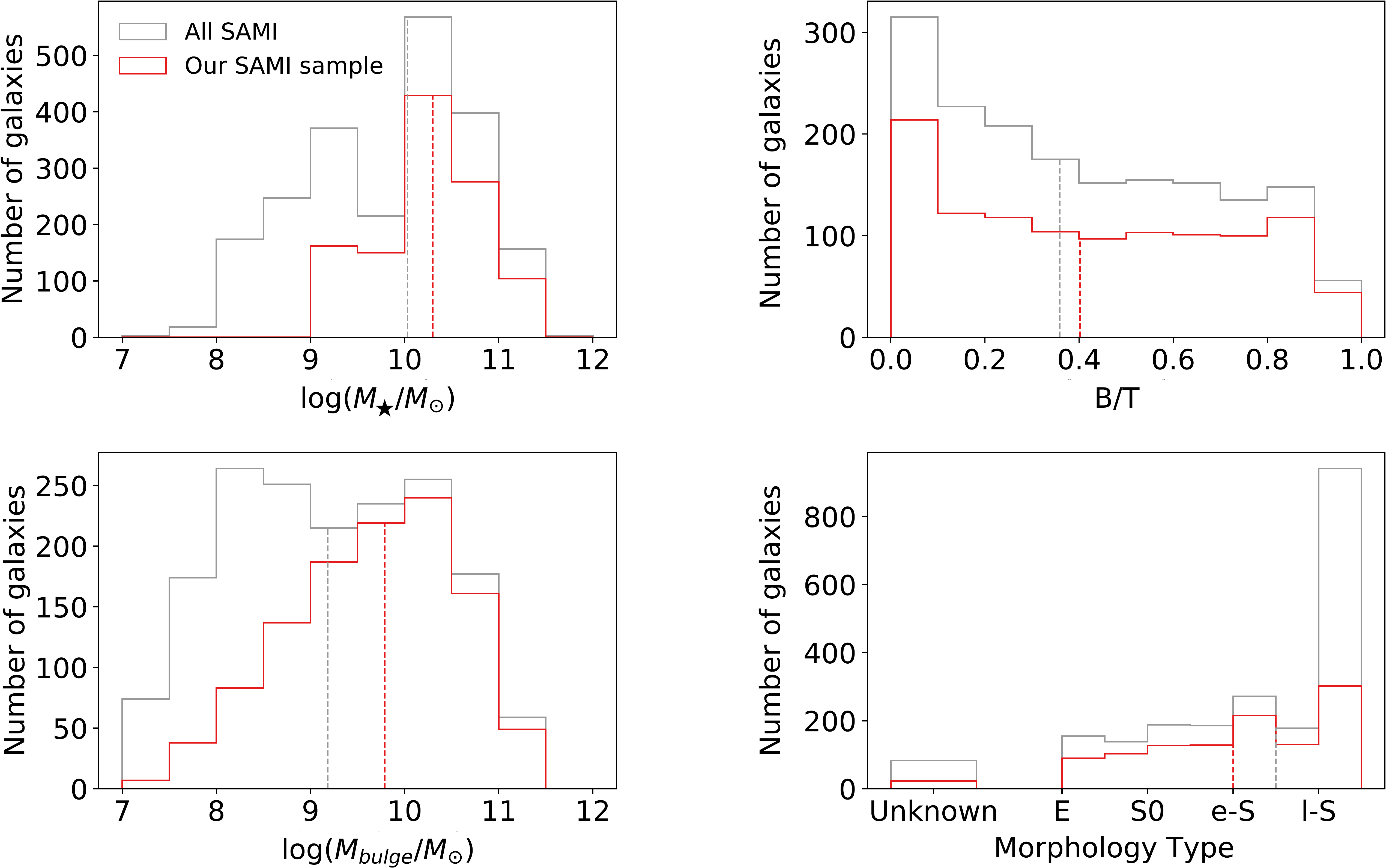}
\caption{The distributions of stellar mass, B/T, bulge mass and morphology for 1121 SAMI galaxies with respect to the SAMI Galaxy Survey sample (excluding cluster galaxies). The red and grey dashed lines show the median values. Our sample includes a wide variety of galaxies, mostly being late-types.}
\label{SAMIproperties}
\end{figure*}

Figure~\ref{SAMIproperties} shows the distributions of stellar mass, B/T, bulge mass and morphology. The median values are $\log(M_\star/M_{\odot}) = 10.30 \pm 0.02$, $\log(M_{\rm bulge}/M_{\odot}) = 9.79 \pm 0.03$, morphology type=2 (early-spiral) and B/T$= 0.40 \pm 0.01$. Our SAMI sample contains a wide variety of galaxies, mostly being late-types; this is typical of the SAMI Galaxy Survey and it is accentuated by excluding galaxies in clusters, which are mainly early-types. 

Gas kinematic PAs are estimated for the 1121 SAMI galaxies. Most of the galaxies ($\sim$80\%) have $\Delta{\rm PA}<30\degree$, indicating that the stellar and gas components are kinematically aligned. Photometric PAs are also measured for the 1121 SAMI galaxies, with about 66\% of them best fitted by the photometric double-component S\'ersic bulge + exponential disc model according to the galaxy characterisation of \citet{Casura2022}.

Of the 1121 SAMI galaxies in our sample, stellar kinematic PAs are measured for 468 bulges and 516 discs, with 196 galaxies having both bulge and disc measurements. We exclude bulges and discs with $\delta{\rm PA}_{\rm bulge}>25$\degree and $\delta{\rm PA}_{\rm disc}>25$\degree, following the same approach used for the galaxy kinematic PAs. Visually inspecting the spatially-resolved velocity maps of the bulges and the discs after fitting the PAs, we find that 8--10\% of the bulges might not have well-resolved PAs, while only 2 discs are flagged. We include these cases, since their exclusion does not change our results.

\subsection{The GAMA survey}
\label{GAMA galaxy survey}

In order to reconstruct the cosmic web, we take advantage of the Galaxy And Mass Assembly (GAMA; \citealp{Driver2011}) survey. GAMA is a spectroscopic and photometric survey of $\sim$300,000 galaxies down to $r < 19.8$\,mag that covers $\sim$286\,deg$^{2}$ in 5 regions called G02, G09, G12, G15 and G23. The redshift range of the GAMA sample is $0<z<0.5$, with a median value of $z\sim0.25$. Most of the spectroscopic data were obtained using the AAOmega multi-object spectrograph at the Anglo-Australian Telescope, although GAMA also incorporates previous spectroscopic surveys such as SDSS \citep{York2000}, 2dFGRS \citep{Colless2001,Colless2003}, WiggleZ \citep{Drinkwater2010} and the Millennium Galaxy Catalogue \citep{Driver2005}. The deep and highly complete spectroscopic redshift data (98.5\% in the equatorial regions; \citealp{Liske2015}), combined with the wide area, make the GAMA survey an ideal galaxy redshift sample for extracting the filaments of the cosmic web.

We make use of the DR3 data release of the GAMA survey \citep{Baldry2018}. We select 35882 galaxies with secure redshifts and stellar masses in the SAMI redshift range ($0<z<0.13$) lying within the G09, G12 and G15 regions (where SAMI galaxies have been observed). Most galaxies have stellar masses between $10^{8}\,M_{\odot}$ and $10^{12}\,M_{\odot}$; the median stellar mass and redshift are $M_\star \sim 10^{9.7}\,M_{\odot}$ and $z \sim 0.09$.

\section{Methods}
\label{Methods}

\subsection{Orientation of the spin axes}
\label{Orientation of the spin axes}

To identify the orientation of the spin axes of galaxies, bulges and discs, we follow the 3D thin-disc approximation implemented by \citet{Lee2007} and used in \citet{Kraljic2021} (see Section~2.6 of \citealp{Kraljic2021} for a summary of the formulae). This technique requires position angles and inclination angles. Our results are obtained using stellar kinematic PAs as described in Section~\ref{Stellar kinematics} for galaxies and in Section~\ref{Stellar kinematics of bulges and discs} for bulges and discs. For comparisons we also make use of photometric PAs (Section~\ref{Photometric position angles}) and gas kinematic PAs (Section~\ref{Gas kinematics}). The inclination angle is computed according to \citet{Haynes1984}, where for the intrinsic flatness parameter we use the mean value 0.171 of our SAMI sample. Using values of the intrinsic flatness parameter as a function of the galaxy morphology does not alter our results. The galaxy axial ratio is measured from the ellipticity (Section~\ref{Stellar kinematics}), while the bulge and disc axial ratios are measured from the 2D photometric bulge/disc decomposition (Section~\ref{Photometric position angles}). 
The inclination angle is set to 90\degree\ if the axial ratio is lower than the intrinsic flatness parameter. The sign of the inclination angle is ambiguous, since it is not possible to determine whether the spin axis is pointing in projection towards or away from us. Following \citet{Kraljic2021}, we chose to consider only the positive sign for the cosine of the inclination angle. Our conclusions do not change if we take into account the two-fold ambiguity as proposed by \citet{Lee2011}, by assigning to each galaxy both signs (see also Appendix~A of \citealp{Kraljic2021}). 

\citet{Lee2007} and \citet{Kraljic2021} applied this modelling to disc-dominated galaxies, however it can be extended to bulge-dominated galaxies assuming that the projected short axis and the spin axis are parallel \citep{Pahwa2016}. The assumption is based on the fact that the short and spin axes of most bulge-dominated galaxies are found to be aligned \citep{Franx1991}. In this work we extend the 3D modelling of the spin axis to bulge-dominated galaxies since most have regular and ordered velocity maps similar to those of disc-dominated galaxies \citep{Emsellem2011,Krajnovic2011,Cappellari2011}. Moreover, most passive SAMI galaxies are observed to be rotating oblate spheroids by applying orbit-superposition Schwarzschild models \citep{Santucci2022}.

\subsection{Reconstruction of the cosmic web}
\label{Reconstruction of the cosmic web}

We reconstruct the filaments of the cosmic web using the Discrete Persistent Structure Extractor public code ({\sc DisPerSe}; \citealp{Sousbie2011a,Sousbie2011b}). {\sc DisPerSe} has been widely used in the literature to map the cosmic web, both using simulations (e.g., \citealp{Dubois2014,Welker2018,Codis2018}) and spectroscopic data, including the GAMA survey (e.g., \citealp{Kraljic2018, Duckworth2019, Welker2020}). Its strength for astrophysical applications resides in the fact that 3D structures of the cosmic web are identified starting from a point-like distribution, without making any assumption about its geometry or homogeneity. The 3D density field is built from the discrete distribution using the Delaunay Tessellation Field Estimator technique (DTFE; \citealp{Schaap2000, Cautun2011}), and it represents the input to the geometric 3D ridge extractor. {\sc DisPerSe} is a parameter-free and scale-free topologically motivated algorithm, based on discrete Morse and persistence theories. Voids, walls, and filaments are identified as individual components of the cosmic web and defined as distinct regions in the geometrical segmentation of space. The most significant structures can be selected according to their persistence ratio, which traces the significance of the topological connection between individual pairs of critical points and can be expressed in terms of number of standard deviations $\sigma$.

\begin{figure}
\includegraphics[width=\columnwidth]{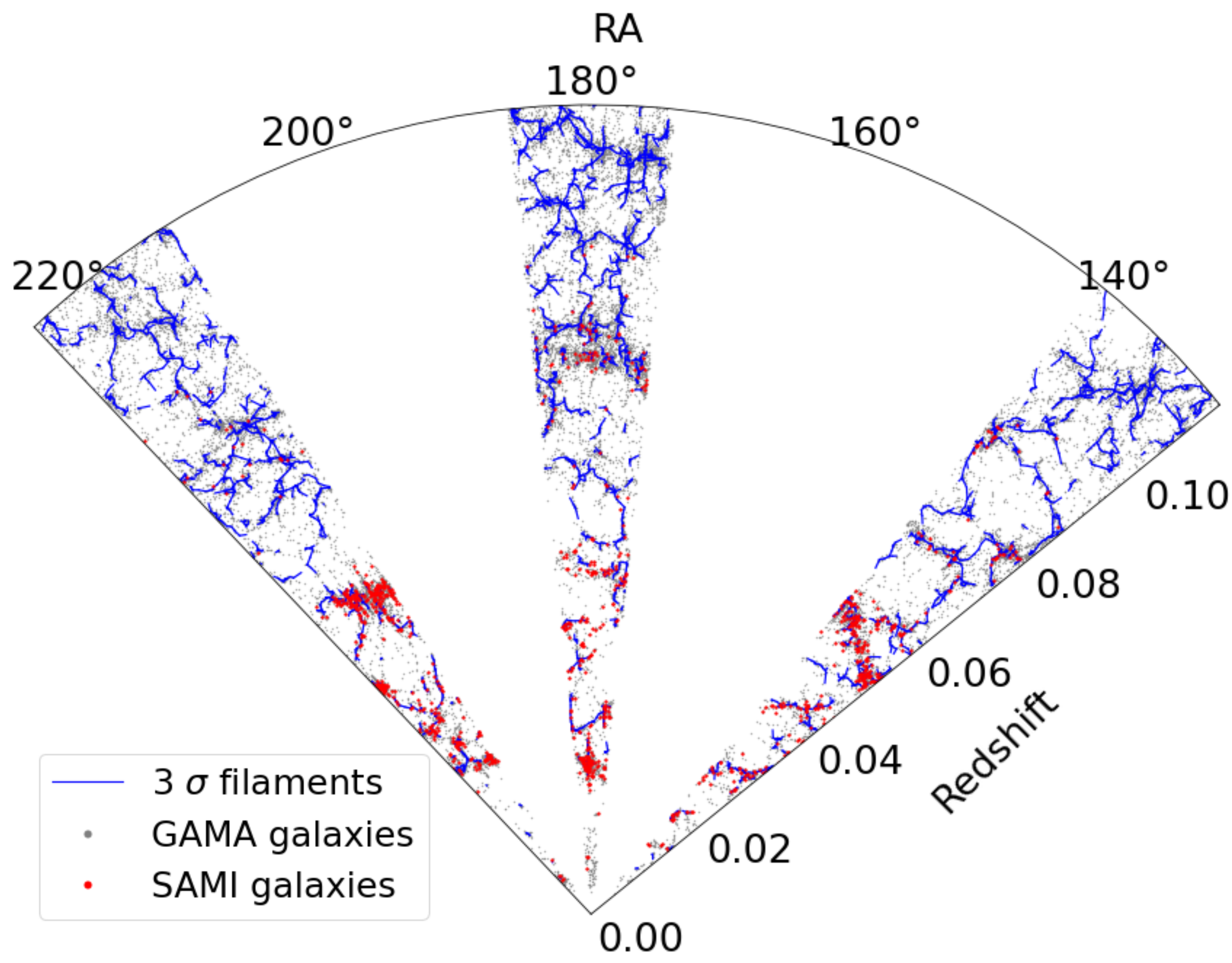}
\caption{Projected network of cosmic filaments (blue lines) for the GAMA G09, G12 and G15 regions. Grey points are the 35882 GAMA galaxies used to reconstruct the spine of the cosmic web, while red points are the 1121 SAMI galaxies.}
\label{PolarPlotSAMIGAMAFilaments}
\end{figure}

For the galaxy distribution, we use the 35882 GAMA galaxies selected in Section~\ref{GAMA galaxy survey}, taking advantage of the right ascension, declination and spectroscopic redshift measurements. We run {\sc DisPerSe} with a 3$\sigma$ persistence threshold, in accord with previous studies that investigate galaxy spin--filament alignments \citep{Welker2020,Kraljic2021}. A reconstruction of the 3D density fields using the Python package {\sc pyvista} \citep{Sullivan2019} and displaying the typical tetrahedrons for the GAMA G09, G12 and G15 regions can be found \href{https://skfb.ly/o9MXv}{here}. The 3D filamentary structure is shown as interactive plot at
\href{https://skfb.ly/o9MXz}{this URL}. Figure~\ref{PolarPlotSAMIGAMAFilaments} shows the projected network of filaments (blue lines), the 35882 GAMA galaxies (grey points) used to reconstruct the cosmic web, and the 1121 SAMI galaxies for which we aim to study the spin--filament alignments (red points). 

\begin{figure}
\includegraphics[width=\columnwidth]{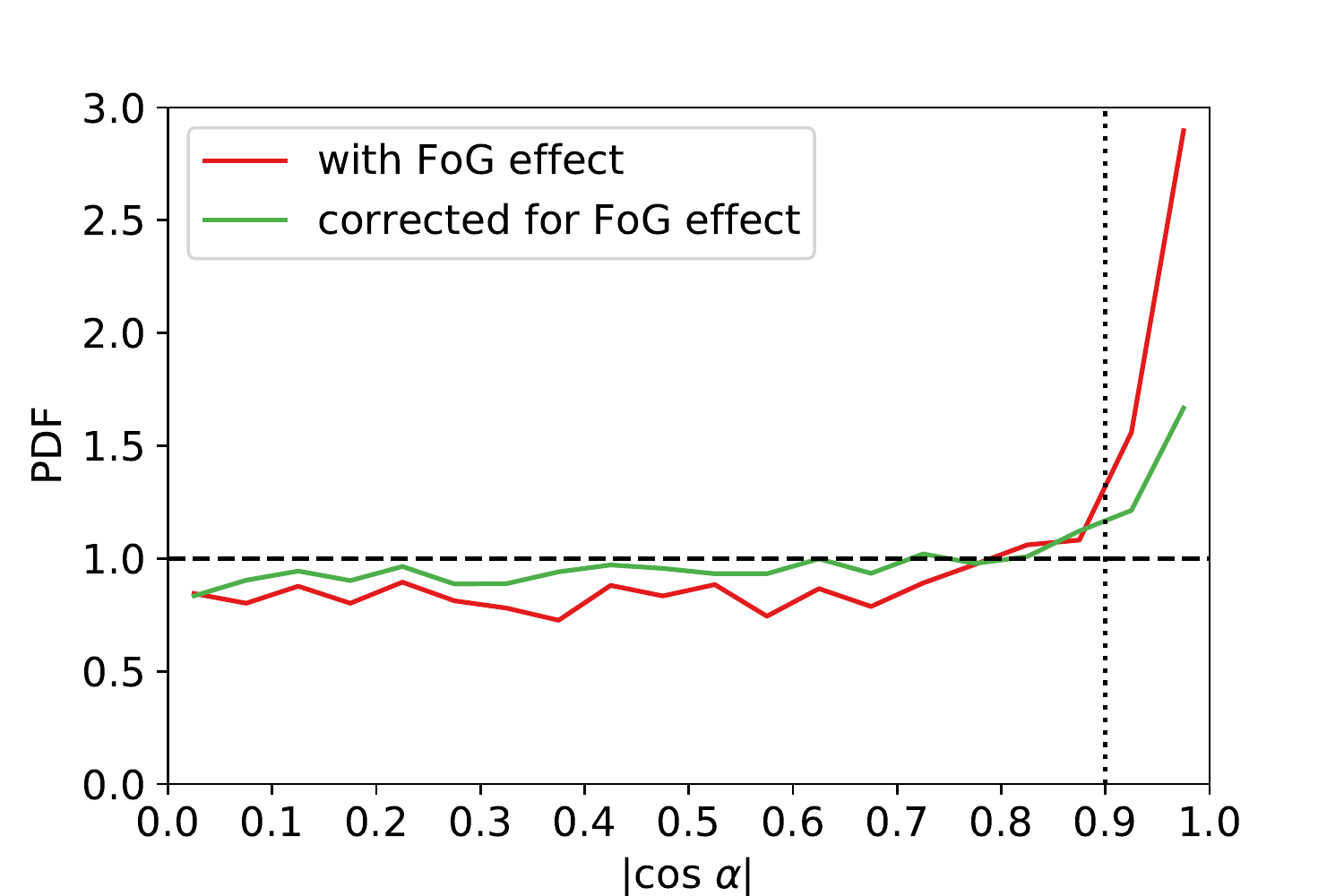}
\caption{PDF of |$\cos\alpha$|, where $\alpha$ is the angle between the filament and the GAMA line-of-sight, with and without the FoG effect (red and green lines, respectively). The dashed line represents the uniform distribution. An increase of aligned filaments is found for |$\cos\alpha$|\,>\,0.9 (dotted line), even when correcting for the FoG effect.}
\label{GAMA_LOS_filaments}
\end{figure}

The reconstruction of the cosmic filaments is affected by the `Fingers of God' effect (FoG; \citealp{Jackson1972}). This distortion effect is due to the random motions of galaxies within virialised groups and clusters, and it causes the elongation of halos in redshift space, possibly leading to the identification of spurious filaments (see Figure~1 of \citealp{Kraljic2018}). We investigate the impact of the FoG effect within the SAMI region of the GAMA survey, implementing a compression correction by making galaxies isotropically distributed around their group centres, as in \citet{Kraljic2018}. Figure~\ref{GAMA_LOS_filaments} shows the probability distribution function (PDF) for |$\cos\alpha$|, where $\alpha$ is the angle between the filament and the GAMA line-of-sight (similar to Figure 2 of \citealp{Welker2020}). For |$\cos\alpha$|\,<\,0.9, the PDF with the FoG effect is consistent with the PDF corrected for the FoG effect according to the two-sample Kolmogorov-Smirnov test (K-S test; \citealp{Lederman1984}). An increase of aligned filaments is found for |$\cos\alpha$|\,>\,0.9, even when correcting for the FoG effect. The percentage of these aligned filaments decreases by only $\sim8$\% when applying the correction, from $\sim22$\% with FoG effect to $\sim14$\% after correction, and it still shows an increase at |$\cos\alpha$|\,>\,0.9.

We conclude that the correction for the FoG effect does not make significant changes to the cosmic web for the SAMI Galaxy Survey, largely because a relatively small fraction of galaxies are affected (see also next Section). This in turn is due to the SAMI volume probing only the nearby region of the GAMA survey at $0< z < 0.12$, where the number of rich and massive groups, for which the FoG effect is most important, is very limited \citep{Barsanti2018}. Finally, it is worth noting that the FoG compression does not correct for boundary effects that can also lead to spurious filaments along the line-of-sight \citep{Welker2020}, and it has been found to perform poorly when applied to groups individually \citep{Kuchner2021}.

\subsection{Cosmic web metrics}
\label{Cosmic web metrics}

Each SAMI galaxy is assigned to the closest filament using the smallest 3D Euclidean distance, $D_{\rm fil}$; if the projection point is beyond the start or end of the filament segment, we use the closest distance to the node. 

To exclude spurious filaments identified along the GAMA line-of-sight, when assigning each SAMI galaxy to the closest filament we follow Method~1 of \citet{Welker2020}: SAMI galaxies assigned to filaments with |$\cos\alpha$|\,>\,0.9 are reassigned to the closest filament with |$\cos\alpha$|\,<\,0.9, where $\alpha$ is the angle between the filament and the GAMA line-of-sight. If we follow the other two methods proposed by \citet{Welker2020}---i.e.\ Method~0 (all SAMI galaxies and filaments are taken into account) or Method~2 (we disregard SAMI galaxies assigned to filaments with |$\cos\alpha$|\,>\,0.95)---we find consistent results, with differences only in the statistical significance of the spin--filament correlation signals. This is in agreement with the conclusions regarding the three methods by \citet{Welker2020}. 

\begin{figure}
\centering
\includegraphics[width=8cm]{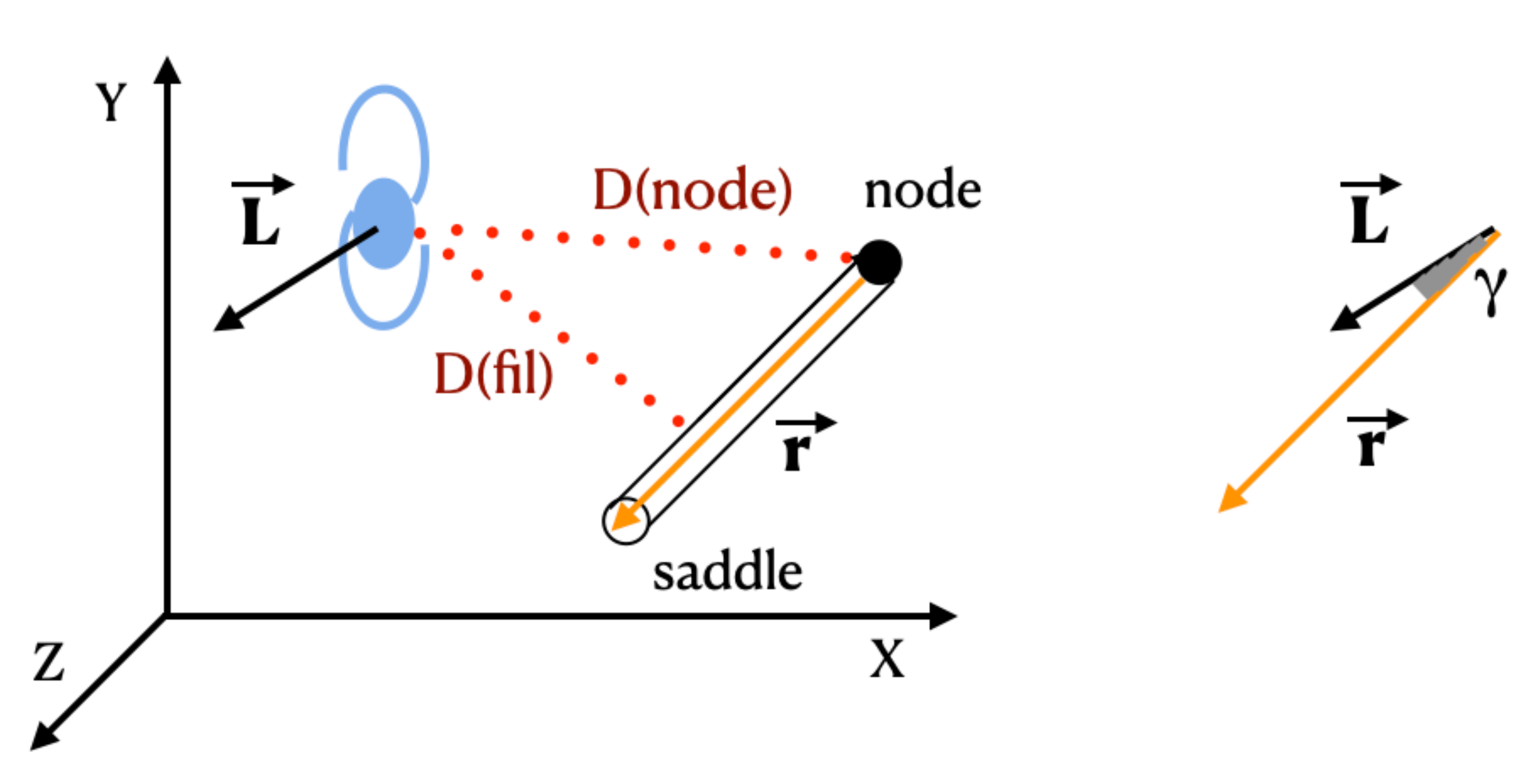}
\caption{Schematic representation of the cosmic web metrics, the galaxy spin axis, and the orientation vector of the filament. We show a filament's segment between a node and a saddle point (a critical point that is neither a maximum nor a minimum).}
\label{CosmicWebMetrics}
\end{figure}

The galaxy spin--filament alignment is parametrised as the absolute value of the cosine of the angle between the galaxy spin axis and the closest filament in 3D Cartesian coordinates (e.g., \citealp{Tempel2013b,Kraljic2021}):
\begin{equation}
|\cos\gamma|=\frac{|\mathbf{L} \cdot \mathbf{r}|}{|\mathbf{L}| \cdot |\mathbf{r}|}
\end{equation}
where \textbf{L} is the galaxy spin axis identified as described in Section~\ref{Orientation of the spin axes} and \textbf{r} is the orientation vector of the filament. The angle |$\cos\gamma$| varies in the range [0,1], with |$\cos\gamma$|=1 meaning the galaxy spin axis is parallel to the filament while |$\cos\gamma$|=0 means the galaxy spin axis is perpendicular to the filament. The same parameter is used to quantify the separate bulge and disc spin--filament alignments, in which case \textbf{L} represents the respective bulge/disc spin axis. A schematic view of the cosmic web metrics and the associated vectors is given in Figure~\ref{CosmicWebMetrics} (see also Figure~4 of \citealp{Kraljic2018} and Figure~1 of \citealp{Winkel2021}).    

Comparing the $|\cos\gamma|$ values obtained for the cosmic web with and without the FoG effect, we find that the values deviate significantly from a 1:1 relation for only $\sim5$\% of the 1121 SAMI galaxies. The exclusion of these galaxies does not alter our conclusions and so we choose to study the $|\cos\gamma|$ values of the whole sample of 1121 SAMI galaxies without correcting the cosmic web for the FoG effect. 

\begin{figure*}
\includegraphics[scale=0.33]{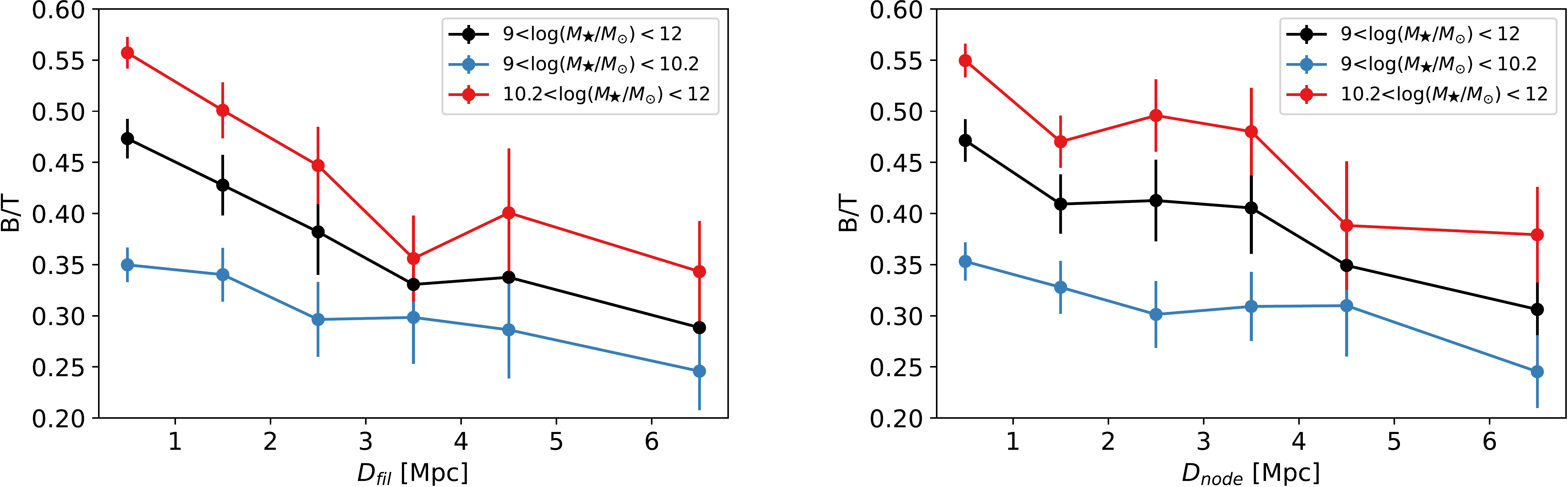}
\caption{Average bulge-to-total flux ratio as a function of distance to the closest filament (left) and to the closest node (right) for 1121 SAMI galaxies (black points). We show 1$\sigma$ error bars on the means. More bulge-dominated galaxies are found closer to the spine of the filamentary structure and closer to the nodes. The trend is seen also when dividing the galaxy sample according to $M_{\star}$.}
\label{BT_DfilDistribution}
\end{figure*}

Finally, we show the B/T distribution of the 1121 SAMI galaxies as a function of the distance to the closest filament and to the closest node in Figure~\ref{BT_DfilDistribution}. More bulge-dominated galaxies are found closer to the spine of the filament with respect to disc-dominated galaxies and also closer to the nodes, resembling the morphology-density relation of \citet{Dressler1980}. These trends are expected, since B/T correlates with stellar mass which traces $D_{\rm fil}$ and $D_{\rm node}$, which are, by definition, regions of higher density and stronger collapse \citep{Kraljic2018,Welker2020}. Dividing the galaxy sample into 489 low-mass galaxies with $9<\log{(M_\star/M_{\odot)}}<10.2$ and 632 high-mass galaxies with $10.2<\log{(M_\star/M_{\odot)}}<12$, we find that the increasing of B/T with lower values of $D_{\rm fil}$ and $D_{\rm node}$ is reproduced for both ranges in stellar mass. Overall, it suggests a scenario where galaxies migrate along the filament spine and towards the nodes, growing the bulge component via mergers.  

\section{Results}
\label{Results}

Our aim is to identify which galaxy properties are mostly related to the flipping of the galaxy spin--filament alignments, in order to understand the involved physical processes. We focus on stellar mass, bulge-to-total flux ratio and their product, the mass of the bulge $M_{\rm bulge}=M_\star\times({\rm B/T})$. In fact, $M_\star$ plays a key role in the alignments \citep{Codis2015,Welker2020}. However, $M_\star$ alone might not be able to explain the dependence on morphology \citep{Kraljic2020}. Thus, B/T might also be an independent driver, since it is identified as the main tracer of gas-rich major mergers that most efficiently build-up the bulge component and change the galaxy angular momentum \citep{Welker2014,Welker2017,Lagos2018}. We also investigate the spin--filament alignments as a function of morphological, star formation, kinematic and environmental properties. These parameters correlate with $M_\star$ and/or B/T, so they are expected to show secondary correlations. 

As stated in Section~\ref{Orientation of the spin axes}, the identification of the spin axis is based on stellar kinematic PAs. The descriptions of how the galaxy properties are measured, together with the primary references, are given in Section~\ref{Galaxy properties}. By exploiting the spatially-resolved kinematic bulge/disc decomposition for the SAMI galaxies, we are able to study separately the spin--filament alignments of bulges and discs.

These analyses help us understand the different formation scenarios for galaxies, bulges, and discs. 

\subsection{\texorpdfstring{$\boldsymbol{M_{\rm bulge}=M_\star\times(B/T)}$}{} is the primary parameter}
\label{Correlations with galaxy properties}

\begin{figure*}
\includegraphics[width=17.25cm]{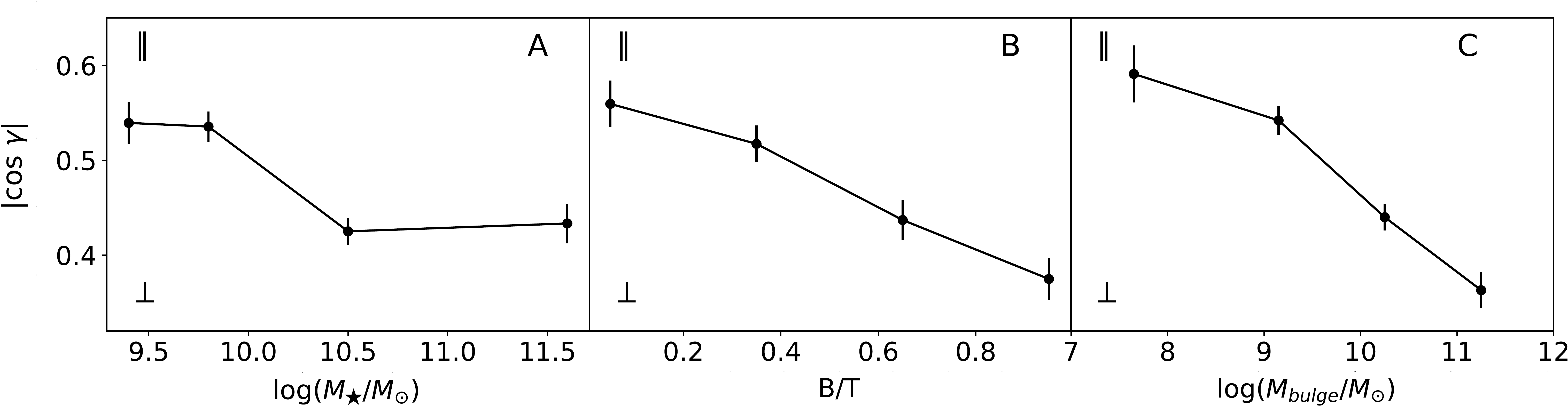}
\vspace{0.3cm} \\
\includegraphics[width=17.25cm]{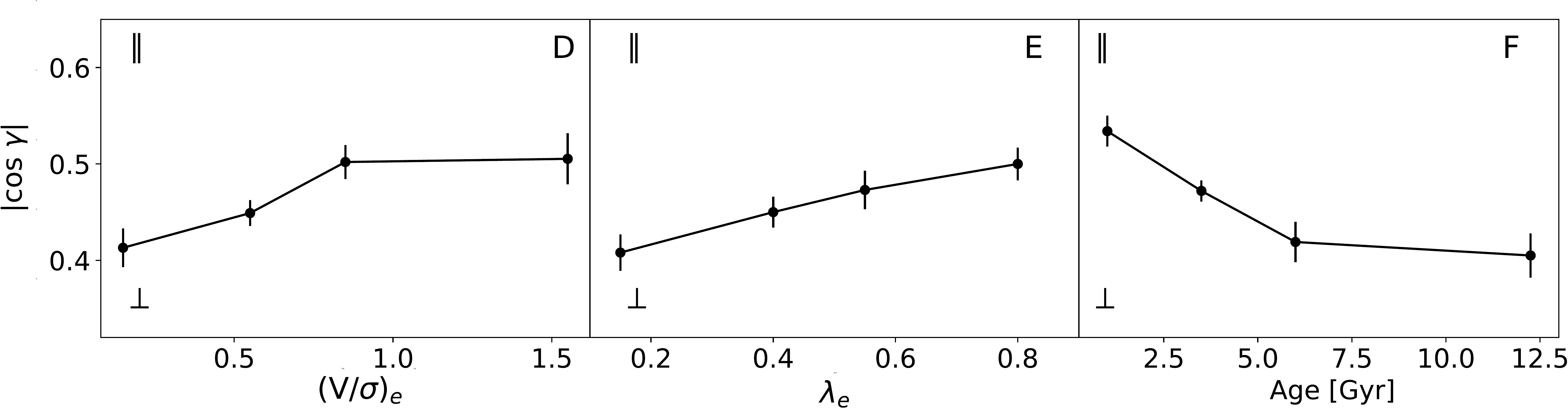}
\vspace{0.3cm} \\
\includegraphics[width=17.25cm]{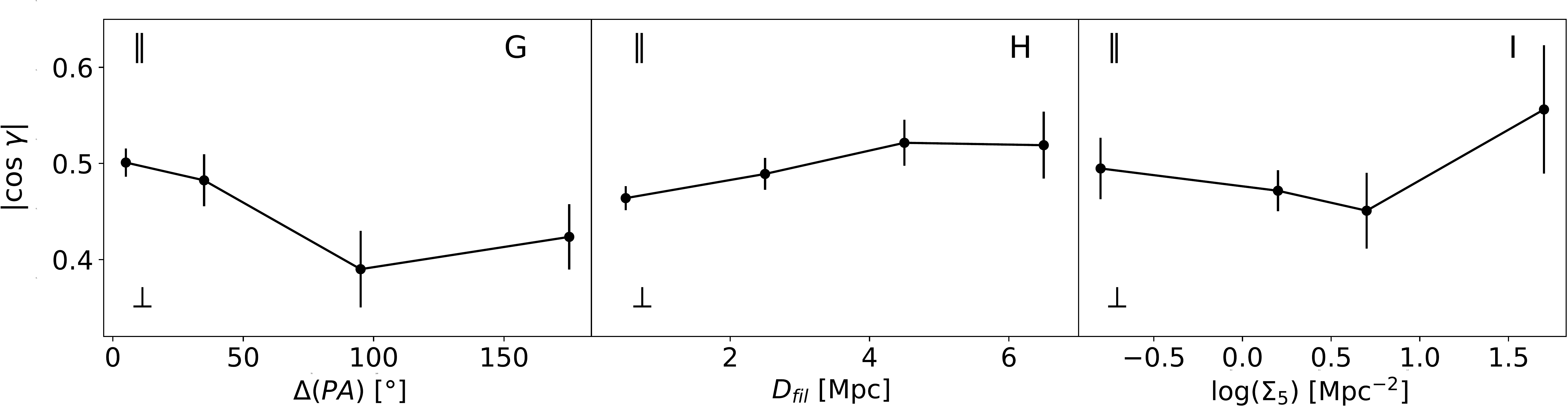}
\caption{Average |$\cos\gamma$| values as a function of bins in stellar mass (panel A), bulge-to-total flux ratio (panel B), bulge mass (panel C), degree of stellar rotation (panels D and E), average light-weighted age (panel F), kinematic misalignment between stars and gas (panel G), distance from the closest filament (panel H), and local galaxy density (panel I). Higher values of |$\cos\gamma$| indicate the distribution has relatively more parallel spin--filament alignments; lower values of |$\cos\gamma$| indicate the distribution has relatively more perpendicular spin--filament alignments. Standard errors on the mean are shown. The mean |$\cos\gamma$| values decrease with increasing $M_\star$, B/T, $M_{\rm bulge}$ and age, and with decreasing $(V/\sigma)_e$, $\lambda_e$ and $D_{\rm fil}$; there is no significant change in |$\cos\gamma$| as a function of $\Delta{\rm PA}$ or $\Sigma_5$.}
\label{AnglevsGalaxyParameters}
\end{figure*}

We explore whether galaxy spin--filament alignments depend on different galaxy properties, with the goal of understanding their statistical significance and possible physical linkages. Figure~\ref{AnglevsGalaxyParameters} shows the average |$\cos\gamma$| values as a function of stellar mass (panel A), bulge-to-total flux ratio (panel B), bulge mass (panel C), degree of stellar rotation (panels D and E), average light-weighted age (panel F), kinematic misalignment between stars and gas (panel G), distance from the closest filament (panel H), and local galaxy density (panel I). The panels display that the mean |$\cos\gamma$| {\em decreases} with increasing $M_\star$, B/T, $M_{\rm bulge}$ and age, indicating a relative shift from parallel to perpendicular spin--filament alignments as these quantities increase. By contrast, the mean |$\cos\gamma$| {\em increases} with increasing $(V/\sigma)_e$, $\lambda_e$ and $D_{\rm fil}$, indicating a relative shift from perpendicular to parallel alignments as these quantities increase. No significant trends are found for |$\cos\gamma$| as a function of $\Delta{\rm PA}$ or $\Sigma_5$.

We look for possible dependencies performing Spearman rank correlation test for individual galaxies (correlation coefficients, $\rho$, and $p$-values, $p_{\rm S}$, are listed in Table~\ref{SpearmanResults}). We adopt $p_{\rm S}<0.05$ as the criterion for rejecting the null hypothesis of no correlation. With this criterion, statistically significant correlations with |$\cos\gamma$| are detected for $M_\star$, B/T, $M_{\rm bulge}$, age and $D_{\rm fil}$. The result is marginal for $\lambda_e$, while for $\Delta{\rm PA}$ and $\Sigma_5$ there is no correlation. The strongest correlations are found for $M_{\rm bulge}$ and B/T.

We explore whether one of these parameters primarily correlates with the spin--filament alignments, following a similar method to that used in \citet{Oh2022}. We fit linear relations between |$\cos\gamma$| and the galaxy properties. Choosing a galaxy property as the tested primary parameter, we estimate the expected  |$\cos\gamma$|$_{\rm exp}$ values from the linear fit between the tested primary parameter and the observed |$\cos\gamma$|$_{\rm obs}$ values. We define the difference $\Delta\cos\gamma=|\cos\gamma|_{\rm obs}-|\cos\gamma|_{\rm exp}$, removing the dependency on the tested primary parameter from the other galaxy properties. Then, we use the Spearman test to check whether correlations are still present between $\Delta\cos\gamma$ and the other galaxy properties. Assuming $M_\star$ as the primary parameter, significant ($p_{\rm S}<0.05$) residual correlations are found for B/T and $M_{\rm bulge}$, implying that $M_\star$ alone cannot explain the dependence of galaxy spin--filament alignments on B/T and $M_{\rm bulge}$. Assuming B/T as primary parameter, we detect a significant residual correlations for $M_\star$ and $M_{\rm bulge}$. Finally, considering $M_{\rm bulge}$ as primary parameter, we find no residual dependence on any galaxy property. 

\begin{figure*}
\centering
\includegraphics[scale=0.25]{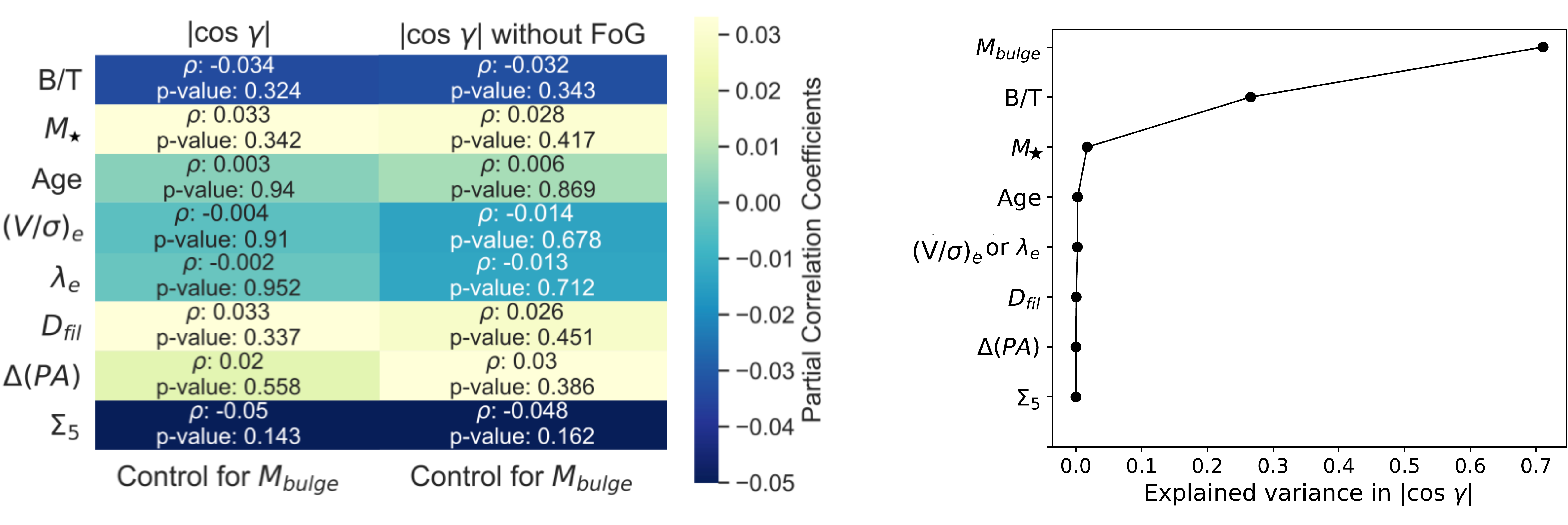}
\caption{Left panel: partial correlation coefficients from the Spearman test between |$\cos\gamma$| and the galaxy properties while controlling for $M_{\rm bulge}$. The right column is for |$\cos\gamma$| values where the cosmic web is corrected for the FoG effect. Right panel: the fraction of the variance in |$\cos\gamma$| that is explained by each galaxy property according to the partial least squares (PLS) regression method. Most ($\sim$70\%) of the variance in |$\cos\gamma$| is explained by $M_{\rm bulge}$.}
\label{ExplainedVariance}
\end{figure*}

These results are confirmed by the estimate of the partial correlation coefficients, which allow us to explore the true correlation between two parameters while controlling for a third quantity, avoiding the cross-correlation driven by their dependency on the third property \citep{Lawrance1976,Baker2022}. The left panel of Figure~\ref{ExplainedVariance} shows the partial correlation coefficients from the Spearman test between |$\cos\gamma$| and the studied galaxy properties while controlling for $M_{\rm bulge}$. No significant correlations remain once the correlation with $M_{\rm bulge}$ is taken into account. Similar results are obtained for the |$\cos\gamma$| values where the cosmic web is corrected for the FoG effect. 

An alternative approach is to apply the partial least squares regression technique (PLS; \citealp{Wold1966,Hoskuldsson1988}) to estimate the contribution to the |$\cos\gamma$| variance by each galaxy parameter. We use the PLSRegression Python function \citep{Pedregosa2012} and follow the approach described in \citet{Oh2022}. The right panel of Figure~\ref{ExplainedVariance} and Table~\ref{SpearmanResults} show the fraction of the variance in |$\cos\gamma$| explained by each galaxy parameter. The highest contribution ($\sim$70\%) is found for $M_{\rm bulge}$, suggesting that this parameter most significantly correlates with the galaxy spin--filament alignments. B/T accounts for $\sim$27\%, $M_\star$ for $\sim$2\%, and the remaining 1\% is explained by the other galaxy properties. These results are consistent with the findings from the analysis above, and suggest that $M_{\rm bulge}$ is the primary parameter to correlate with spin--filament alignments. 

\begin{table}
\centering
\caption{Results from Spearman rank correlation test and PLS technique for galaxy spin--filament alignments $|\cos\gamma|$ as a function of various galaxy properties. Column~1 lists the analysed galaxy property, column~2 the number of galaxies, column~3 the Spearman correlation coefficient, column~4 the $p$-value from the Spearman test with significant $p$-values ($<$0.05) highlighted in bold, and column~5 the explained variance in $|\cos\gamma|$.}
\label{SpearmanResults}
\begin{tabular}{@{}lcccc@{}}
\toprule
Galaxy property        & $N_{\rm gal}$ & $\rho$  & $p_{\rm S}$& Variance (\%)\\
\midrule
$\log(M_\star/M_{\odot})$       & 1121 & $-$0.07 & \textbf{0.014} & 1.73 \\
B/T                             & 1121 & $-$0.11 & $\mathbf{10^{-4}}$ &26.58  \\
$\log(M_{\rm bulge}/M_{\odot})$ & 1121 & $-$0.13 & $\mathbf{10^{-5}}$ &71.09\\
\midrule
$(V/\sigma)_e$                    & 1071 & $+$0.06 & \textbf{0.047} & 0.07 \\
$\lambda_e$                     & 1071 & $+$0.06 & \textbf{0.048} &0.07 \\
Age                             & 1121 & $-$0.10 & \textbf{0.001} &0.27 \\
$\Delta{\rm PA}$                & 1121 & $-$0.02 & 0.481 & 0.01\\
\midrule
$D_{\rm fil}$                   & 1121 & $+$0.07 & \textbf{0.024}  & 0.24 \\
$\log(\Sigma_5)$                & 1110 & $-$0.05 & 0.096 & 0.01 \\
\bottomrule
\end{tabular}
\end{table}

\subsection{Trends with \texorpdfstring{$\boldsymbol{M_\star}$}{}, B/T and \texorpdfstring{$\boldsymbol{M_{\rm bulge}}$}{}}
\label{Trends with stellar mass, B/T and mass bulge}

We find significant correlations of the galaxy spin--filament alignments with $M_\star$, B/T and $M_{\rm bulge}$. We now explore this signal further, dividing the SAMI galaxies into $M_\star$, B/T and $M_{\rm bulge}$ ranges, and analysing the tendency of the alignment for each sub-sample. 

\begin{figure}
\includegraphics[scale=0.225]{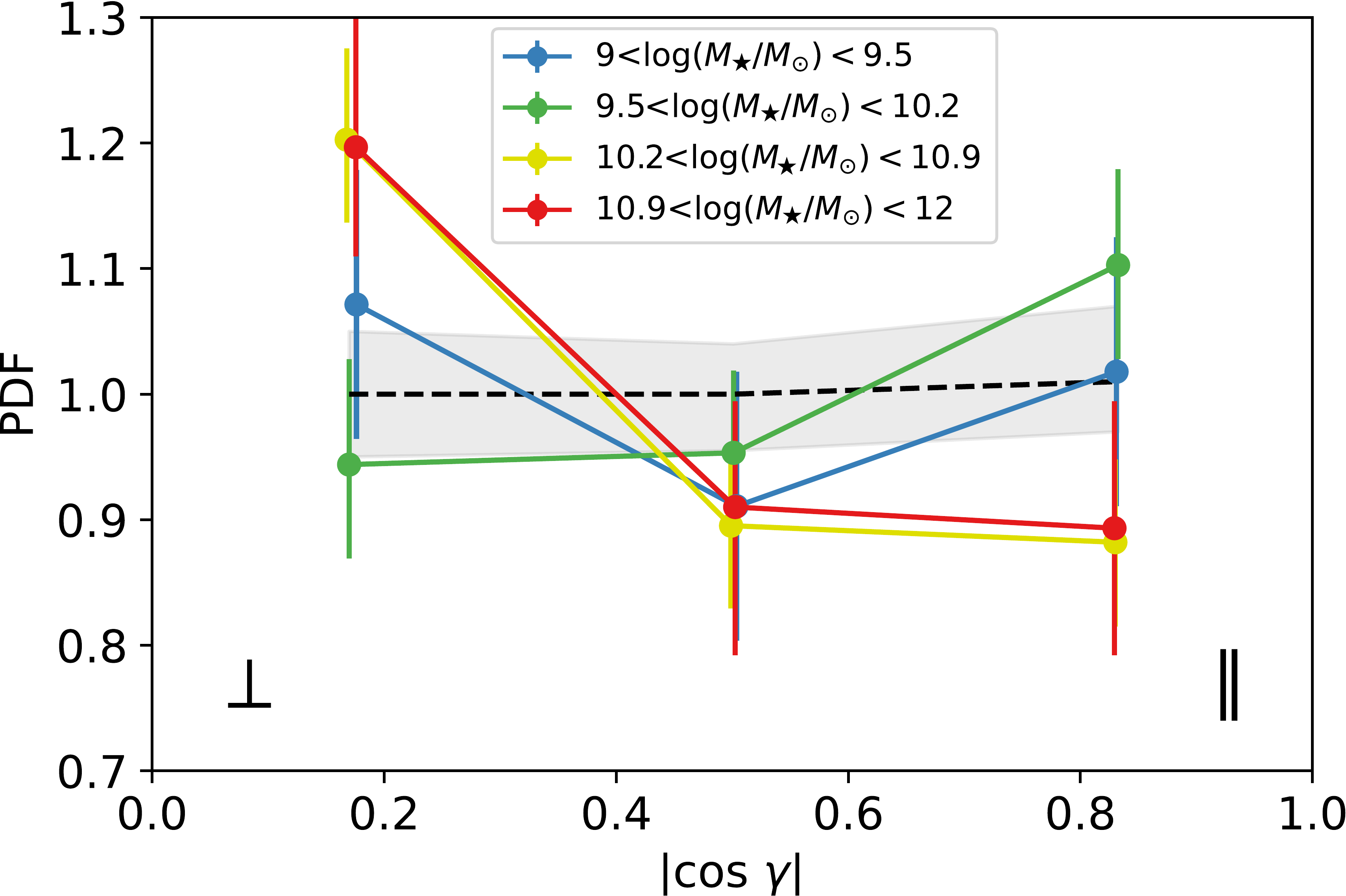}
\caption{PDF of the spin--filament alignments for 1121 SAMI galaxies divided in stellar mass. We show bootstrap error bars from 1000 random samples. The black dotted line and the shaded region represent the constructed null hypothesis with its 95\% confidence intervals. The PDF is skewed towards more perpendicular spin--filament alignments (i.e.\ lower values of |$\cos\gamma$|) for more massive galaxies and towards more parallel spin--filament alignments (i.e.\ higher values of |$\cos\gamma$|) for less massive galaxies.}
\label{TrendsStellarMass}
\end{figure}

Following \citet{Welker2020}, we divide the 1121 SAMI galaxies into four stellar mass ranges: $9<\log{(M_\star/M_{\odot)}}<9.5$, $9.5<\log{(M_\star/M_{\odot)}}<10.2$, $10.2<\log{(M_\star/M_{\odot)}}<10.9$ and $10.9<\log{(M_\star/M_{\odot)}}<12$. Figure~\ref{TrendsStellarMass} shows the probability distribution function (PDF) for |$\cos\gamma$|, the absolute value of the cosine of the angle between the galaxy spin axis and the orientation of the closest filament. The data are grouped in three bins of |$\cos\gamma$| and the PDFs are normalised such that the mean value over the bins is unity. The error bars are estimated from the bootstrap method using 1000 sample realizations. To assess the statistical significance of each trend, we apply the K-S test to test the null hypothesis that |$\cos\gamma$| has a uniform distribution. In order to account for possible observational bias, we build the null hypothesis by generating 3000 randomised samples where the galaxy spins are fixed, but the galaxy positions (and thus the identification of their closest filaments) are shuffled \citep{Tempel2013a,Tempel2013b,Kraljic2021}. The median from the 3000 random samples is used as null hypothesis and the reconstructed distributions are nearly uniform. We take $p$-values ($p_{\rm K-S}$) smaller than 0.05 to indicate that the distribution is significantly different from the null  hypothesis. Table~\ref{PAstellResults} lists the galaxy property of interest, the selection of galaxy sub-samples, the number of galaxies for each sub-sample, the mean |$\cos\gamma$| value, and the $p$-value from the K-S test. We find that the PDF is skewed towards more perpendicular spin--filament alignments (i.e.\ lower values of |$\cos\gamma$|) for more massive galaxies and towards more parallel spin--filament alignments (i.e.\ higher values of |$\cos\gamma$|) for less massive galaxies. With these four mass bins, significant statistical results are found for galaxies with $\log{(M_\star/M_{\odot})}>10.2$.

\begin{figure*}
\includegraphics[scale=0.33]{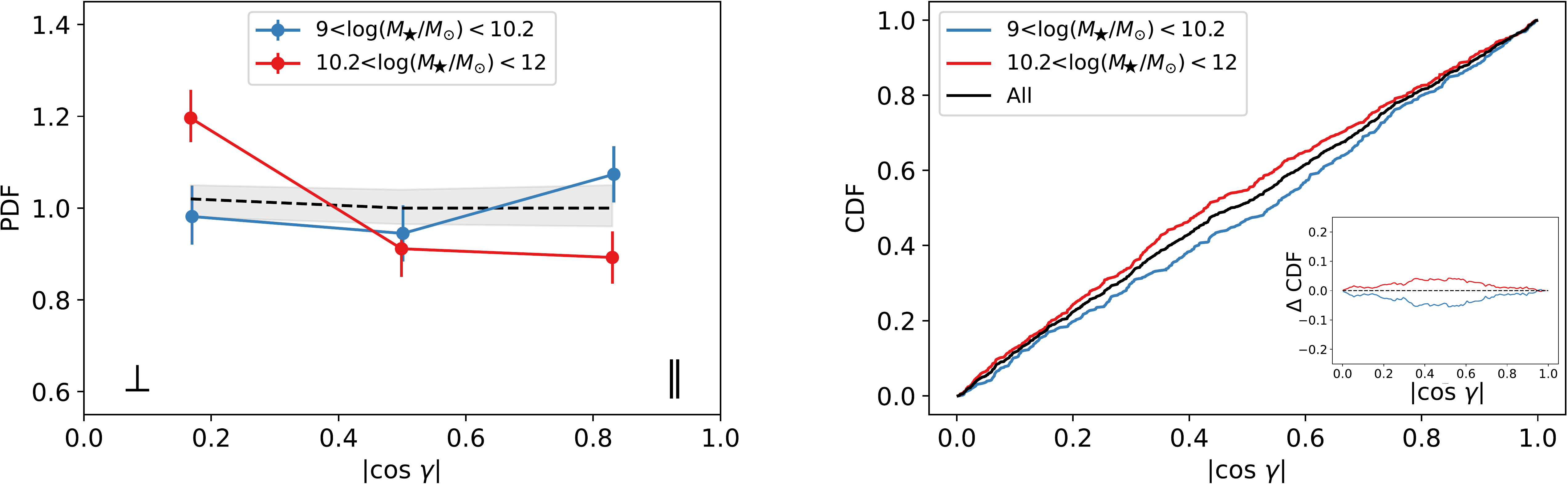}
\vspace{2.5mm} \\
\includegraphics[scale=0.33]{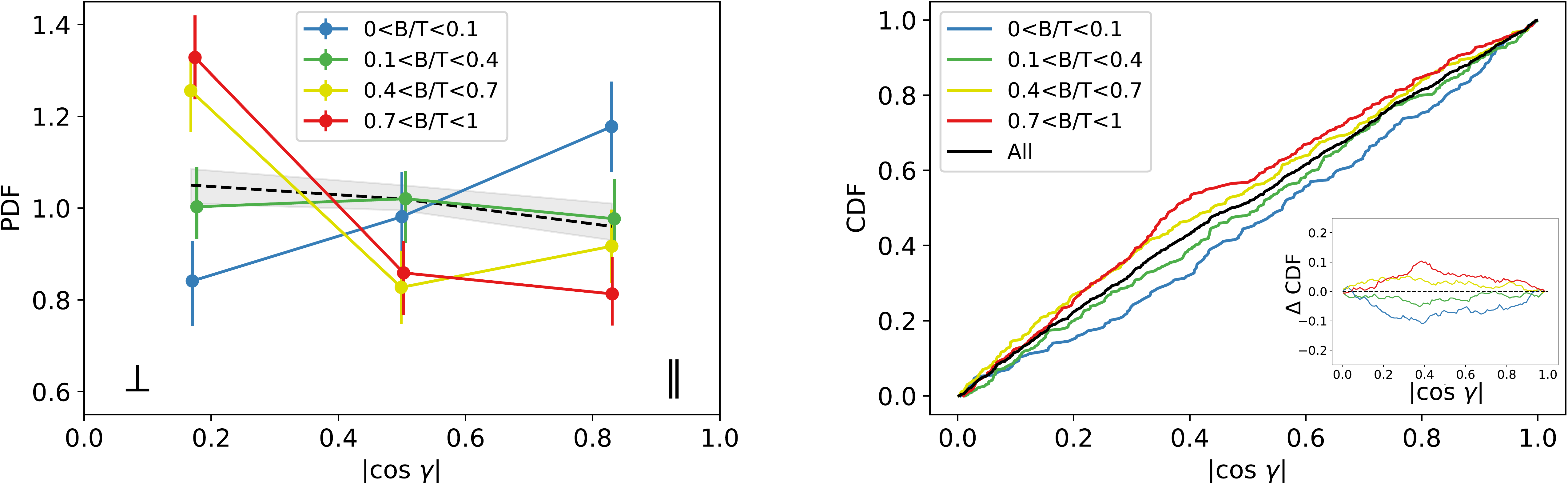}
\vspace{2.5mm} \\
\includegraphics[scale=0.33]{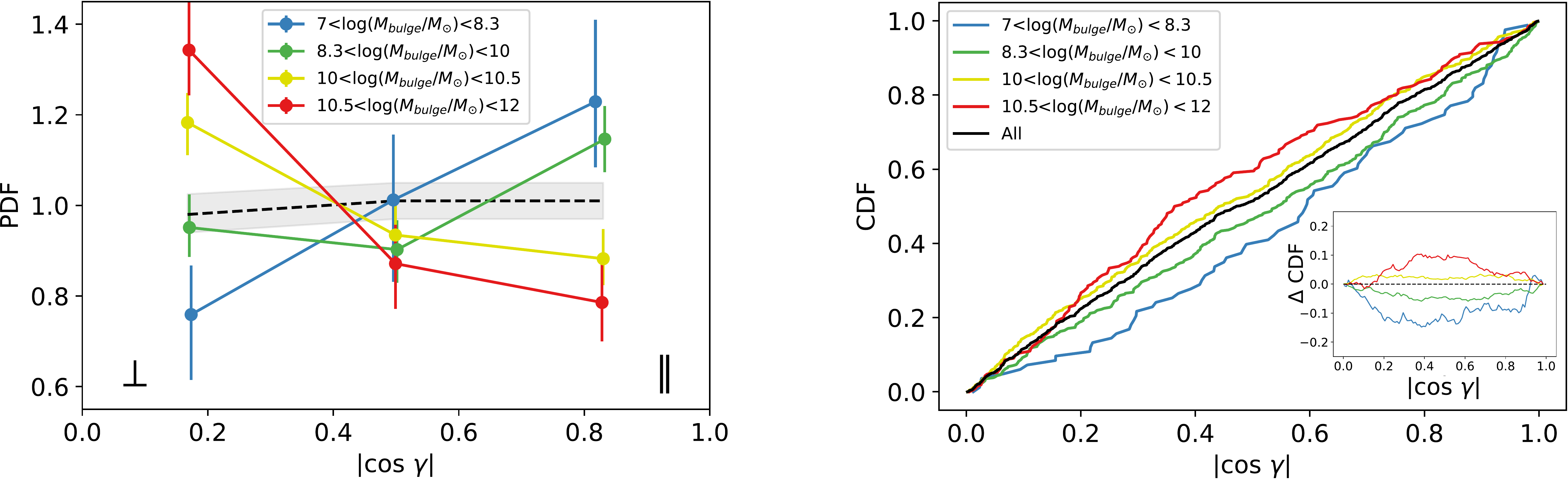}
\caption{PDFs (left) and CDFs (right) of the spin--filament alignments for 1121 SAMI galaxies divided in ranges of $M_\star$ (top), B/T (middle) and $M_{\rm bulge}$ (bottom). Galaxies with higher values of $M_\star$, B/T and $M_{\rm bulge}$ have spins aligned preferentially perpendicular to the closest filament, whereas galaxies with lower values of $M_\star$, B/T and $M_{\rm bulge}$ tend to have spins parallel to the closest filament. Significant differences are detected between the |$\cos\gamma$| distributions of the galaxies belonging to the extreme ranges in $M_\star$, B/T and $M_{\rm bulge}$.}
\label{TrendsGalaxyProperties}
\end{figure*}

For the purpose of our subsequent analyses, we group the SAMI galaxies into two mass bins: low-mass galaxies, with $9<\log{(M_\star/M_{\odot})}<10.2$, and high-mass galaxies, with $10.2<\log{(M_\star/M_{\odot})}<12$. The top left panel of Figure~\ref{TrendsGalaxyProperties} shows the PDFs of the spin--filament alignments for these two sub-samples. We find $p_{\rm K-S}<0.05$ only for the high-mass galaxies, which show a tendency to perpendicular alignment. Using the two-sample K-S test, we check the null hypothesis that the |$\cos\gamma$| distributions of the low-mass and high-mass galaxy sub-samples are drawn from the same parent population. This test returns $p_{\rm 2\,K-S}=0.006$, listed in Table~\ref{PAstellResults} and indicating that the two distributions are unlikely to be drawn from the same population. The top right panel of Figure~\ref{TrendsGalaxyProperties} shows the cumulative distribution functions (CDFs) for the two sub-samples and the entire population. The insert displays the difference $\Delta$(CDF)$\,=\,$CDF(sub-sample)\,$-$\,CDF(all) as a function of |$\cos\gamma$|. 
 
The PDFs and CDFs of the spin--filament alignments for the SAMI galaxies divided into B/T and $M_{\rm bulge}$ ranges are displayed in the middle and bottom panels of Figure~\ref{TrendsGalaxyProperties}, respectively. The selection of the ranges, the number of galaxies for each sub-sample, the average |$\cos\gamma$| values and the $p$-values from the K-S tests are reported in Table~\ref{PAstellResults}. Galaxies with the lowest values of B/T and $M_{\rm bulge}$ tend to have spins parallel to the closest filament, while galaxies with highest B/T and $M_{\rm bulge}$ show preferentially perpendicular orientations. These deviations from a uniform PDF are statistically significant. Comparing the |$\cos\gamma$| distributions of the galaxies in the most extreme ranges, the pairs of sub-samples in B/T and $M_{\rm bulge}$ have statistically different distributions. Finally, the comparison of the CDF plots shows that the largest deviations of the sub-samples from the population as a whole is displayed for $M_{\rm bulge}$.

\begin{table*}
  \caption{Galaxy spin--filament alignments for various galaxy properties. Column~1 lists the galaxy property, column~2 the set(s) of galaxy sub-samples for that property, column~3 the number of galaxies in each sub-sample, column~4 the average |$\cos\gamma$|, column~5 the $p$-value from the K-S test, and column~6 the tendency of the alignment (where there is a significant deviation from the reconstructed uniform distribution). Columns~7 and~8 give the sub-samples for which |$\cos\gamma$| distributions are compared with a two-sample K-S test; the associated $p$-value is listed in column~9. Significant $p$-values (those less than 0.05) are highlighted in bold.}
\makebox[\textwidth][c]{
    \begin{tabular}{lccccc|ccc}
\toprule
  Galaxy Property  & Selection & $N_{\rm gal}$& <|$\cos\gamma$|> & $p_{\rm K-S}$ & Alignment &  \multicolumn{1}{c}{Sample 1} & Sample 2 & $p_{\rm 2\,K-S}$\\
 \midrule
$\log{(M_\star/M_{\odot})}$ & [9; 9.5] & 168 & 0.521$\pm$0.022 & 0.934 &  &  & & \\
 & [9.5; 10.2] & 321 & 0.551$\pm$0.016 & 0.155 & & & & \\
 & [10.2; 10.9] & 454 & 0.425$\pm$0.014 & \textbf{0.002} & $\perp$ & & & \\
 & [10.9; 12] & 178 & 0.433$\pm$0.021& \textbf{0.002}& $\perp$ & & & \\
 &  & &  & & & & & \\
& [9; 10.2] & 489 & 0.539$\pm$0.013 & 0.102
 & & [9; 10.2] &[10.2; 12]  &\textbf{0.006}  \\
& [10.2; 12] & 632 & 0.427$\pm$0.011 & $\mathbf{1\times10^{-4}}$& $\perp$ & & & \\
&  & &  & & & & & \\
& ${\rm PA}_{\rm gas}$, [8; 9] & 180 & 0.561$\pm$0.042 & 0.954 & &[8; 9] & [10.2; 12] & \textbf{0.022}\\
& ${\rm PA}_{\rm gas}$, [9; 10.2] & 489 & 0.482$\pm$0.012 & 0.646 & & & &\\
& ${\rm PA}_{\rm gas}$, [10.2; 12] & 632 & 0.441$\pm$0.010 & $\mathbf{8\times10^{-5}}$& $\perp$& & &\\
\midrule
B/T & [0; 0.1] & 214 & 0.559$\pm$0.020 &  \textbf{0.001} & $\parallel$ & [0; 0.1] & [0.7; 1] & $\mathbf{7\times10^{-5}}$\\
& [0.1; 0.4] & 344 & 0.517$\pm$0.016 & 0.991 & & & & \\
& [0.4; 0.7] & 301 & 0.437$\pm$0.017 & \textbf{0.017} &$\perp$ & & & \\
& [0.7; 1] & 262 & 0.375$\pm$0.018 & $\mathbf{7\times10^{-5}}$& $\perp$& & & \\
\midrule
$\log{(M_{\rm bulge}/M_{\odot})}$ & [7; 8.3] & 83 & 0.591$\pm$0.030 & \textbf{0.008} & $\parallel$ & [7; 8.3]  & [10.5; 12] &  $\mathbf{3\times10^{-5}}$\\
& [8.3; 10] & 369 & 0.542$\pm$0.015 & 0.113 & & & & \\
& [10; 10.5] & 459 & 0.440$\pm$0.014& \textbf{0.015} & $\perp$ & & & \\
& [10.5; 12] & 210 & 0.363$\pm$0.019 & $\mathbf{1\times10^{-5}}$ & $\perp$& & & \\
\midrule
$(V/\sigma)_e$ & [0; 0.3] & 207 & 0.413$\pm$0.021 & \textbf{0.037} &  $\perp$ & [0; 0.3] &[1; 1.7] & 0.433\\
& [0.3; 0.7] & 459 & 0.449$\pm$0.014 & \textbf{0.020}& $\perp$ & & &  \\
& [0.7; 0.1] & 287 & 0.502$\pm$0.018 & 0.582& & & &  \\
& [1; 1.7] & 118 & 0.468$\pm$0.027& 0.984 & & & &  \\
\midrule
$\lambda_e$ & [0; 0.3] & 229 & 0.408$\pm$0.019 & \textbf{0.011} &  $\perp$ & [0; 0.3] &[0.6; 1] & 0.245\\
& [0.3; 0.5] & 330 & 0.450$\pm$0.016 & \textbf{0.015} & $\perp$ & & &  \\
& [0.5; 0.6] & 214 & 0.473$\pm$0.020 & 0.993& & & &  \\
& [0.6; 1] & 298 & 0.500$\pm$0.017& 0.989 & & & &  \\
\midrule
Age [Gyr] & [0; 2] & 285 & 0.534$\pm$0.016 & 0.881 & &[0; 2] & [7; 17.5]& \textbf{0.018}\\
& [2; 5] & 537 & 0.472$\pm$0.011 & 0.437 & & & & \\
& [5; 7] & 160 & 0.419$\pm$0.021& \textbf{0.006}& $\perp$ & & & \\
& [7; 17.5] & 139 & 0.405$\pm$0.023 & \textbf{0.014} & $\perp$& & &\\
\midrule
Visual  & Elliptical & 90 & 0.429$\pm$0.026 & 0.099  & & Elliptical+S0 & Late-type& \textbf{0.023} \\
Morphology & S0 & 358 & 0.433$\pm$0.014 &
\textbf{0.004}& $\perp$  & & &\\
& Late-type & 647 & 0.513$\pm$0.011 & 0.872 & & & &\\
&  & &  & & & & & \\
& Late-type, $\scriptstyle{7<\log{(M_{\rm bulge}/M_{\odot})}<8.3}$ & 75 & 0.638$\pm$0.034 & \textbf{0.016} & $\parallel$ &  & & \\
& Late-type, $\scriptstyle{10.5<\log{(M_{\rm bulge}/M_{\odot})}<12}$ & 94 & 0.441$\pm$0.031& \textbf{0.038} & $\perp$ & & &
\\
\midrule
Spectral  & Passive & 394 & 0.409$\pm$0.015 & $\mathbf{1\times10^{-4}}$& $\perp$& Passive& Star-forming & \textbf{0.003} \\
Classification & Star-forming & 714 & 0.502$\pm$0.011 & 0.990& & & &\\
&  & &  & & & & & \\
& Star-forming, $\scriptstyle{7<\log{(M_{\rm bulge}/M_{\odot})}<8.3}$ & 72 & 0.606$\pm$0.032 & \textbf{0.038}& $\parallel$ & & & \\
& Star-forming, $\scriptstyle{10.5<\log{(M_{\rm bulge}/M_{\odot})}<12}$ & 114&0.423$\pm$0.027 & \textbf{0.003}& $\perp$& & &\\
\midrule
Kinematic  & Slow rotator & 68 & 0.408$\pm$0.032 & 0.240& & Slow rotators & Fast rotators  & 0.730\\
Morphology & Fast rotator & 864 & 0.471$\pm$0.009 & 0.100& & & &\\
&& & & & & & &\\
&  Fast rotator, $\scriptstyle{7<\log{(M_{\rm bulge}/M_{\odot})}<8.3}$ & 49 & 0.693$\pm$0.041 & \textbf{0.004}& $\parallel$ & & & \\
&  Fast rotator, $\scriptstyle{10.5<\log{(M_{\rm bulge}/M_{\odot})}<12}$ & 247& 0.406$\pm$0.018&\textbf{0.002} & $\perp$& & & \\
\midrule
Local  & Central & 385 & 0.441$\pm$0.013& \textbf{0.036}& $\perp$& Central & Isolated& 0.439 \\
environment & Satellite & 328 & 0.495$\pm$0.015 & 0.092& & & &\\
& Isolated & 408 & 0.484$\pm$0.013 & 0.758& & & &\\
& &  & & & & & &\\
& Isolated, $\scriptstyle{7<\log{(M_{\rm bulge}/M_{\odot})}<8.3}$ & 35 & 0.643$\pm$0.050 & \textbf{0.047}& $\parallel$ & & & \\
& Isolated, $\scriptstyle{10.5<\log{(M_{\rm bulge}/M_{\odot})}<12}$ & 147 & 0.429$\pm$0.024&\textbf{0.017} & $\perp$& & & \\

\bottomrule
\end{tabular}
}
\label{PAstellResults}
\end{table*}

\subsection{Secondary tracers}
\label{Secondary tracers}
We investigate the spin--filament alignments as a function of galaxy properties that show weaker correlations with respect to $M_{\rm bulge}$. We focus on light-weighted stellar age and the degree of ordered stellar rotation. We also classify galaxies according to visual morphological type, star formation properties, kinematic morphology and membership in a galaxy group. The correlation of these properties with $|\cos\gamma$| are expected due to their dependency on $M_{\star}$ and B/T (e.g., \citealp{vandeSande2017,vandeSande2018}), and thus $M_{\rm bulge}$. The results are listed in Table~\ref{PAstellResults}, while the figures similar to Figure~\ref{TrendsGalaxyProperties}, containing the PDFs and CDFs, are reported as supplementary material to this article. We also report the $\lambda_e$ versus ellipticity plot with points colour-coded according to $M_{\rm bulge}$ showing the expected dependency of $\lambda_e$ on $M_{\rm bulge}$: galaxies with lower $\lambda_e$ values have higher $M_{\rm bulge}$.

Significant results are found only for tendencies towards more perpendicular alignments for galaxies with $(V/\sigma)_e$\,<\,0.7 ($p_{\rm K-S}=0.020$), $\lambda_e$\,<\,0.5 ($p_{\rm K-S}=0.015$), age\,>\,5\,Gyr ($p_{\rm K-S}=0.014$), S0 morphology ($p_{\rm K-S}=0.004$), passive spectral type ($p_{\rm K-S}=1\times10^{-4}$) and central group membership ($p_{\rm K-S}=0.036$). Similar results are obtained using inclination-corrected $\lambda_e$ values from \citet{vandeSande2021b}. We also note that similar PDFs are found as a function of $(V/\sigma)_e$ and $\lambda_e$.

\begin{figure*}
\includegraphics[width=\columnwidth]{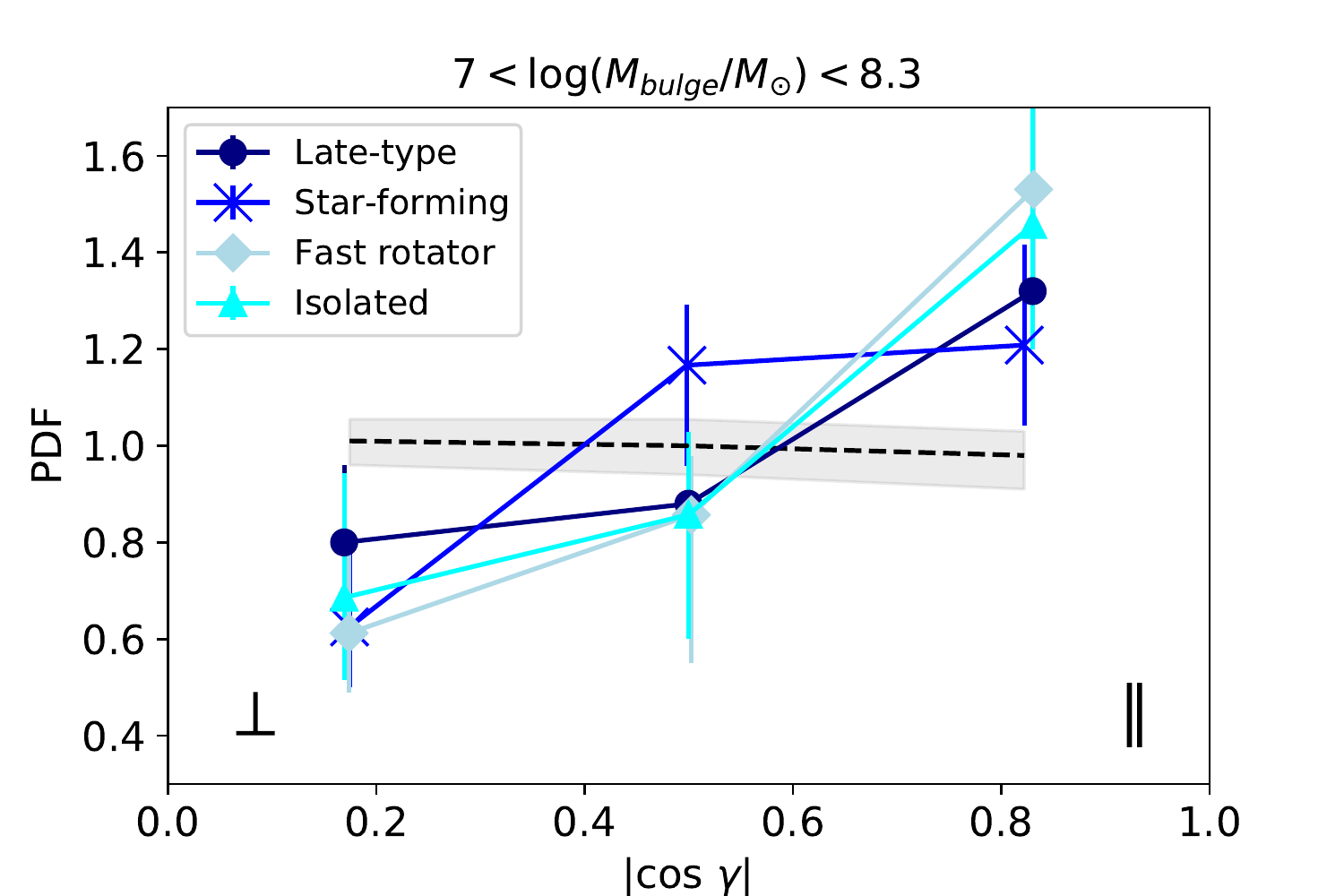}
\includegraphics[width=\columnwidth]{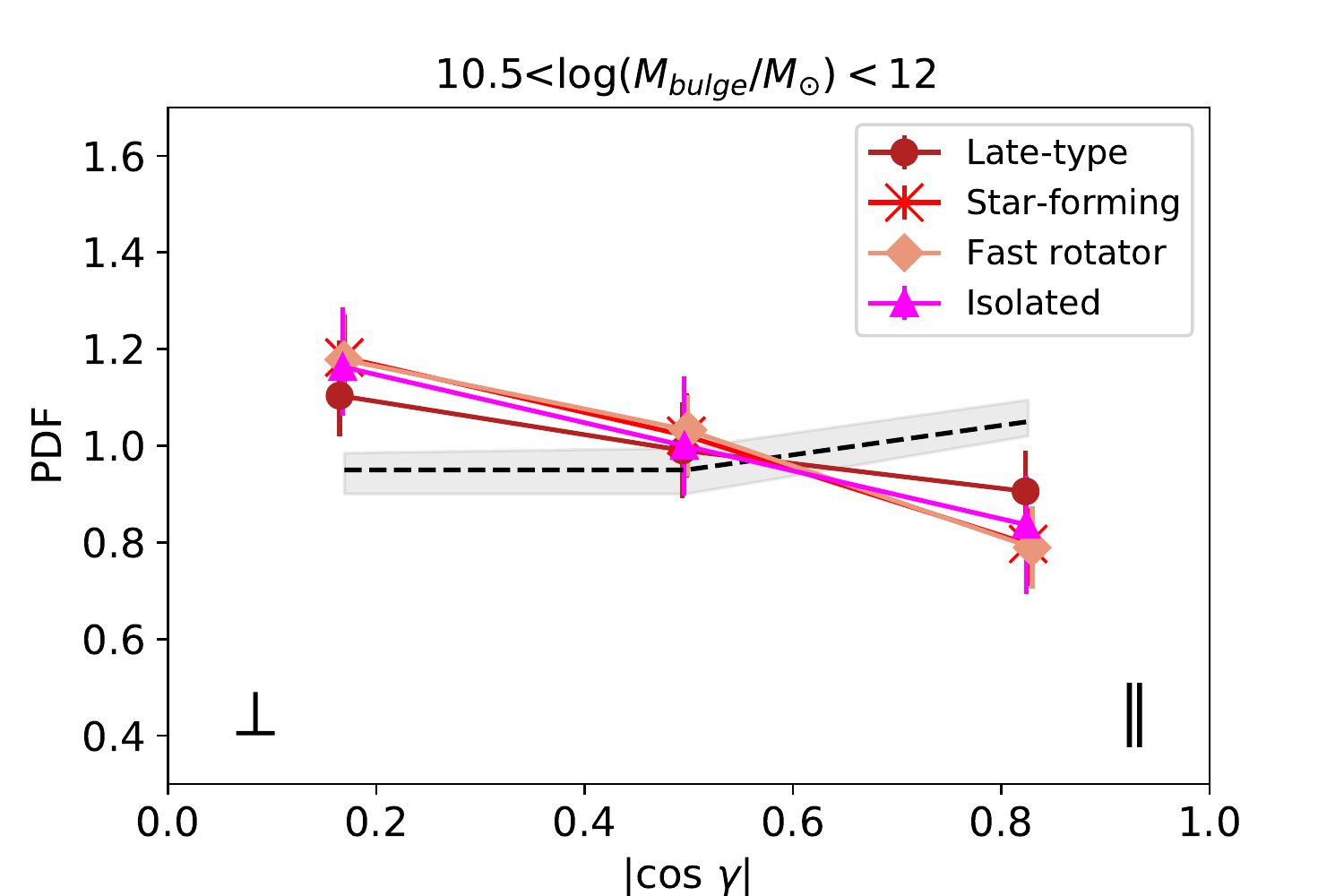}
\caption{PDFs of spin--filament alignments for late-type galaxies, star-forming galaxies, fast rotators and isolated galaxies with $7<\log{ (M_{\rm bulge}/M_{\odot})}<8.3$ (left panel) and with $10.5<\log{(M_{\rm bulge}/M_{\odot})}<12$ (right panel). Low-$M_{\rm bulge}$ galaxies show significant tendencies to parallel alignments, while high-$M_{\rm bulge}$ galaxies have more perpendicular alignments, in agreement with the observed dependency of the signal on $M_{\rm bulge}$.}
\label{TrendsVisualKinematicMorphologyMassBulge}
\end{figure*}

We check whether late-type galaxies, star-forming galaxies, fast rotators and isolated galaxies with low-bulge masses ($7<\log{(M_{\rm bulge}/M_{\odot})}<8.3$) show a preferential parallel alignment, in agreement with the observed dependency of the signal on $M_{\rm bulge}$. Similarly, we test whether these classified galaxies with high-bulge masses ($10.5<\log{(M_{\rm bulge}/M_{\odot})}<12$) tend to be aligned more perpendicularly. The PDFs for these sub-samples are shown in Figure~\ref{TrendsVisualKinematicMorphologyMassBulge}. Late-types, star-forming galaxies, fast rotators and isolated galaxies with low-$M_{\rm bulge}$ have statistically significant tendencies towards parallel alignments (with $p_{\rm K-S}=0.016$, 0.038, 0.004 and 0.047 respectively), while those with high-$M_{\rm bulge}$ have a more perpendicular tendency (with $p_{\rm K-S}=0.038$, 0.003, 0.002 and 0.017 respectively). This confirms that morphological, star formation, kinematic and environmental properties are secondary tracers of the spin--filament alignments with respect to $M_{\rm bulge}$.

\subsection{S0 galaxies}
\label{S0 galaxies}

Using the analysis of the spin--filament alignments, we aim to better understand the formation of S0 galaxies. A significant preference for perpendicular alignments for S0 galaxies ($p_{\rm K-S}=0.004$) is detected in the previous Section. We now explore this trend as a function of $M_\star$ and the kinematic misalignment between stars and gas, in order to understand which sub-samples dominate the perpendicular signal for S0 galaxies. The choice of these galaxy properties is motivated by the fact that \citet{Kraljic2021} found that the perpendicular alignment of S0 MaNGA galaxies is dominated by low-mass galaxies (at apparent odds with the predicted mass dependency of the signal) or by misaligned galaxies. 

\begin{figure}
\centering
\includegraphics[scale=0.32]{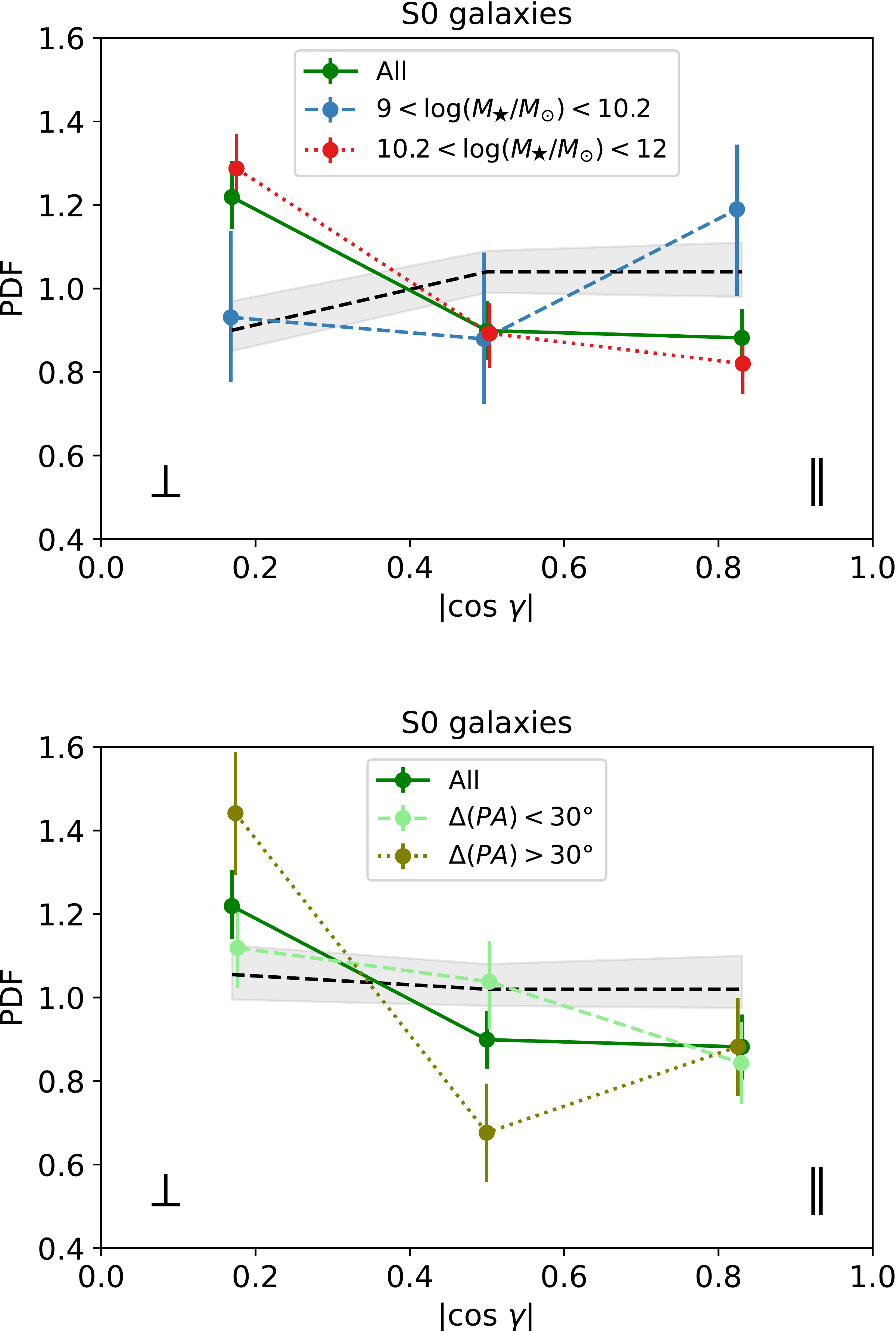}
\caption{PDFs of the spin--filament alignments for the selected 216 SAMI S0 galaxies divided into ranges of $M_\star$ (top) and $\Delta{\rm PA}$ (bottom). The strongest preferences for perpendicular alignments are seen for S0 galaxies that are high-mass or misaligned.}
\label{TrendsS0}
\end{figure}

According to the visual morphology classification, there are 469 S0 galaxies. Of this sample, we select only fast rotators and galaxies best fitted by a double-component S\'ersic bulge plus exponential disc model in the $r$-band (see Section~\ref{Photometric position angles}). This allows us to exclude possible ellipticals and late-type galaxies that might contaminate the signal. The final S0 sample comprises 216 galaxies. The top and bottom panels of Figure~\ref{TrendsS0} show the PDFs of the spin--filament alignments for the 216 S0 galaxies divided into $M_\star$ and $\Delta{\rm PA}$ ranges, respectively; the results are reported in Table~\ref{PAstellResultsS0}. More statistically significant perpendicular alignments are observed for high-mass S0 galaxies and kinematically-misaligned S0 galaxies (with $\Delta{\rm PA}>30\degree$) relative to low-mass or aligned S0s. We chose $\Delta{\rm PA}=30\degree$ as the threshold between misaligned and aligned galaxies in line with previous studies (e.g., \citealp{Bryant2019,Duckworth2019}), but our results do not change if we use 20$\degree$ or 40$\degree$ as the threshold. 

\begin{figure*}
\includegraphics[width=18cm]{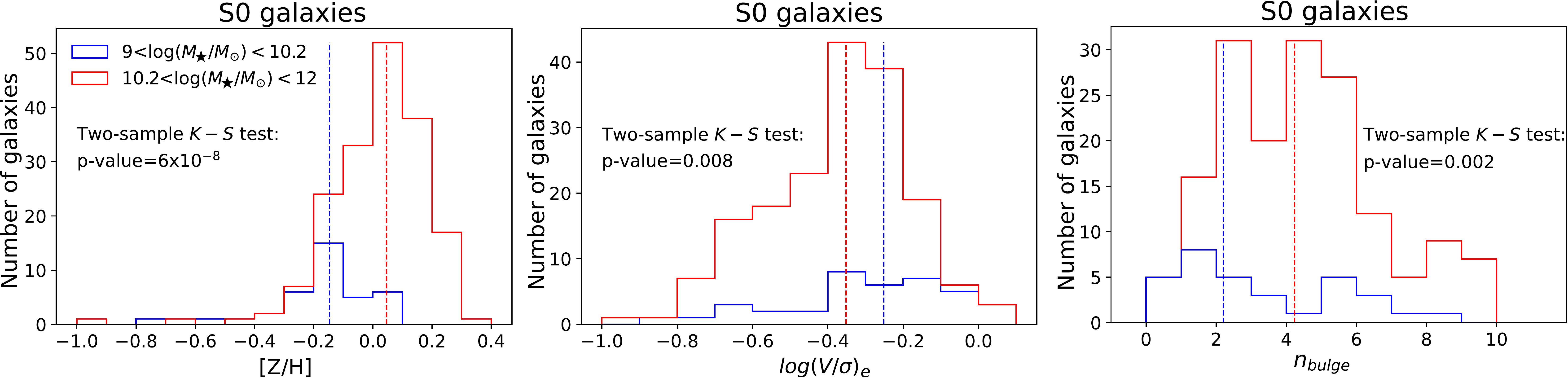}
\caption{Distributions as a function of [Z/H] (left), $(V/\sigma)_{e}$ (middle), and S\'ersic index of the bulge (right) for 178 high-mass S0 galaxies (red) and 38 low-mass S0 galaxies (blue). The dotted lines show the median values. High-mass S0 galaxies are significantly more metal-rich and dispersion-dominated, and tend to have higher S\'ersic indices for their bulges relative to low-mass S0 galaxies.}
\label{PropertiesS0}
\end{figure*}

Using the two-sample K-S test, we find that the |$\cos\gamma$| distribution of the high-mass S0 galaxies is marginally different to the |$\cos\gamma$| distribution of low-mass S0 galaxies ($p_{\rm 2\,K-S}=0.043$). On the other hand, the |$\cos\gamma$| distributions of aligned and misaligned S0 galaxies are consistent ($p_{\rm 2\,K-S}=0.185$). High-mass S0 galaxies are also significantly more metal-rich ($p=6\times10^{-8}$), dispersion-dominated ($p=0.008$), and tend to have a more classical \citet{deVaucouleurs1948,deVaucouleurs1956} bulge profile ($p=0.002$) compared to low-mass S0 galaxies (high-mass S0 galaxies have median $n_{\rm bulge}=4.24\pm0.17$, while low-mass S0 galaxies have median $n_{\rm bulge}=2.20\pm0.42$). The distributions as a function of [Z/H], $(V/\sigma)_{e}$ and S\'ersic index of the bulge are shown in Figure~\ref{PropertiesS0}. These findings suggest that high-mass and low-mass S0 galaxies represent different galaxy populations.

Overall, we find that the perpendicular tendency for S0 galaxies is dominated by high-mass galaxies and not by low-mass S0 galaxies as in \citet{Kraljic2021}. They studied 114 low-mass S0 galaxies and 155 high-mass S0 galaxies, while our samples count 38 and 178, respectively. The discrepant finding might be due to the different criteria applied to select S0 galaxies, or could be explained in terms of massive bulges for the low-mass S0 MaNGA galaxies giving rise to a strong perpendicular trend.

\begin{table*}
\caption{Galaxy spin--filament alignments for 216 S0 galaxies as a function of $M_\star$ and $\Delta{\rm PA}$. The columns are the same as in Table~\ref{PAstellResults}.}
\makebox[\textwidth][c]{
\begin{tabular}{lccccc|ccc}
\toprule
Galaxy Property  & Selection & $N_{\rm gal}$& <|$\cos\gamma$|> & $p_{\rm K-S}$ & Alignment & 
\multicolumn{1}{c}{Sample 1} & Sample 2 & $p_{\rm 2\,K-S}$\\
\midrule
& All & 216 & 0.433$\pm$0.016 & \textbf{0.007} &$\perp$ & & & \\
\midrule
$\log{(M_\star/M_{\odot})}$ & [9; 10.2] & 38 & 0.621$\pm$0.040 & 0.891 & &[9; 10.2] &[10.2: 12] & \textbf{0.043}\\
& [10.2; 12] & 178 & 0.402$\pm$0.017 & $\mathbf{8\times10^{-4}}$  &$\perp$ & & &  \\
\midrule
$\Delta{\rm PA}$ & <30\degree & 116 & 0.443$\pm$0.021 &  0.489 & & <30\degree& >30\degree & 0.185\\
& >30\degree & 100 & 0.342$\pm$0.029 & \textbf{0.013} &$\perp$ & & &  \\
\bottomrule
\end{tabular}
}
\label{PAstellResultsS0}
\end{table*}
 
\subsection{Spin--filament alignments of bulges and discs}
\label{spin--filament alignments of bulges and discs}

\begin{figure}
\includegraphics[width=\columnwidth]{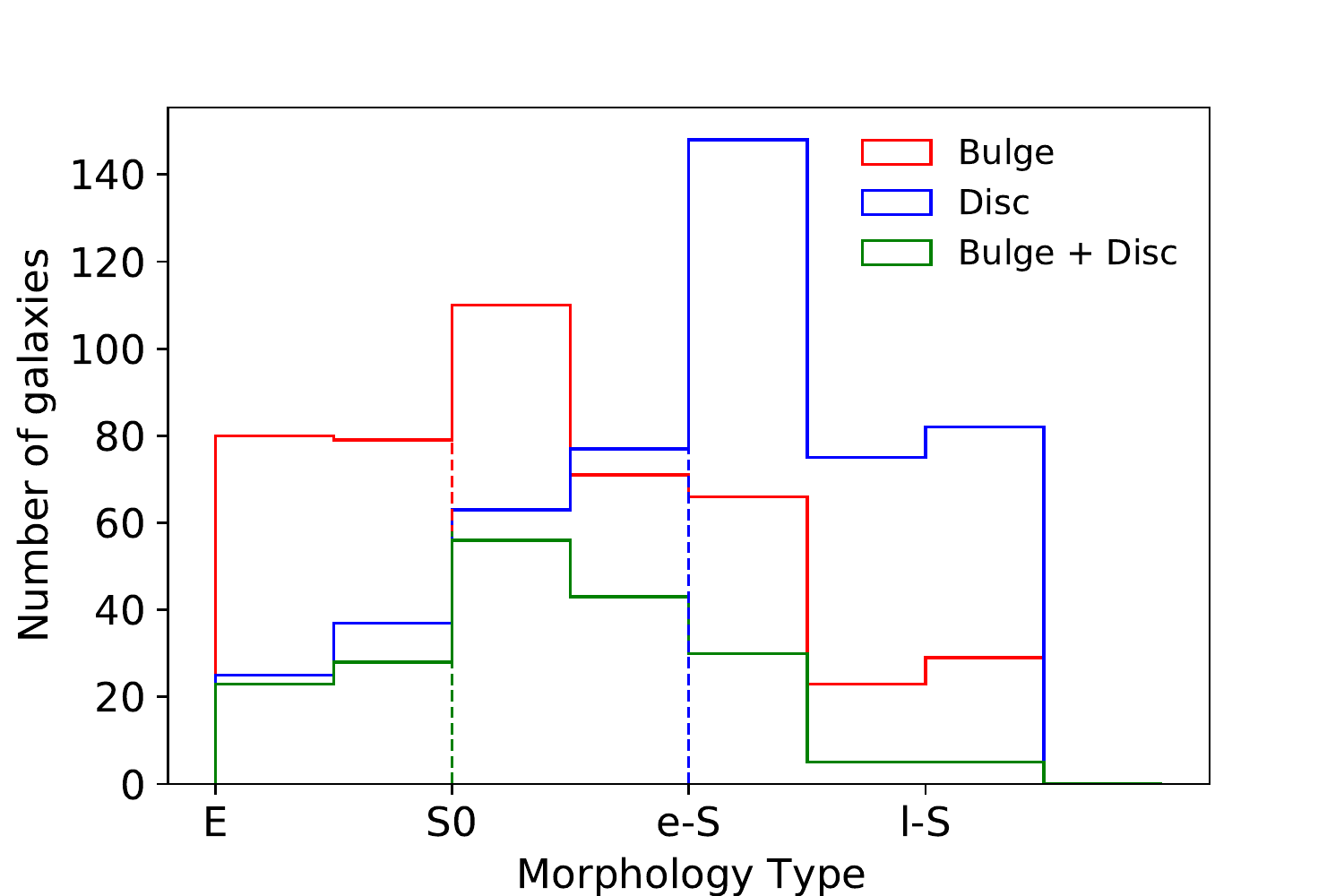}
\caption{Morphology distributions for 468 bulges, 516 discs and 196 galaxies having both components. The dashed lines mark the median morphological types.}
\label{HistogramBulgeDisc}
\end{figure}

We next investigate the separate spin--filament alignments of bulges and discs to understand the formation of these galaxy components. We take advantage of the 2D kinematic bulge/disc decomposition described in Section~\ref{Stellar kinematics of bulges and discs}, which allows us to measure the separate stellar kinematic PAs of bulges and discs. Out of the 1121 SAMI galaxies, kinematic PAs are available for 468 bulges and 516 discs, with 196 galaxies having both bulge and disc measurements (see Section~\ref{Galaxy sample and selection criteria}). Bulge PAs have an average 1$\sigma$ error of $\sim$5\degree, while it is $\sim$2\degree\ for disc PAs, highlighting the larger uncertainty in estimating PAs from the bulge velocity maps. To identify the orientation of the bulge and disc spin axes, we apply the method of Section~\ref{Orientation of the spin axes}. Figure~\ref{HistogramBulgeDisc} shows the visual morphology distributions for bulges, discs and galaxies with both components. Bulges belong mainly to early-type galaxies, while discs mainly trace late-type galaxies. Galaxies with both components are mostly S0s.

\begin{figure*}
\includegraphics[scale=0.33]{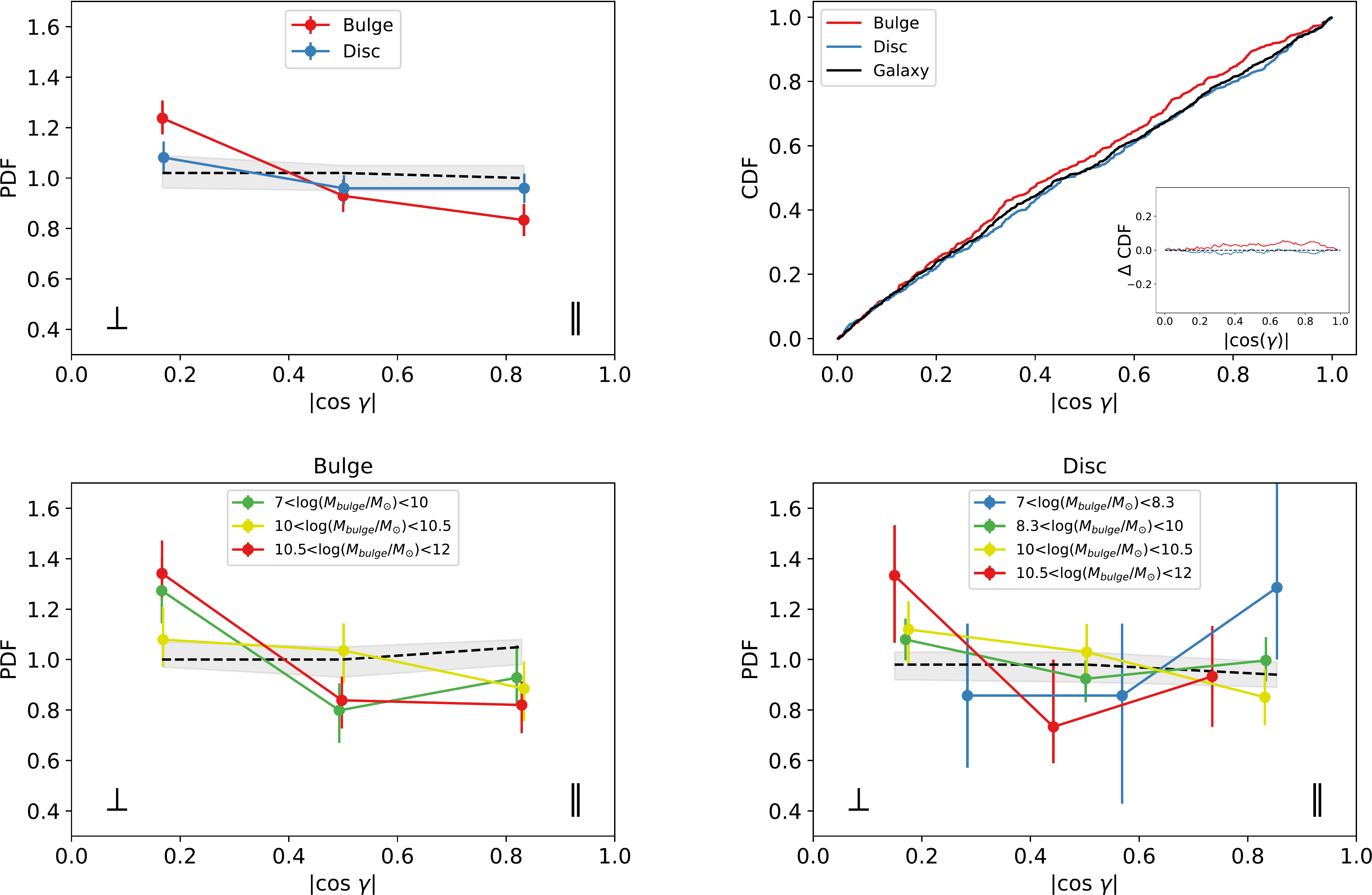}
\caption{Top panels: PDFs (left) and CDFs (right) of the spin--filament alignments for 468 bulges and 516 discs. Bottom panels: spin--filament alignments of bulges (left) and discs (right) divided into $M_{\rm bulge}$ ranges. Bulges show more perpendicular alignments, while discs tend to be parallel aligned for low $M_{\rm bulge}$ and perpendicularly aligned for high $M_{\rm bulge}$.}
\label{TrendsBulgeDisc}
\end{figure*}

In Figure~\ref{TrendsBulgeDisc} we display the PDFs and CDFs of the spin--filament alignments for bulges and discs. We also investigate the tendencies according to $M_{\rm bulge}$. Table~\ref{PAstellResultsBulgeDisc} lists the number of galaxies, the average |$\cos\gamma$| and the $p$-values from the K-S tests. Bulges have a significant perpendicular tendency, while discs show a uniform PDF. The |$\cos\gamma$| distributions of the bulges and the discs are not statistically different ($p_{\rm 2\,K-S}=0.139$), and they are consistent with the galaxy spin--filament alignments. The alignment for bulges tends to be more perpendicular at all $M_{\rm bulge}$ values (there is only one bulge component with $\log(M_{\rm bulge}/M_{\odot})<8.3$). Discs show a significant parallel tendency for galaxies with $7<\log(M_{\rm bulge}/M_{\odot})<8.3$, while the alignment is significantly perpendicular at $10.5<\log(M_{\rm bulge}/M_{\odot})<12$. The two disc |$\cos\gamma$| distributions are statistically different ($p_{\rm 2\,K-S}=0.005$). 

\begin{figure*}
\includegraphics[scale=0.33]{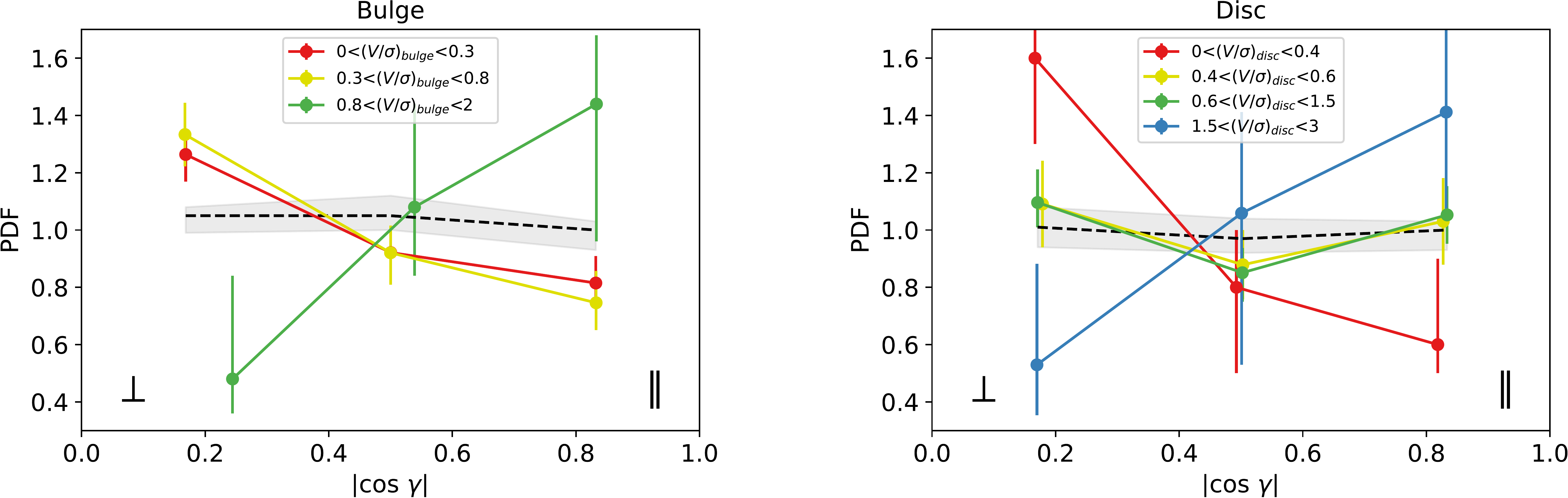}
\caption{Spin--filament alignments as a function of $(V/\sigma)$ for 468 bulges (left panel) and 516 discs (right panel). Dispersion-dominated bulges and discs tend to have perpendicular alignments, while rotation-dominated bulges and discs show a parallel tendency.}
\label{TrendsBulgeDisc2}
\end{figure*}

To better understand the kinematic characteristics of the bulges and discs, we measure $V/\sigma$ for both components as the flux-weighted means within 1\,$R_e$ of the galaxies \citep{Cappellari2007,Oh2020}. Bulges tend to have lower $V/\sigma$ values with respect to discs, with median $(V/\sigma)_{\rm bulge}=0.30$ and median $(V/\sigma)_{\rm disc}=0.84$. Figure~\ref{TrendsBulgeDisc2} shows the PDFs of the spin--filament alignments as a function of $(V/\sigma)$ for bulges (left panel) and discs (right panel). Perpendicular tendencies are seen for dispersion-dominated bulges with $(V/\sigma)_{\rm bulge}<0.8$ ($p_{\rm K-S}=0.004$), while rotation-dominated bulges with $(V/\sigma)_{\rm bulge}>0.8$ show a more parallel trend ($p_{\rm K-S}=0.016$). Rotation-dominated discs tend to show a parallel alignment ($p_{\rm K-S}=0.037$), while dispersion-dominated discs tend to be perpendicular ($p_{\rm K-S}=0.024$). 

\begin{figure*}
\includegraphics[scale=0.33]{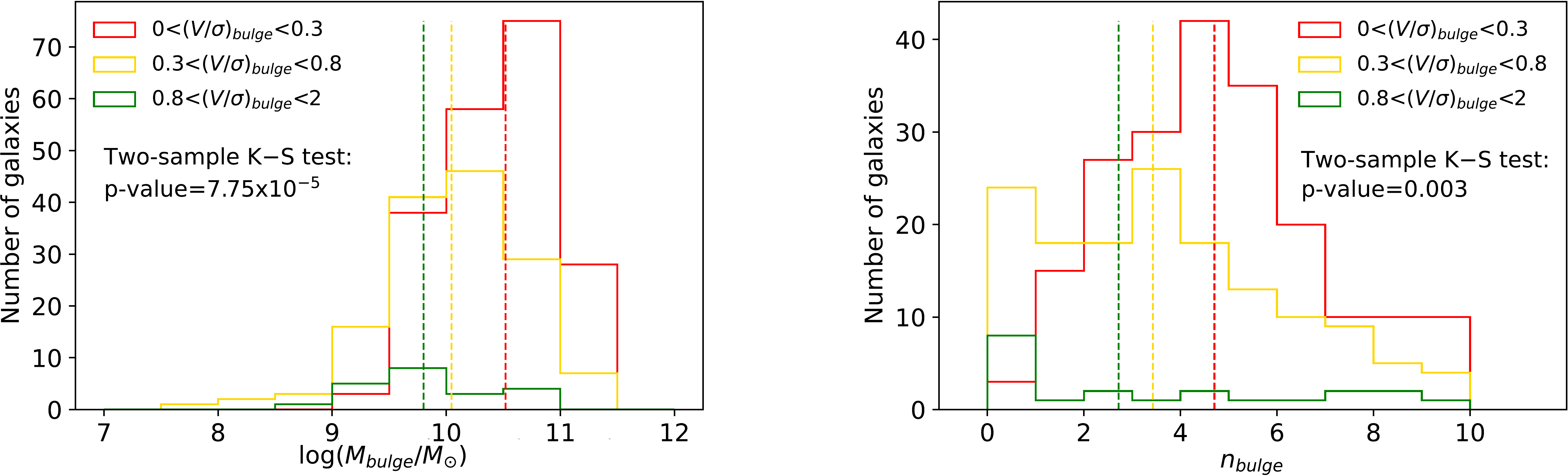}
\caption{Bulge mass (left) and bulge S\'ersic index (right) distributions divided into $(V/\sigma)_{\rm bulge}$ ranges. Dispersion-dominated bulges tend to have high-bulge masses and median $n_{\rm bulge}=4.70$, while rotation-dominated bulges show low-bulge masses and lower median n$_{\rm bulge}$.}
\label{HistogramsBulge} 
\end{figure*}

Since we find a parallel tendency for rotation-dominated bulges that we do not detect for bulges with $7<\log(M_{\rm bulge}/M_{\odot})<8.3$, we inspect in Figure~\ref{HistogramsBulge} the bulge mass distributions (left panel) and the bulge S\'ersic index distributions (right panel) according to $(V/\sigma)_{\rm bulge}$. Rotation-dominated bulges tend to have low $M_{\rm bulge}$ values with respect to dispersion-dominated bulges (two-sample K-S test: $p$-value $=7.75\times 10^{-5}$ ). Thus, the trends of the PDFs in Figure~\ref{TrendsBulgeDisc2} are in agreement with the dependency of the signal on $M_{\rm bulge}$. The PLS technique confirms $M_{\rm bulge}$ as the primary correlation parameter over $(V/\sigma)_{\rm bulge}$. Dispersion-dominated bulges show a peak around the de~Vaucouleurs profile, with median $n_{\rm bulge}=4.70\pm0.15$. On the other hand, rotation-dominated bulges have median $n_{\rm bulge}=2.72\pm0.67$. The two $n_{\rm bulge}$ distributions are also significantly different according to the two-sample K-S test ($p$-value $=$ 0.003). 

\begin{figure}
\includegraphics[scale=0.225]{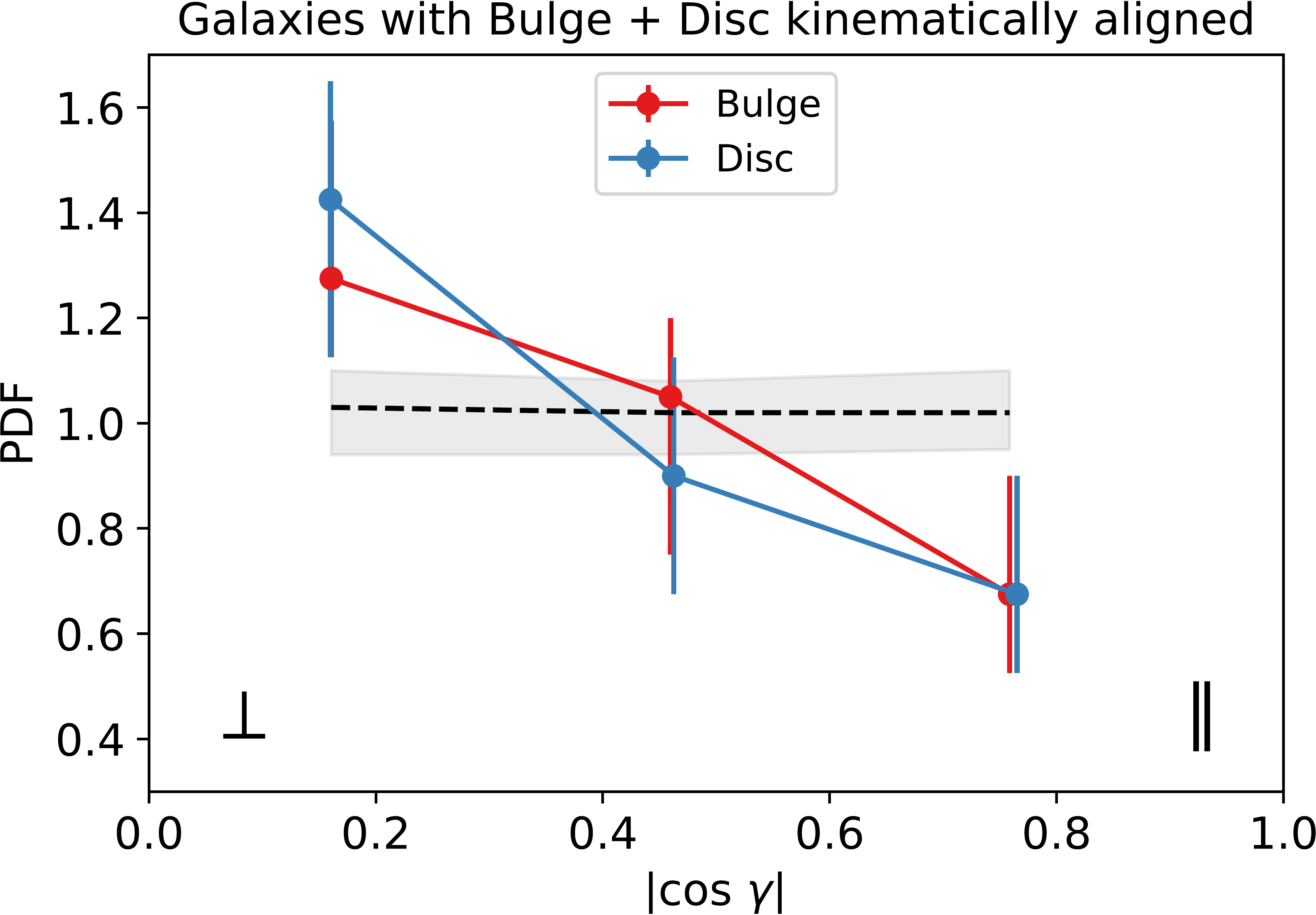}
\caption{PDFs of the bulge and disc spin--filament alignments for 137 SAMI galaxies having the components kinematically aligned. Both bulges and discs show more perpendicular alignments.}
\label{TrendsBulgeDiscBothComponents}
\end{figure}

Finally, we analyse the 196 SAMI galaxies with available |$\cos\gamma$| for both the bulge and the disc components. About 65\% of these galaxies are visually classified as S0s. Accounting for any kinematic misalignment, we investigate the tendencies as a function of the absolute difference between the stellar kinematic PAs of the bulges and the discs, $\Delta{\rm PA}_{\rm bulge - disc}=|\rm PA_{\rm bulge}-PA_{\rm disc}|$, and the associated uncertainties $\delta{\rm PA}_{\rm bulge - disc}=\Delta{\rm PA}_{\rm bulge - disc}/(\delta{\rm PA}_{\rm bulge}+\delta{\rm PA}_{\rm disc})$. Most (70\%) galaxies have aligned components: $\Delta{\rm PA}_{\rm bulge - disc}<30$\degree or $\delta{\rm PA}_{\rm bulge - disc}<3$, i.e. ${\rm PA}_{\rm bulge}$ and ${\rm PA}_{\rm disc}$ are less than 3$\sigma$ different. For these galaxies both 
bulges and discs show a tendency to perpendicular alignments, as shown in Figure~\ref{TrendsBulgeDiscBothComponents}, and in agreement with the perpendicular trend found for S0 galaxies in Section~\ref{S0 galaxies}. The remaining (30\%) galaxies have bulge and disc significantly misaligned (30\degree$<\Delta{\rm PA}_{\rm bulge - disc}<90$\degree and $\delta{\rm PA}_{\rm bulge - disc}>3$) and no significant tendencies. 

In conclusion, although these results are affected by the limitations of the bulge-disc decompositions and are based on small samples (see Section~\ref{Caveats}), they show how the study of the bulge and disc spin--filament alignments provides further clues on the tendencies for the various galaxy populations.

\begin{table*}
\caption{Spin--filament alignments of bulges and discs. Column 1 specifies the galaxy component, while columns 2--10 are the same as in Table~\ref{PAstellResults}.}
\makebox[\textwidth][c]{
\begin{tabular}{lcccccc|ccc}
\toprule 
Galaxy Component  & Galaxy Property  & Selection & $N_{\rm gal}$& <|$\cos\gamma$|> & $p_{\rm K-S}$ & Alignment & \multicolumn{1}{c}{Sample 1} & Sample 2 & $p_{\rm 2\,K-S}$\\
\midrule
Bulge & & All & 468& 0.428$\pm$0.013 & \textbf{0.001} & $\perp$ &  Bulge & Disc & 0.139 \\
Disc & & All & 516 & 0.459$\pm$0.013 & 0.244 & & & &\\
\midrule
Bulge & $\log{(M_{\rm bulge}/M_{\odot})}$ & [7; 10] & 139 & 0.486$\pm$0.025 & \textbf{0.020}& $\perp$ & [7; 10] &[10.5; 12] & 0.403\\
& & [10; 10.5] & 139 & 0.457$\pm$0.024 & 0.506 & & & &\\
& & [10.5; 12] & 161 & 0.383$\pm$0.022 & \textbf{0.004} & $\perp$ & & &\\
& & & & & & & & &\\
& $(V/\sigma)_{\rm bulge}$ & [0; 0.3] & 254 & 0.404$\pm$0.018 & \textbf{0.004}& $\perp$ & [0; 0.3] & [0.8; 2] & \textbf{0.003}\\
& & [0.3; 0.8] & 189 & 0.420$\pm$0.021 & \textbf{0.008} & $\perp$ & & &\\
& & [0.8; 2] & 25 & 0.685$\pm$0.047 & \textbf{0.016} & $\parallel$ & & &\\
\midrule
Disc & $\log{(M_{\rm bulge}/M_{\odot})}$ & [7; 8.3] & 21 & 0.685$\pm$0.056 & \textbf{0.030}& $\parallel$ & [7; 8.3] & [10.5; 12] & \textbf{0.005}\\
& & [8.3; 10] & 289 & 0.458$\pm$0.018 & 0.329 & & & &\\
& & [10; 10.5] & 134 & 0.486$\pm$0.025 & 0.706 & & & &\\
& & [10.5; 12] & 45 & 0.348$\pm$0.037 & \textbf{0.025} & $\perp$ & & &\\
& & & & & & & & &\\
& $(V/\sigma)_{\rm disc}$ & [0; 0.4] & 30 & 0.265$\pm$0.056 & \textbf{0.024}& $\perp$ & [0; 0.3] &[1.5; 3] & \textbf{0.007}\\
& & [0.4; 0.6] & 99 & 0.499$\pm$0.028 & 0.915 & & & & \\
& & [0.6; 1.5] & 208 & 0.459$\pm$0.021 & 0.584 & & & & \\
& & [1.5; 3] & 17 & 0.655$\pm$0.069 & \textbf{0.037} & $\parallel$ & & &\\
\midrule
Bulge & $\Delta{\rm PA}_{\rm bulge - disc}$ or $\delta{\rm PA}_{\rm bulge - disc}$& <30$\degree$ or <3$\sigma$ & 137  &0.431$\pm$0.024 & \textbf{0.018} & $\perp$ & Bulge & Disc & 0.989\\
Disc & $\Delta{\rm PA}_{\rm bulge - disc}$ or $\delta{\rm PA}_{\rm bulge - disc}$& <30$\degree$ or <3$\sigma$ & 137  &0.410$\pm$0.026 & \textbf{0.022} & $\perp$ & &  & \\
\bottomrule
\end{tabular}
}
\label{PAstellResultsBulgeDisc}
\end{table*}

\subsection{Gas spin--filament alignments}
\label{Gas spin--filament alignments}

\begin{figure}
\includegraphics[scale=0.225]{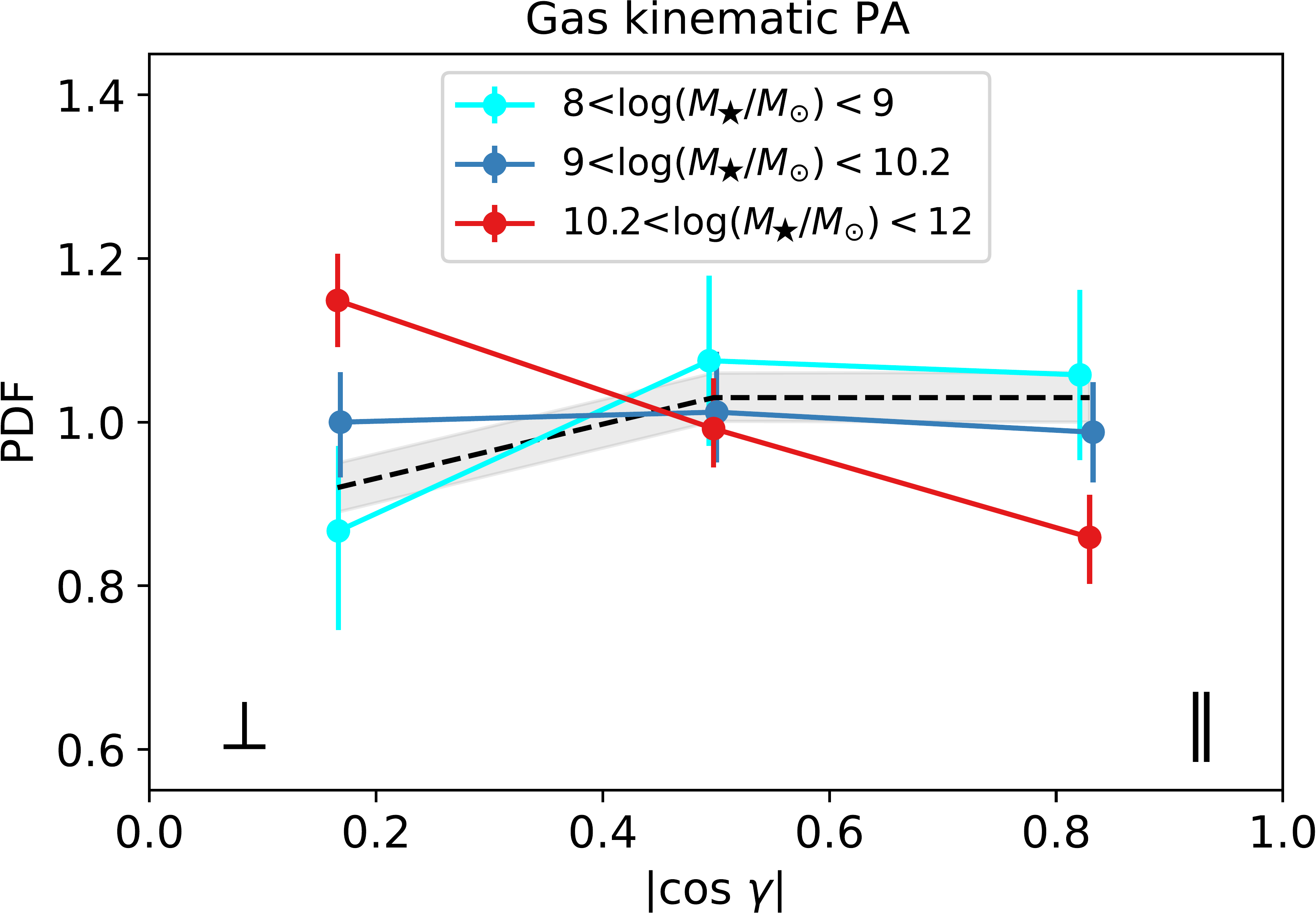}
\caption{PDFs of the spin--filament alignments for 1301 SAMI galaxies divided in stellar mass. The galaxy spin axes are identified using gas kinematic PAs. Galaxies with $8<\log(M_\star/M_{\odot})<9$ show a tendency to more parallel alignments, but it is not statistically significant.}
\label{TrendsStellarMassGasPAs}
\end{figure}

We explore the orientation of galaxy spin axes identified using the gas kinematic PAs in Section~\ref{Gas kinematics} and the method described in Section~\ref{Orientation of the spin axes}. This allows us to study the spin--filament alignments for 180 low-mass SAMI galaxies with $8<\log(M_\star/M_{\odot})<9$ in addition to the 1121 SAMI galaxies with $9<\log(M_\star/M_{\odot})<12$. Figure~\ref{TrendsStellarMassGasPAs} shows the PDFs of the spin--filament alignments for the 1301 SAMI galaxies divided into $M_\star$ ranges. The results from the K-S tests are listed in Table~\ref{PAstellResults} and marked with ${\rm PA}_{\rm gas}$. Galaxies with $8<\log(M_\star/M_{\odot})<9$ tend to show a parallel tendency, however the distribution is not statistically different from uniform. Galaxies with $9<\log(M_\star/M_{\odot})<10.2$ are also consistent with a uniform distribution, while the result is statistically significant for the most massive galaxies, which have more perpendicular alignments. The |$\cos\gamma$| distribution for $8<\log(M_\star/M_{\odot})<9$ is significantly different to the one for $10.2<\log(M_\star/M_{\odot})<12$ ($p_{\rm 2\,K-S}=0.022$). 

These findings are in agreement with those obtained in Section~\ref{Trends with stellar mass, B/T and mass bulge} based on stellar kinematic PAs (see the top panels of Figure~\ref{TrendsGalaxyProperties} and Table~\ref{PAstellResults}). This is expected, since $\sim80$\% of the galaxies have aligned ${\rm PA}_{\rm stars}$ and ${\rm PA}_{\rm gas}$, and we do not detect any correlation between |$\cos\gamma$| and $\Delta{\rm PA}$ in Section~\ref{Correlations with galaxy properties}. Overall, these results further tie spin--filament alignments to accretion. 

\section{Discussion}
\label{Discussion}

In this Section we discuss our results with respect to previous work, the tracers of the physical processes regulating spin--filament alignments and possible formation scenarios for galaxies, bulges and discs. Finally, we address the caveats of this study.

\subsection{Comparison with previous studies}

We find that the mass of the bulge, defined as $M_{\rm bulge}=M_\star\times$ (B/T), is the primary parameter of correlation with the flipping of the spin--filament alignment from parallel to perpendicular: galaxies with lower-$M_{\rm bulge}$ tend to have their spin aligned in parallel with respect to the closest filament, while galaxies with higher-$M_{\rm bulge}$ show a more perpendicular tendency. This result highlights that neither $M_\star$ alone nor B/T alone can fully unravel spin--filament alignments. Accordingly \citet{Kraljic2020} found that the correlation of the signal with the galaxy morphology cannot be totally explained by its dependence on stellar mass.

The analyses of the spin--filament alignments as a function of $\lambda_e$, stellar age, visual morphology, star formation characteristics, kinematic morphology and local environment show similar trends compared to those with $M_{\rm bulge}$, although with lower statistical significance (see Table~\ref{PAstellResults}). These galaxy properties are secondary tracers of the spin--filament alignments and their correlation is explained by their dependency with $M_{\rm bulge}$. Overall, our results are in agreement with the previous studies based on shape as a proxy for spin \citep{Tempel2013a,Tempel2013b,Pahwa2016,Hirv2017,Chen2019}, in projection for the SAMI survey \citep{Welker2020} and from 3D spins for the MaNGA survey \citep{Kraljic2021}. We also analyse shape--filament alignments for the 1121 SAMI galaxies and for 735 bulges and discs in Appendix~\ref{shape--filament alignments}, where the spin axes are identified using photometric PAs. We observe similar results to spin--filament alignments, but with weaker statistical significance. 

Focusing on 216 S0 galaxies (Section~\ref{S0 galaxies}), we find a preference for perpendicular alignments, with the signal dominated by high-mass galaxies and galaxies with kinematically misaligned stellar and gas components. \citet{Kraljic2021} found a more perpendicular tendency for low-mass S0 MaNGA galaxies, at odds with our result. The discrepancy might be explained by the different criteria used to select S0 galaxies, underlying the importance of sample selection, as well as by other galaxy properties, such as massive bulges giving rise to more perpendicular alignments.

\subsection{Tracers of physical processes}

Our findings suggest that the flipping of the spin--filament alignment is most strongly related to the growth of a bulge in the galaxy, traced by $M_{\rm bulge}$. Mergers have been found to be responsible for both driving the alignment flips  \citep{Codis2012,Dubois2014,Welker2014} and causing the formation of the bulge \citep{Sales2012,Wilamn2013}. $M_{\rm bulge}$ is constructed as the product of $M_{\star}$ and B/T, and it is the primary correlation parameter in a statistical sense. Since the combination of two distinct drivers is expected to show an even stronger correlation, $M_{\star}$ and B/T can be identified as distinct physical tracers of the spin--filament alignments, in agreement with theoretical studies \citep{Welker2014,Codis2015,Welker2017,Lagos2018,Kraljic2020}. B/T shows strongest correlations with respect to $M_{\star}$ (see the right panel of Figure~\ref{ExplainedVariance}, Tables~\ref{SpearmanResults} and \ref{PAstellResults}), in agreement with the fact that B/T traces mergers, in particular gas-rich major mergers, which are the driving mechanism of the flipping. However, a residual independent trend is still detected for $M_{\star}$ (see Section~\ref{Correlations with galaxy properties}), which is small due to its dependency on B/T. From a physical perspective, $M_{\star}$ traces the galaxy position in the cosmic web and the overall accretion independently from merger activity, and it brings additional information about the direction of accretion.

$M_{\rm bulge}$ has also been identified as a key parameter for tracing star formation quenching mechanisms \citep{Lang2014,Bluck2014,Bluck2022,Dimauro2022}: these works found that quenching processes are most strongly related to the bulge. Velocity dispersion is even a stronger tracer \citep{Bluck2020,Brownson2022,Bluck2022}, suggesting that AGN feedback is the main quenching mechanism via feeding the black hole with gas and preventing gas accretion into the galaxy. We find that the spin--filament alignments correlate with the galaxy velocity dispersion estimated within 1\,$R_e$ ($\rho=-$0.08, $p_{\rm S}=0.012$), but $M_{\rm bulge}$ is still the primary parameter. This highlights a different dynamic, where the flipping of the spin--filament alignment is mainly related to star accretion. The result is in agreement with the finding from simulations that galaxies with AGNs show stronger perpendicular tendencies because they tend to belong to more massive pressure-supported galaxies \citep{Soussana2020}. We aim to investigate the role of AGNs in spin--filament alignments within the SAMI Galaxy Survey in an upcoming study (Barsanti et al., in preparation).

We note that the tendency to perpendicular spin--filament alignments is reproduced for all the galaxy properties, while significant results for parallel alignments are only found for $M_{\rm bulge}$ and B/T (see Table~\ref{PAstellResults}). The perpendicular spin--filament alignment is expected to be the most robust tendency at low redshift \citep{Pahwa2016,Chen2019}, since most galaxies will have undergone mergers of some kind, causing a flip from the strong parallel spin--filament alignment acquired at high redshift \citep{Dubois2014,Laigle2015,Codis2018}. The stronger perpendicular tendencies, especially for central galaxies, might also be due to the fact that $\sim64$\% of our SAMI galaxies belong to galaxy groups, which are high galaxy density environments where the larger fraction of mergers is more likely to cause the flipping (e.g., \citealp{Bett2012,Welker2014}). However, our results stand even when selecting only the 408 isolated galaxies ($M_{\rm bulge}$ is the primary correlation parameter with $\rho=-$0.12, $p_{\rm S}=0.006$), highlighting that filament-related processes are the dominant players in regulating the spin alignments.

\subsection{Formation scenarios for galaxies, bulges and discs}

Our findings point to a scenario where bulge-dominated galaxies, showing more perpendicular tendencies, are predominantly assembled through mergers occurring along the filament. Disc-dominated galaxies have more parallel alignments and they are mainly formed from gas accretion. S0 galaxies might belong to different galaxy populations according to $M_\star$ ($p_{\rm 2\,K-S}=0.043$): high-mass S0s show perpendicular alignments and have more de~Vaucouleurs-like bulges, suggesting to be preferentially formed via mergers. These results are in agreement with the findings of \citet{FraserMcKelvie2018} for the MaNGA survey, who suggested processes such as mergers for the formation of high-mass S0 galaxies and a faded spiral scenario for low-mass S0 galaxies.

In Section~\ref{spin--filament alignments of bulges and discs} we explore the separate spin--filament alignments of 468 bulges and 516 discs. To our knowledge, this is the first time that such a study has been conducted in the observations. Bulges, especially more massive or dispersion-dominated mainly having a de~Vaucouleurs profile, show more perpendicular alignments. Thus, they are more likely to be formed via mergers, in agreement with the expected formation channel for bulges (e.g., \citealp{Barnes1988}). Rotation-dominated bulges, typically having $0<n_{\rm bulge}<2$, show a more parallel tendency and they could represent pseudo-bulges formed via secular processes, in agreement with the formation scenario proposed by \citet{Kormendy2004}. They tend to have low bulge masses compared to dispersion-dominated bulges, as expected from the dependency of the signal on $M_{\rm bulge}$. This finding highlights how the alignments are influenced by sample selection, and that a larger number of bulges is needed in order to further investigate the impact of kinematics and morphology on their spin--filament alignments. We also note that strongly rotation-dominated bulges might be mis-classified discs due to the degeneracies of the 2D bulge/disc decompositions. We address this caveat in the next Section.

Our results for discs are consistent with multiple formation and evolution mechanisms at low redshift. Discs in low-bulge mass galaxies or rotation-dominated discs have mainly parallel tendencies, suggesting gas accretion as formation channel. Discs in high-bulge mass galaxies or with low $(V/\sigma)_{\rm disc}$ values show perpendicular alignments, indicating mergers. Several previous works, particularly in simulations, have shown that the formation and the evolution of the disc are heavily affected by tidal forces, mergers, and in situ instabilities \citep{Ostriker1989,Okamoto2005,RomanoDiaz2009,Scannapieco2009}. These processes can lead to the destruction of the disc and its formation at a later stage, as well as they can flip the disc spin--filament alignment. 

The kinematic misalignment between bulges and discs for 59 galaxies might suggest different physical processes acting on the components (e.g., \citealp{Chilingarian2009}), especially since environmental mechanisms affect mainly discs compared to bulges (e.g., \citealp{Barsanti2021}). However, we do not find significant tendencies for the spin-filament alignments of both components, which are consistent with uniform distributions. In order to understand this misalignment it is crucial to take into account the limitations of bulge-disc decomposition and the large uncertainties of the bulge PAs (see Section~\ref{Caveats}).

Overall, studying separately the spin--filament alignments of bulges and discs provides further clues in terms of the pathways that gave rise to the corresponding galaxies. This outcome is in accord with the conclusion of \citet{Jagvaral2022}, who investigated the intrinsic alignments, i.e. the tendency of galaxies to coherently align with the density field and produce correlations among galaxy shapes, separately for bulges and discs. Intrinsic alignments might bias weak lensing measurements and the estimate of cosmological parameters. \citet{Jagvaral2022} concluded that the stellar dynamics of the two galaxy components play a significant role in determining the intrinsic alignments.

\subsection{Caveats}
\label{Caveats}
We address the three main caveats to the results obtained in this study: (i) the definitions of bulges and discs, (ii) the 3D modelling used to identify the spin axes
and (iii) the small galaxy samples and weak statistical significance.

We assume that galaxies are characterised by two components: a central bulge and a surrounding disc. However, bulge/disc decompositions do not respect the entire galaxy complexity. In particular, the photometric profiles show limitations and degeneracies \citep{Head2015,Fischer2019,Barsanti2021a,Papaderos2022,Sonnenfeld2022}. The disc exponential component is dominated by the galaxy outskirts and it is not able to completely capture the central region \citep{Mendez2019b, Breda2020}. The bulge S\'ersic component might wrongly identify parts of discs or bars. In particular, bulges could be inner parts of discs, since even a late-type galaxy can have $(V/\sigma)<0.8$ for the central region \citep{vandeSande2018}. We explore more deeply the 25 strongly rotation-dominated bulges with $0.8<(V/\sigma)_{\rm bulge}<1$ by visually inspecting the photometric bulge-disc decomposition matched to the kinematic bulge-disc decomposition and by comparing the effective radii from the double-component S\'{e}rsic bulge plus exponential disc fit: 8/25 bulges are suspected discs. Thus, we need to be cautious in the interpretations of the galaxy components and their formation scenarios. 

\begin{figure}
\centering
\includegraphics[scale=0.28]{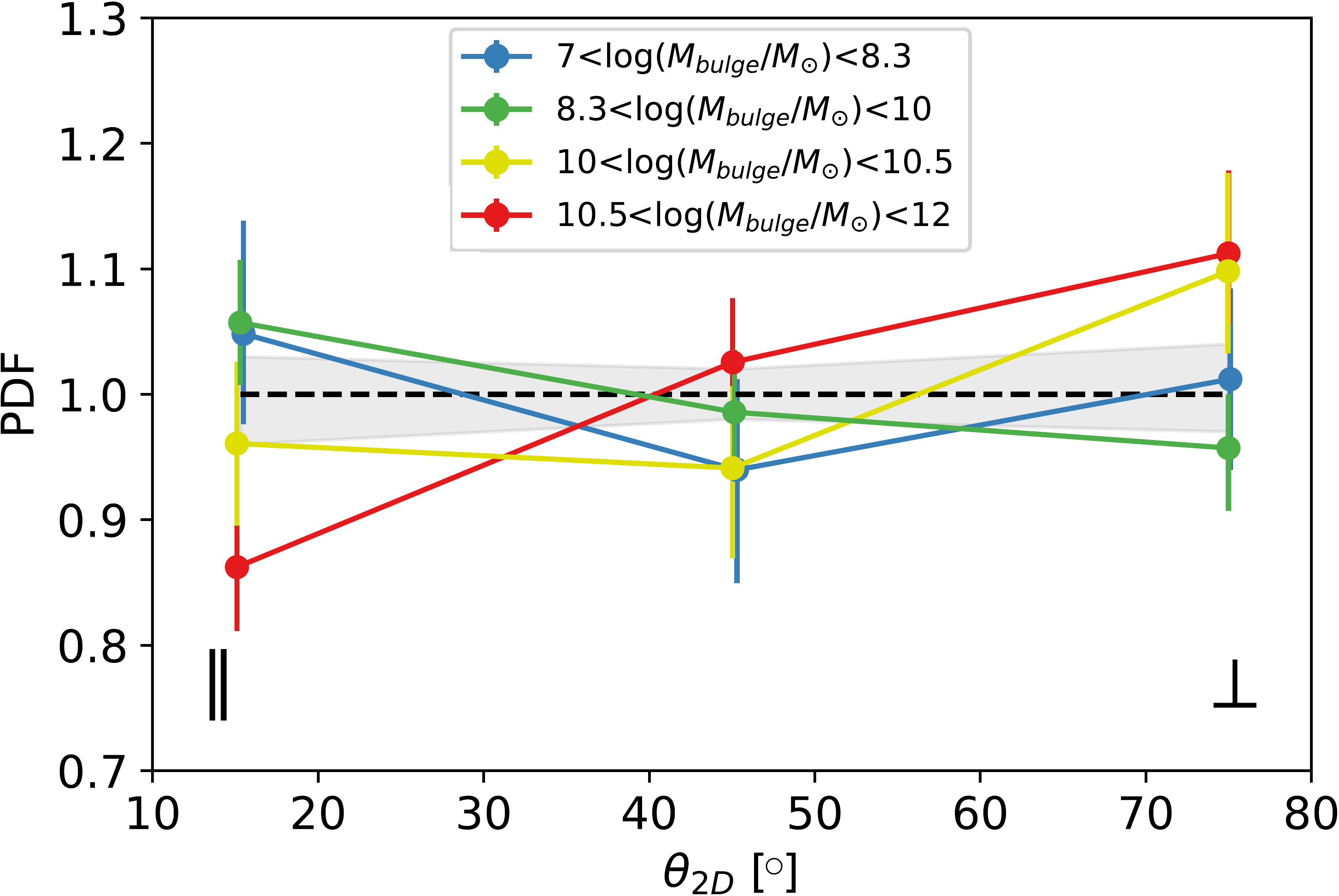}
\caption{PDFs of the spin--filament alignments  estimated via a 2D modelling, i.e. the 2D angles between the kinematic position angles and the projected directions of the associated filaments to the galaxies. The 1121 SAMI galaxies are divided according to $M_{\rm bulge}$: galaxies with high $M_{\rm bulge}$ tend to have the spins perpendicularly aligned with respect to the closest filament, while low-$M_{\rm bulge}$ galaxies show a tendency towards parallel alignments. These findings are consistent with our results for the 3D modelling of spin--filament alignments.}
\label{MassBulgeTrends2D}
\end{figure}

To identify the spin axes we apply the 3D thin-disc approximation, in order to consistently compare results for galaxies, bulges and discs. Our conclusions do not change if we follow a 2D modelling of the spin--filament alignments as in \citet{Welker2020}, by estimating the angle between the kinematic position angle and the projected direction of the associated filament to the galaxy. The mass of the bulge is still found to be the primary parameter of correlation, although with weaker statistical significance ($\rho=-0.07$; $p_{\rm S}=0.020$) with respect to the result from the Spearman test for the 3D thin-disc approximation. The PDFs of the 2D angles as a function of $M_{\rm bulge}$ are shown in Figure~\ref{MassBulgeTrends2D}. The ranges of the PDFs are consistent with those of Figure 6 in \citet{Welker2020}. `Classical' bulges (i.e.\ completely dispersion-dominated bulges) are excluded in our analysis of the bulge spin-filament alignments, since they do not show regular velocity maps and kinematic PAs. However, in a 2D analysis we expect their projected minor axis to be perpendicularly aligned with respect to the closest cosmic filament, in agreement with the previous studies on shape for elliptical galaxies \citep{Tempel2013b,Pahwa2016}.

Finally, the spin--filament alignments recovered in this study are relatively weak preferences (even when statistically significant) and based on limited galaxy samples. Bulge and disc tendencies are based on  $\sim$200-100 galaxies with a minimum sample of 17 discs. We note that the observational study of spin--filament alignments is still a challenging field at low redshift due to the intrinsic weakness of the signal \citep{Codis2018}. Thus, our results provide only hints for galaxy evolution mechanisms and larger samples are required to disentangle internal effects of stellar mass and morphology, to assess the role of the environment, and to take into account effects of sample selection. Larger samples are also needed to confirm our tendencies for the bulges and discs. Nevertheless, this work highlights that the separate study of the spin--filament alignments for the galaxy components is a powerful tool to constrain their different formation scenarios.

\section{Summary and conclusions}
\label{Summary and conclusions}

We study the alignment of the spin axis of galaxies, bulges, and discs with respect to the orientation of the closest cosmic filament. These analyses shed light on the formation mechanisms of the galaxies and their structural components. We explore the alignments as a function of various galaxy properties, with the goal of understanding which property shows the strongest correlation.  

We take advantage of the SAMI Galaxy Survey to identify galaxy spin axes based on spatially-resolved stellar kinematics. The deep and highly complete GAMA spectroscopic survey is used to reconstruct the underlying cosmic filaments, implementing the {\sc DisPerSe} structure extractor. We exploit the 2D bulge/disc kinematic decomposition (where possible) to identify the separate spin axes of bulges and discs. Our sample comprises 1121 SAMI galaxies with $\log(M_\star/M_{\odot})>9$, amongst which 468 bulges and 516 discs have reliable measurements. We carry out the analyses of the spin--filament alignments in 3D. Our results are summarised as follows:
\begin{enumerate}
    \item[(i)] Using the Spearman test, we find statistically significant correlations of the galaxy spin--filament angle |$\cos\gamma$| with $M_\star$ ($p_{\rm S}=0.014$), B/T ($p_{\rm S}=10^{-4}$) and $M_{\rm bulge}$ ($p_{\rm S}=10^{-5}$). Much of the strongest correlation is detected for $M_{\rm bulge}$, which on its own explains $\sim$70\% of the variance in |$\cos\gamma$| and it can be identified as the primary parameter to correlate with spin--filament alignments.
    \item[(ii)] Galaxies with lower $M_{\rm bulge}$ tend to have their spins aligned parallel to the closest filament ($p_{\rm K-S}=0.008$), while galaxies with higher $M_{\rm bulge}$ have perpendicular alignments ($p_{\rm K-S}=1\times 10^{-5}$). 
    \item[(iii)] $M_\star$ or B/T alone cannot fully unravel spin--filament alignments. Since their product $M_{\rm bulge}=M_\star\times$ (B/T) shows the strongest correlation, $M_\star$ and B/T can be identified as two distinct physical tracers of spin-filament alignments.  
    \item[(iv)] Other galaxy properties, such as visual morphology, kinematic features, stellar age, star formation activity and local environment, correlate with spin--filament alignment due to their dependency on
    $M_{\rm bulge}$ and can be identified as secondary tracers. 
    \item[(v)] S0 galaxies have a significant perpendicular alignment ($p_{\rm K-S}=0.007$). The signal is dominated by high-mass S0 galaxies with $10.2<\log(M_\star/M_{\odot})<12$  ($p_{\rm K-S}=8\times10^{-4}$), in agreement with the expected $M_\star$-dependency, and by S0 galaxies with the stellar and gas components misaligned ($p_{\rm K-S}=0.013$). A two-sample K-S test indicates high-mass S0 galaxies have a different spin--filament alignment ($p_{\rm 2\,K-S}=0.043$) and a more de~Vaucouleurs-like bulge ($p=0.002$) relative to low-mass S0 galaxies.
    \item[(vi)] Analysis of the separate spin--filament alignments of bulge and disc components reveals that bulges tend to be aligned more perpendicular to the closest filament ($p_{\rm K-S}=0.001$). This tendency is seen for galaxies with both low- and high-$M_{\rm bulge}$, and for dispersion-dominated bulges, which mostly have de~Vaucouleurs-like profiles. Rotation-dominated bulges tend to have parallel spin--filament alignments and mostly $n_{\rm bulge}<2$. The discs show a tendency towards parallel alignments for low-$M_{\rm bulge}$ galaxies ($p_{\rm K-S}=0.030$) and for rotation-dominated discs ($p_{\rm K-S}=0.037$), while they have more perpendicular alignments for high-$M_{\rm bulge}$ galaxies ($p_{\rm K-S}=0.025$) and low $(V/\sigma)_{\rm disc}$ values ($p_{\rm K-S}=0.024$). Galaxies with both a bulge and a disc tend to have perpendicular alignments for both components. 
    \item[(vii)] We obtain consistent findings using spatially-resolved gas kinematics (rather than stellar kinematics) for the identification of the galaxy spin axes. Similar results are observed for shape--filament alignments using photometric PAs for galaxies, bulges, and discs. 
\end{enumerate}

In conclusion, we find an observational link between galaxy spin--filament alignments and the growth of the bulge. This link can be explained by mergers, which can cause the flipping and the bulge assembling, as seen in galaxy formation simulations. Both B/T and $M_\star$ are needed to fully unravel spin--filament alignments and from a physical perspective, they are independent tracers of the involved physical processes: B/T traces the amount of mergers a galaxy might have experienced, which are the driving mechanisms of flipping and changing the galaxy angular momentum \citep{Welker2014,Welker2017,Lagos2018}; $M_\star$ traces the galaxy position in the cosmic web, where galaxies need to be very close to filaments to start experiencing lots of mergers along it \citep{Codis2015}.

Additional clues about the processes involved in changing the spin alignment relative to the closest cosmic filament can be discovered by studying the separate spin--filament alignments of the galaxy components, such as bulges and discs. This demonstrates that integral field spectroscopy (IFS) tools, such as spatially-resolved stellar kinematic bulge/disc decomposition and gas kinematics, offer powerful information for the study of spin--filament alignments. However, at present we can only derive suggestive hints in the context of galaxy formation scenarios, although they provide a consistent picture in accord with simulations, due to the relatively small number of galaxies involved in the analyses and the weak statistical significance of the results. Upcoming IFS galaxy surveys, such the Hector survey \citep{Bryant2020}, will be able to draw stronger conclusions from spin--filament alignments regarding the physical mechanisms leading to the formation of galaxies, bulges, and discs, as well as to constrain the roles of local and global environments in determining galaxy spins.

\section*{Acknowledgements}

We thank the referee for the constructive report. SB would like to thank Luca Cortese for insightful comments. This research was supported by the Australian Research Council Centre of Excellence for All Sky Astrophysics in 3 Dimensions (ASTRO 3D), through project number CE170100013. The SAMI Galaxy Survey is based on observations made at the Anglo-Australian Telescope. The Sydney-AAO Multi-object Integral field spectrograph (SAMI) was developed jointly by the University of Sydney and the Australian Astronomical Observatory, and funded by ARC grants FF0776384 (Bland-Hawthorn) and LE130100198. The SAMI input catalogue is based on data taken from the Sloan Digital Sky Survey, the GAMA Survey and the VST/ATLAS Survey. The SAMI Galaxy Survey is supported by the Australian Research Council Centre of Excellence for All Sky Astrophysics in 3 Dimensions (ASTRO 3D), through project number CE170100013, the Australian Research Council Centre of Excellence for All-sky Astrophysics (CAASTRO), through project number CE110001020, and other participating institutions. The SAMI Galaxy Survey website is http://sami-survey.org/. This study uses data provided by AAO Data Central (http://datacentral.org.au/). JJB acknowledges support of an Australian Research Council Future Fellowship (FT180100231). FDE acknowledges funding through the H2020 ERC Consolidator Grant 683184, the ERC Advanced grant 695671
“QUENCH” and support by the Science and Technology Facilities Council (STFC). JvdS acknowledges support of an Australian Research Council Discovery Early Career Research Award (project number DE200100461) funded by the Australian Government.

In this work we use the
\href{http://www.python.org}{Python} programming language
\citep{vanrossum1995}. We acknowledge the use of
{\sc \href{https://pypi.org/project/numpy/}{numpy}} \citep{harris+2020},
{\sc \href{https://pypi.org/project/scipy/}{scipy}} \citep{jones+2001},
{\sc \href{https://pypi.org/project/matplotlib/}{matplotlib}} \citep{hunter2007},
{\sc \href{https://pypi.org/project/astropy/}{astropy}} \citep{astropyco+2013}, {\sc \href{https://pypi.org/project/pyvista/}{pyvista}} \citep{Sullivan2019}, {\sc \href{https://pingouin-stats.org/}{pingouin}} \citep{Vallat2018} and {\sc \href{http://www.star.bris.ac.uk/~mbt/topcat/}{topcat}} \citep{taylor2005}.

\section*{Data availability}

The SAMI reduced data underlying this article are publicly available at \href{https://docs.datacentral.org.au/sami}{SAMI Data Release 3}
\citep{Croom2021}. Ancillary data comes from the \href{http://gama-survey.org}{GAMA Data Release 3}
\citep{Baldry2018}. 

\section*{Supporting information}

Supplementary figures are available online. 

\textbf{Figures S1, S2, S4 and S5}. PDFs and CDFs of the spin--filament alignments for the SAMI galaxies in ranges of $(V/\sigma)_{e}$ and $\lambda_e$, age, divided into group centrals, satellites and isolated galaxies, visual morphology, spectral classification and kinematic morphology.

\textbf{Figure S3}. The distribution of $\lambda_e$ and ellipticity for the SAMI galaxies, colour-coded according to $M_{\rm bulge}$.

\section*{Author Contribution Statement}

SB devised the project and drafted the paper. SO and SC performed kinematic and photometric bulge-disc decompositions, respectively. FDE performed MGE fits. SB reconstructed the cosmic web and measured spins for galaxies, bulges and discs. SB, MC and CW contributed to data analyses and interpretation of the results. JJB, SMC, JSL, SNR and JvdS provided key support to all the activities of the SAMI Galaxy Survey ('builder status'). All authors discussed the results and commented on the manuscript.    


\bibliographystyle{mnras}
\bibliography{biblioSAMI} 


\appendix

\section{Shape--filament alignments}
\label{shape--filament alignments}

We investigate the shape--filament alignments using the photometric PAs to identify the spin axes of galaxies, bulges, and discs. To find the spin orientation we apply the method in Section~\ref{Orientation of the spin axes}. The galaxy photometric PA is measured for the 1121 SAMI galaxies selected in Section~\ref{Galaxy sample and selection criteria} from the $r$-band single-component S\'{e}rsic model, as presented in Section~\ref{Photometric position angles}. 

\begin{figure}
\includegraphics[width=\columnwidth]{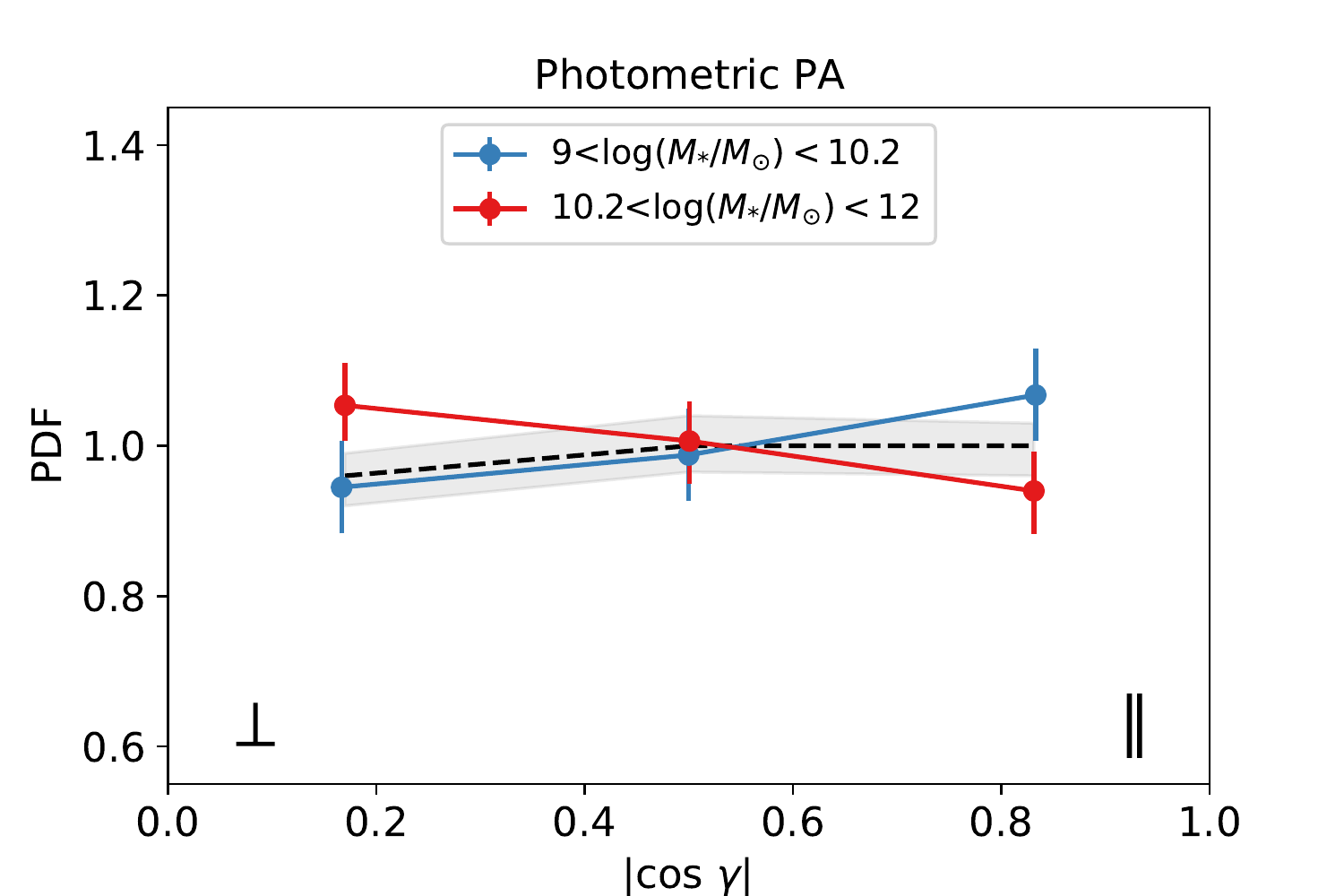}
\includegraphics[width=\columnwidth]{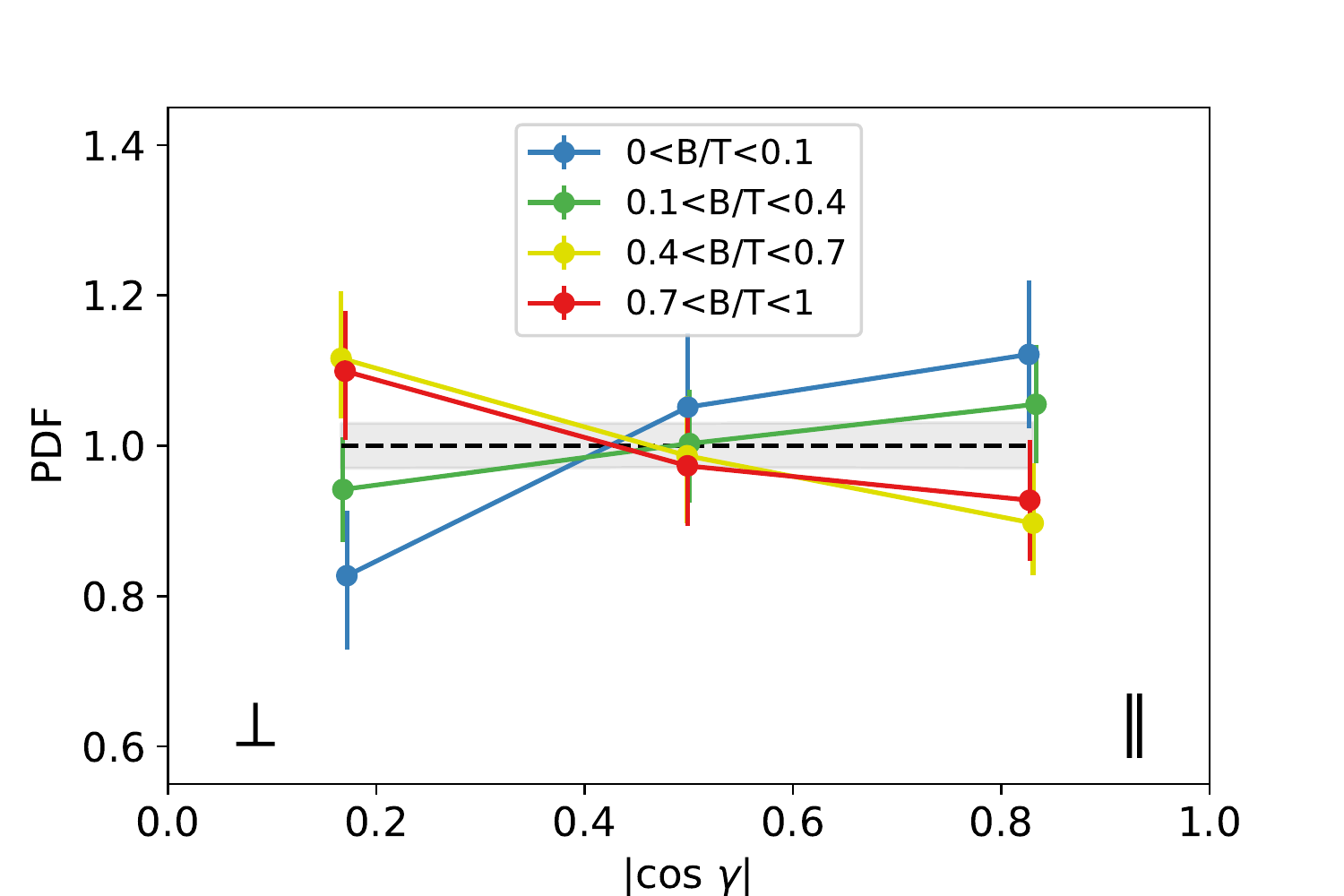}
\includegraphics[width=\columnwidth]{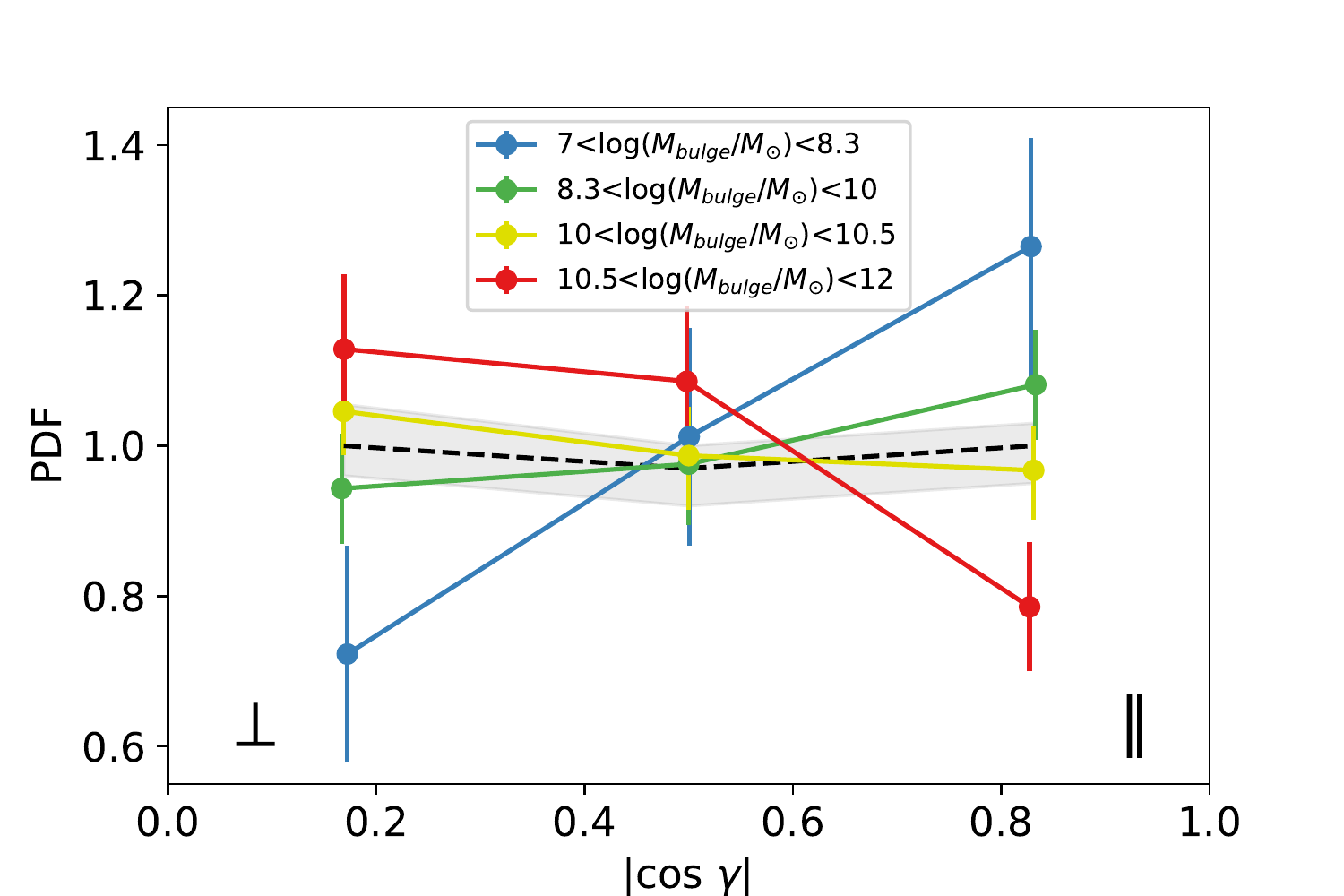}
\caption{PDFs of the shape--filament alignments for 1121 SAMI galaxies divided in $M_\star$ (top), B/T (middle) and $M_{\rm bulge}$ (bottom). We observe similar tendencies to those found for the spin--filament alignments in Figure~\ref{TrendsGalaxyProperties}.}
\label{TrendsGalaxyPropertiesPhotometricPA}
\end{figure}

Applying the Spearman test, we find significant correlations of the galaxy shape--filament alignments with $M_{\star}$ ($\rho=-0.06$; $p_{\rm S}=0.038$), B/T ($\rho=-0.07$; $p_{\rm S}=0.026$) and $M_{\rm bulge}$ ($\rho=-0.08$; $p_{\rm S}=0.005$). Figure~\ref{TrendsGalaxyPropertiesPhotometricPA} shows the shape--filament alignments as a function of these galaxy properties. The PDFs tend towards perpendicular alignments for more massive and bulge-dominated galaxies, while a tendency towards parallel alignments is mainly seen for low values of $M_\star$, B/T and $M_{\rm bulge}$. These findings are consistent with the results based on stellar kinematic PAs in Section~\ref{Trends with stellar mass, B/T and mass bulge}, although they show lower statistical significance when compared with a uniform distribution. A similar result has been found by \citet{Kraljic2021}: when comparing the results for the spin--filament alignments with those based on photometric PAs for the MaNGA survey, they observed qualitatively the same trends and weaker statistical significance for the latter.

\begin{figure}
\includegraphics[width=\columnwidth]{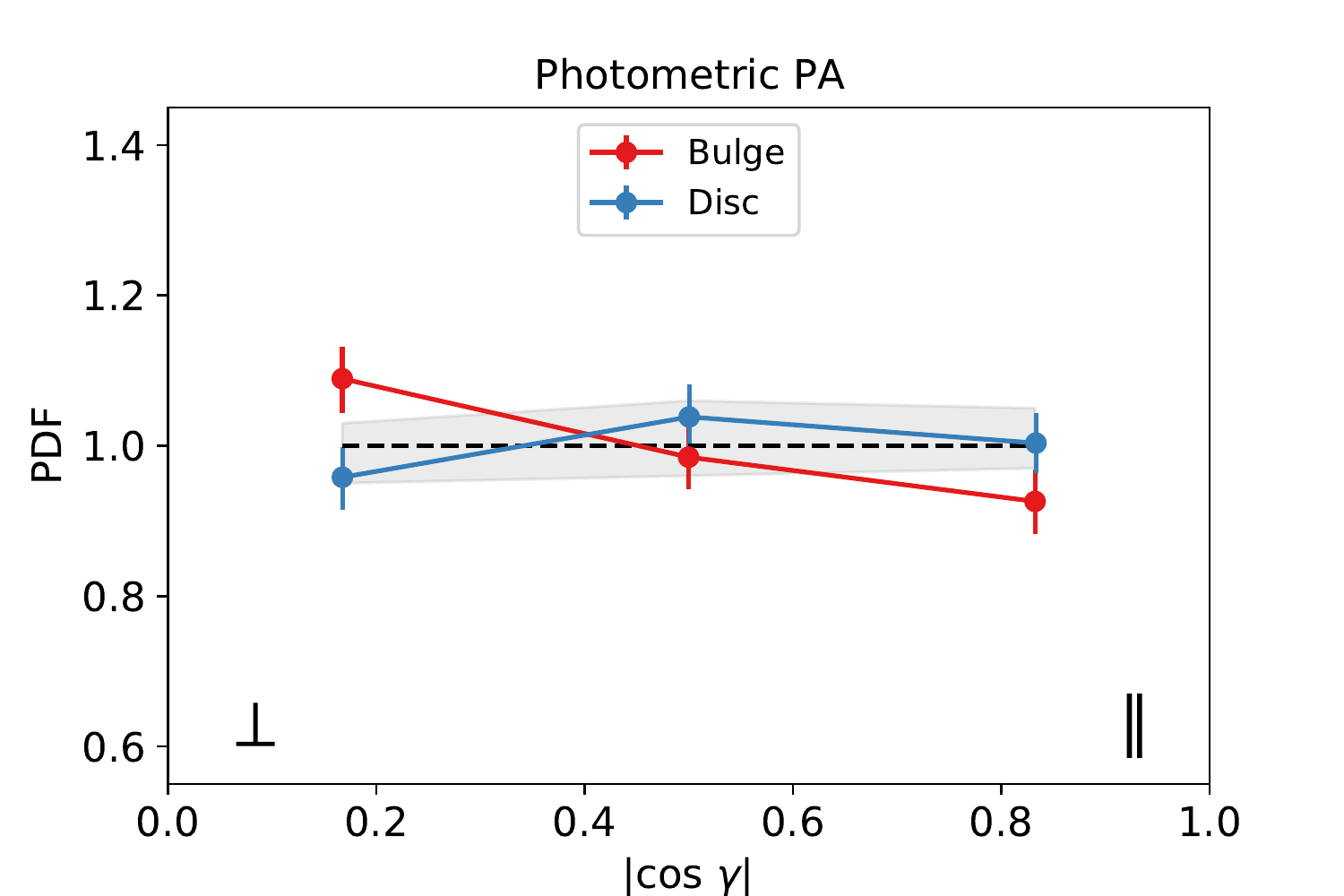}
\includegraphics[width=\columnwidth]{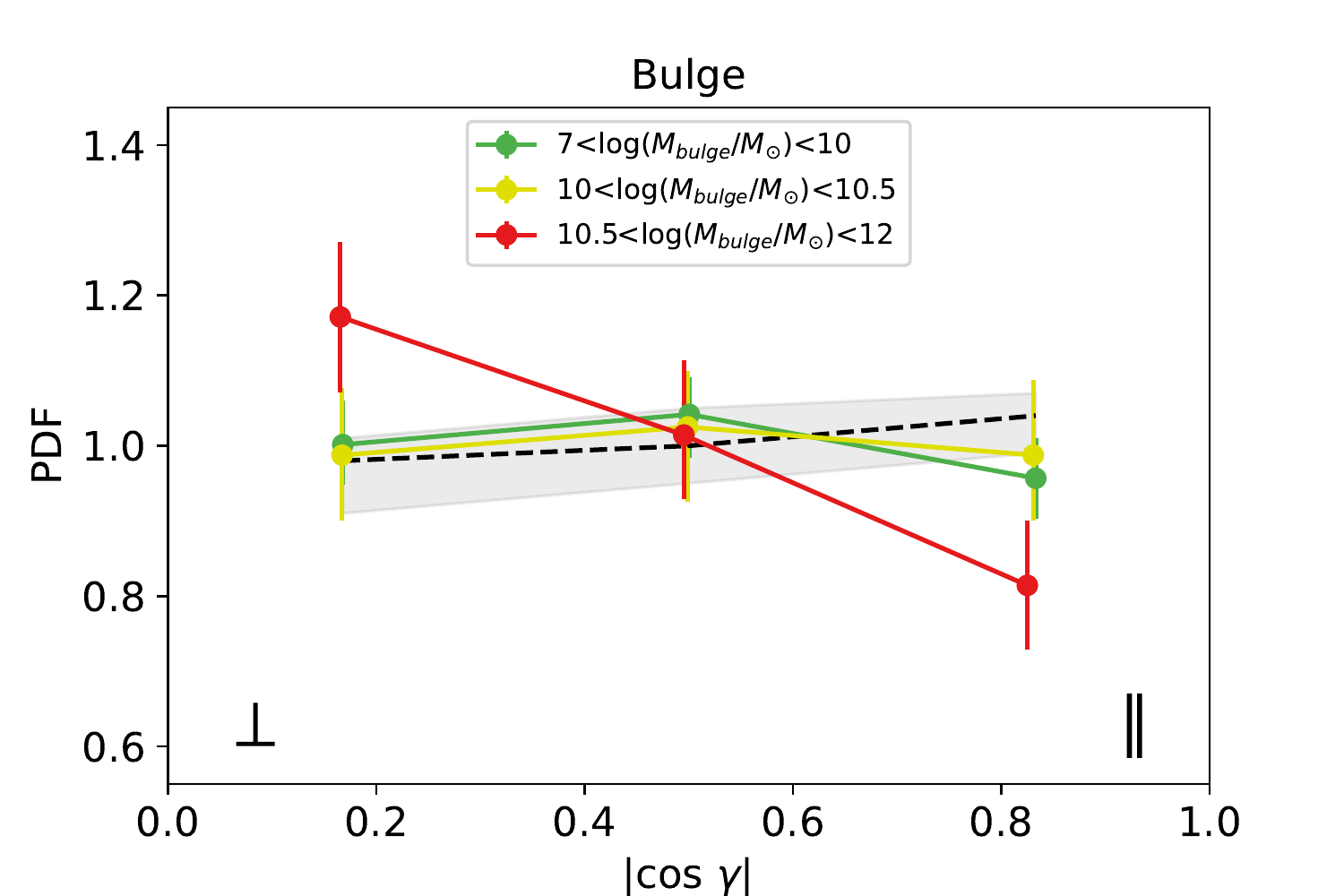}
\includegraphics[width=\columnwidth]{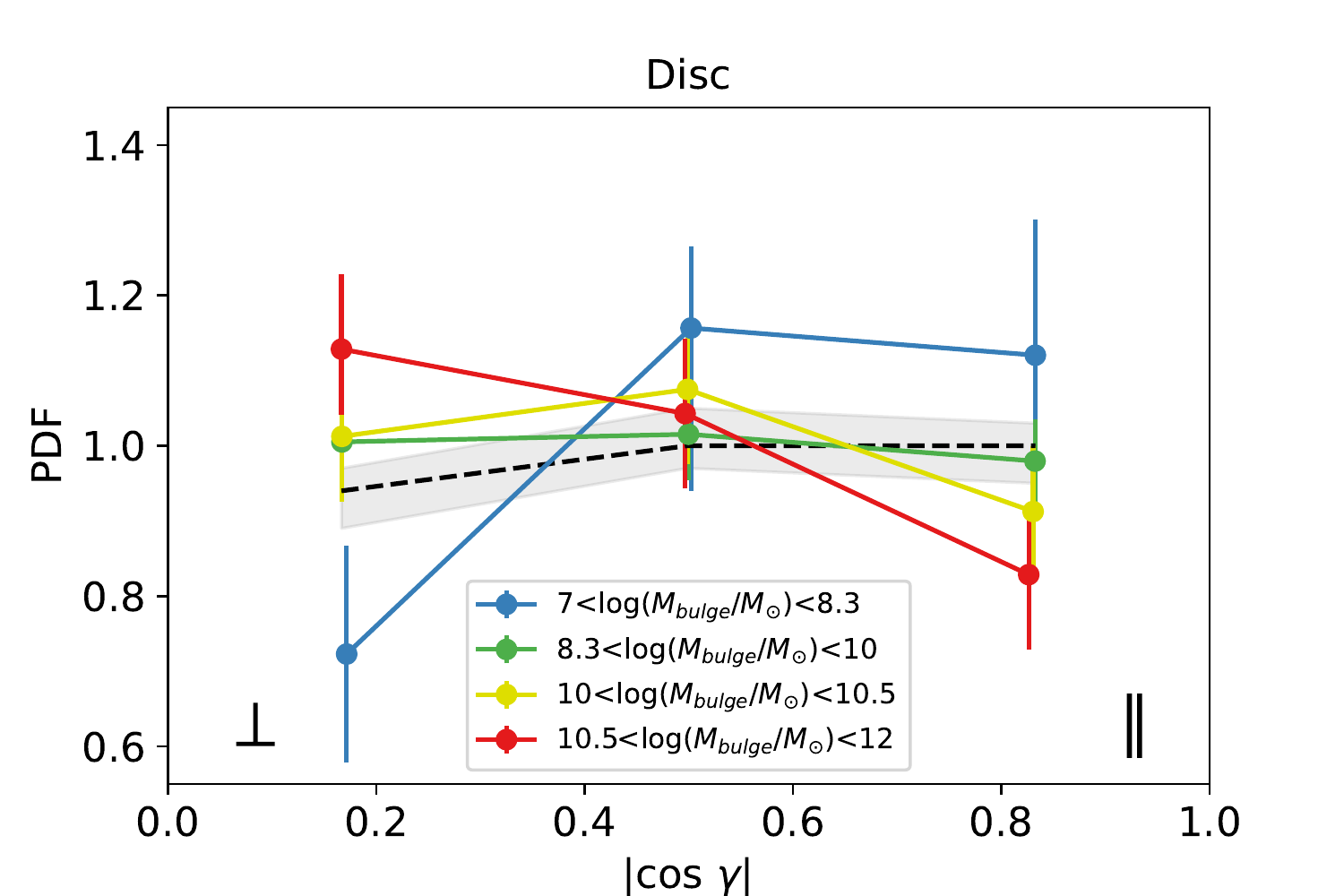}
\caption{PDFs of the shape--filament alignments for 735 bulges and discs, also divided into $M_{\rm bulge}$ ranges. The results are in agreement with the bulge and disc spin--filament alignments seen in Figure~\ref{TrendsBulgeDisc}.}
\label{TrendsBulgeDiscPhotometricPA}
\end{figure}

The photometric PAs for the bulges and the discs are estimated for 735 SAMI galaxies from the $r$-band double-component S\'{e}rsic bulge plus exponential disc model, as described in Section~\ref{Photometric position angles}. We select only galaxies that are best fitted by this model in order to measure reliable photometric PAs for the two galaxy components. Figure~\ref{TrendsBulgeDiscPhotometricPA} shows the PDFs for the shape--filament alignments of the bulges and the discs, also as a function of $M_{\rm bulge}$. Bulges show perpendicular tendencies, with a significant result for $10.5<\log{(M_{\rm bulge}/M_{\odot})}<12$ ($p_{\rm K-S}=0.042$). Discs tend to be aligned in parallel for lowest-$M_{\rm bulge}$ and perpendicularly for highest-$M_{\rm bulge}$ ($p_{\rm 2\,K-S}=0.025$). Overall, the results are consistent with the bulge and disc spin--filament alignments shown in Section~\ref{spin--filament alignments of bulges and discs}, but with lower statistical significance. Further useful constraints on shape--filament alignments for galaxies, bulges and discs could be obtained by exploiting the whole GAMA survey, and not just the limited SAMI sample used here for comparison with the spin--filament alignments.

\bsp	
\label{lastpage}
\end{document}